\documentclass[aps,prd,reprint,superscriptaddress,preprintnumbers,longbibliography]{revtex4-1}
\usepackage{amsmath,amssymb,bm,mathrsfs}
\usepackage[sort&compress]{natbib}
\usepackage{srcltx}
\usepackage{dsfont}
\usepackage{epsfig}
\usepackage{slashed}
\usepackage{bbold}
\usepackage{psfrag}
\usepackage[dvipsnames]{xcolor}
\PassOptionsToPackage{caption=false}{subfig}
\usepackage{subfig}
\usepackage{xfrac}
\usepackage{multirow}
\usepackage{booktabs}
\usepackage[colorlinks=true
,urlcolor=blue
,anchorcolor=blue
,citecolor=blue
,filecolor=blue
,linkcolor=red
,menucolor=blue
,linktocpage=true
,pdfproducer=medialab
,pdfa=true
]{hyperref}
\usepackage{cleveref}
\usepackage{setspace}
\usepackage{relsize}
\usepackage{tikz,tikz-feynman,enumerate,comment}

\usepackage[normalem]{ulem}
\usepackage[caption=false]{subfig}

\newcommand{\be}{\begin{equation}}
\newcommand{\ee}{\end{equation}}
\newcommand{\bea}{\begin{equation}\begin{aligned}}
\newcommand{\eea}{\end{aligned}\end{equation}}

\newcommand{\Fig}[1]{Fig.~\ref{#1}}
\newcommand{\Refc}[1]{Ref.~\cite{#1}}
\newcommand{\Refs}[1]{Refs.~\cite{#1}}
\newcommand{\Eq}[1]{Eq.~\eqref{#1}}

\newcommand{\ga}{g_{a\gamma\gamma}}
\newcommand{\gae}{g_{aee}}
\newcommand{\gaHB}{g_{a\gamma\gamma}^{\scriptscriptstyle \textrm{HB}}}
\newcommand{\DM}{{\scriptscriptstyle \textrm{DM}}}
\newcommand{\CMB}{{\scriptscriptstyle \textrm{CMB}}}
\newcommand{\gaDM}{g_{a\gamma\gamma}^\DM}
\newcommand{\tU}{t_{\scriptscriptstyle U}}
\newcommand{\RH}{{\scriptscriptstyle \textrm{RH}}}
\newcommand{\TRH}{T_\RH}
\newcommand{\dH}{d_{\scriptscriptstyle H}}

\begin{document}

\preprint{CERN-TH-2022-148}

\title{The Irreducible Axion Background}

\author{Kevin Langhoff}
\affiliation{Berkeley Center for Theoretical Physics, University of California, Berkeley, CA 94720, U.S.A.}
\affiliation{Theory Group, Lawrence Berkeley National Laboratory, Berkeley, CA 94720, U.S.A.}

\author{Nadav Joseph Outmezguine}
\affiliation{Berkeley Center for Theoretical Physics, University of California, Berkeley, CA 94720, U.S.A.}
\affiliation{Theory Group, Lawrence Berkeley National Laboratory, Berkeley, CA 94720, U.S.A.}

\author{Nicholas L. Rodd}
\affiliation{Theoretical Physics Department, CERN, 1 Esplanade des Particules, CH-1211 Geneva 23, Switzerland}

%%%%%%%%%%%%%%%%
\begin{abstract}
\noindent Searches for dark matter decaying into photons constrain its lifetime to be many orders of magnitude larger than the age of the Universe.
A corollary statement is that the abundance of any particle that can decay into photons over cosmological timescales is constrained to be much smaller than the cold dark-matter density.
We show that an {\it irreducible} freeze-in contribution to the relic density of axions is in violation of that statement in a large portion of the parameter space.
This allows us to set stringent constraints on axions in the mass range $100\rm \;eV - 100\; MeV$.
At $10\rm \; keV$ our constraint on a photophilic axion is $g_{a\gamma \gamma} \lesssim 8.1 \times 10^{-14}~{\rm GeV}^{-1}$, almost three orders of magnitude stronger than the bounds established using horizontal branch stars; at $100~{\rm keV}$ our constraint on a photophobic axion coupled to electrons is $\gae \lesssim 8.0 \times 10^{-15}$, almost four orders of magnitude stronger than present results.
Although we focus on axions, our argument is more general and can be extended to, for instance, sterile neutrinos. 
\end{abstract}
%%%%%%%%%%%%%%%%

\maketitle

%%%%%%%%%
% Intro %
%%%%%%%%%

The existence of an axion was first predicted to solve the problem of strong CP~\cite{Peccei:1977hh,Peccei:1977ur,Weinberg:1977ma,Wilczek:1977pj}; that the same particle could constitute dark matter (DM) was only realized years later~\cite{Preskill:1982cy,Abbott:1982af,Dine:1982ah}.
Accordingly, experimental searches for the axion are distinguished by whether the axion is all of DM, or instead simply a state that exists in the spectrum of the Universe.
The classic realization of this dichotomy is the haloscope and helioscope of \textcite{Sikivie:1983ip}.
In this \textit{Letter}, we consider the axions at the interface of the two paradigms.
At sufficiently large couplings, axions are produced from the thermal plasma and contribute to an {\it irreducible} fraction of the DM density.
Even when the relic density is gravitationally negligible, the strength of present constraints on DM decays can still probe this contribution and reach parameter space inaccessible to existing searches that consider stellar emission or cooling.
In short, axions that do not constitute the DM of our Universe can still be strongly constrained by DM searches.
We focus on axions that couple to standard model (SM) photons and electrons, through $-\tfrac{1}{4} \ga a (F \tilde{F})$ and $\tfrac{\gae}{2m_e} (\partial_{\mu} a)\, \bar{e} \gamma^{\mu} \gamma_5 e$, in order to demonstrate that this argument can lead to considerably stronger bounds in the mass range $100\rm \;eV - 100\; MeV$.
We emphasize that these are only examples of a broader observation; decaying DM searches can be highly sensitive to states that are not DM. 
The concept can be generalized to additional axion couplings and other light dark particles such as sterile neutrinos or dark photons.

Let us illustrate the essence of our argument with an example.
For $10~{\rm keV}$ axions that are not DM, one of the strongest constraints arises from considering anomalous energy loss from horizontal branch stars~\footnote{For $m_a \lesssim 10~{\rm eV}$ these limits were recently strengthened by a factor of $\sim$1.4~\cite{Dolan:2022kul}, and the constraint in our mass range can likely be similarly improved}, and requires $\ga \lesssim \gaHB \equiv 6.6 \times 10^{-11}~{\rm GeV}^{-1}$~\cite{Ayala:2014pea,Carenza:2020zil}.
As we will establish, there exists an irreducible freeze-in contribution to the relic axion energy density, $\rho_a$, even if the Universe only reheated to just above the temperature associated with Big Bang nucleosynthesis (BBN).
For $m_a = 10~{\rm keV}$, the relic density is approximately given by
\be
  \rho_a/\rho_\DM \simeq 10^{-4}\, (\ga/\gaHB)^2.
  \label{eq:rhoa10keV}
\ee
Although this fraction is small, constraints on $10~{\rm keV}$ axion DM are strong.
One of these constraints comes from observations with XMM-Newton for DM decaying to X-ray photons. These observations require $\tau_\DM \gtrsim 10^{29}~{\rm s} \sim 10^{11}\, \tU$ at this mass~\cite{Foster:2021ngm}, where $\tU$ is the present age of the Universe.
For axion DM, this implies $\ga \lesssim \gaDM\equiv 10^{-18}~{\rm GeV}^{-1}$, almost eight orders of magnitude smaller than $\gaHB$.
If the irreducible axion contribution has a decay rate to photons of $\tau_a^{-1}$, then to be consistent with the X-ray observations, it must satisfy $\rho_a/\tau_a \lesssim \rho_\DM/ \tau_\DM$, or $\rho_a/\rho_\DM \lesssim \tau_a/\tau_\DM = (\gaDM/\ga)^2$.
Combined with \Eq{eq:rhoa10keV}, this requires $\ga \lesssim 10^{-3}\, \gaHB$, considerably stronger than the original bound.

%%%%%%%%%%%%
% Figure 1 %
%%%%%%%%%%%%

\begin{figure*}[!t]
\centering
\includegraphics[width=1\textwidth]{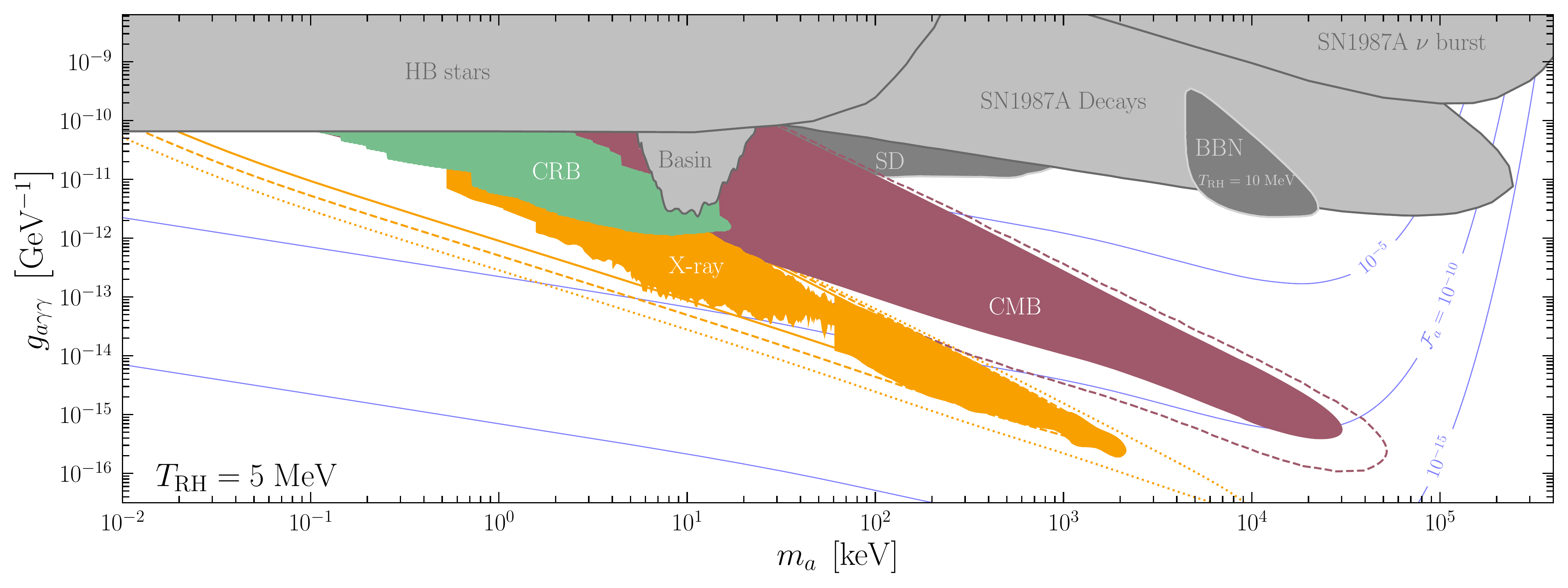}
\vspace{-0.8cm}
\caption{Constraints on the \textit{photophilic} axion parameters space, for the irreducible case of $\TRH=5~{\rm MeV}$.
The {\color[rgb]{0.972549,0.627451,0.}\textbf{mustard}} region represents the local X-ray constraints, rescaled to the case of irreducible freeze-in axion density~\cite{Boyarsky:2006fg,Boyarsky:2006ag,Boyarsky:2007ay,Foster:2021ngm,Ng:2019gch,Roach:2019ctw,Roach:2022lgo,Calore:2022pks}.
The sensitivity of possible future instruments on the lifetime of DM decaying locally to two photons, at the level of $10^{29}$, $10^{30}$, and $10^{31}\,{\rm s}$, are indicated as solid, dashed, and dotted lines respectively.
(Existing searches already achieve such sensitivities at several masses.)
In {\color[rgb]{0.52941176, 0.7372549 , 0.56862745}\textbf{lime}} we show the parameters constrained by not overproducing the CRB~\cite{Hill:2018trh}.
The {\color[rgb]{0.623529, 0.34902, 0.419608}\textbf{velvet}} region is the irreducible constraint derived from CMB anisotropies~\cite{Slatyer:2016qyl,Poulin:2016anj,Cang:2020exa,Bolliet:2020ofj}, which can be extended to the dashed region with the Simons array~\cite{Cang:2020exa}.
The {\color[HTML]{808080}\textbf{dark gray}} region shows the irreducible constraints determined from CMB spectral distortions~\cite{Balazs:2022tjl} and BBN~\cite{Baumholzer:2020hvx} (the latter was derived for $\TRH = 10~{\rm MeV}$).
{\color[HTML]{C0C0C0}\textbf{Light gray}} regions represent astrophysical constraints on the production of axions, in horizontal branch (HB) stars~\cite{Ayala:2014pea,Carenza:2020zil} (cf. \Refc{Dolan:2022kul}), in Supernova 1987A~\cite{Balazs:2022tjl,Jaeckel:2017tud,Masso:1995tw,Lucente:2020whw,Caputo:2022mah}, and in the Solar Basin~\cite{DeRocco:2022jyq}.
For reference, the {\color[rgb]{0.50196078, 0.50196078, 0.98039216}\textbf{blue}} contours indicate the irreducible axion fraction ${\cal F}_a$.
}
\vspace{-0.5cm}
\label{fig:Constraints}
\end{figure*}

\begin{figure}[!t]
\centering
\includegraphics[width=0.49\textwidth]{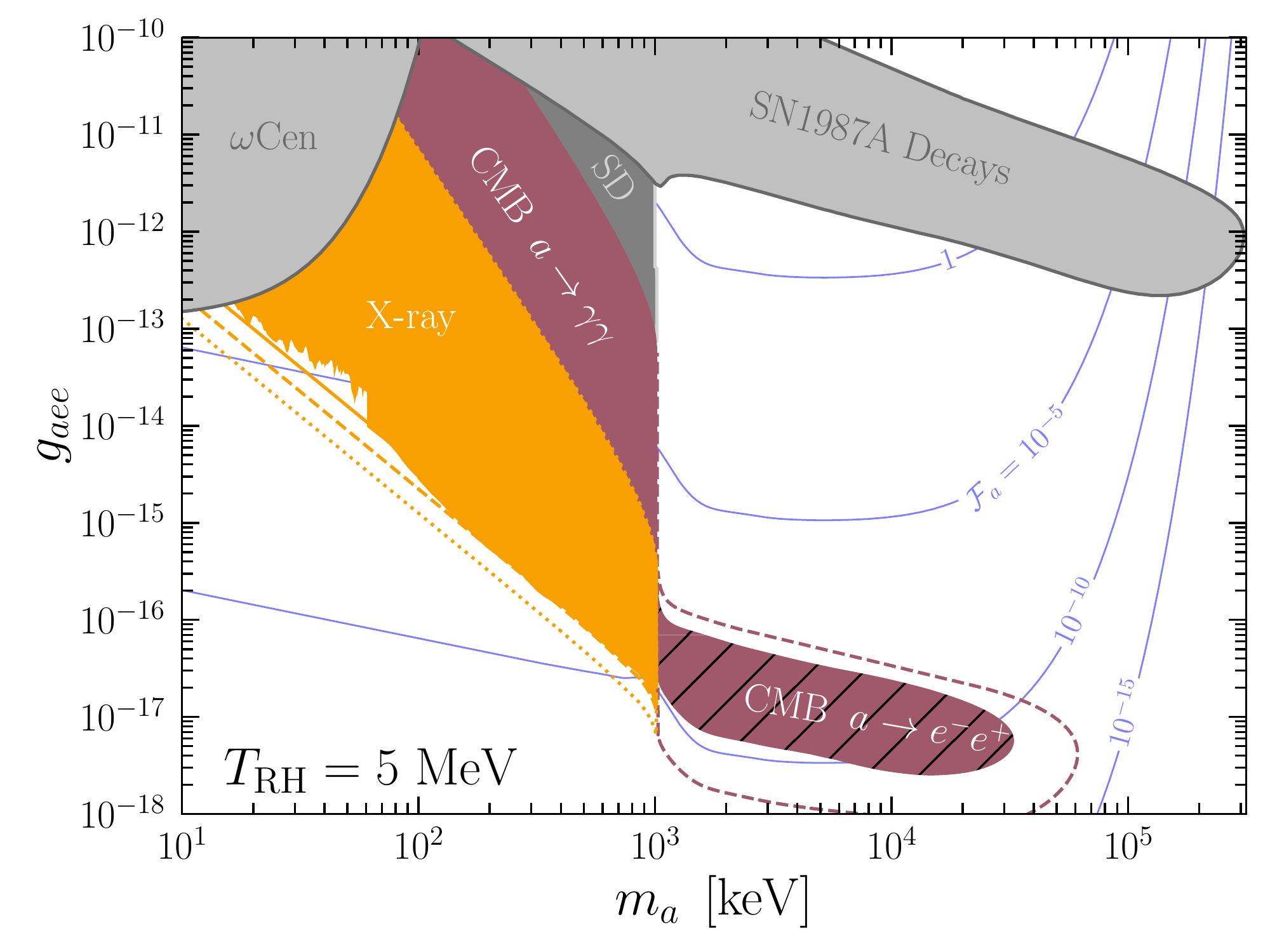}
\vspace{-0.8cm}
\caption{The analogue of \Fig{fig:Constraints} for a \textit{photophobic} axion.
The constraints and contours are as shown there, although here we also distinguish constraints from CMB anisotropies by whether they arise from decays to photons or electrons, with the latter shown as a black hashed region.
There is a clear impact at $m_a = 2 m_e$; just above threshold, the decays to $e^- e^+$ dominate, which shuts off the photon constraints by depleting the axion abundance. However, at lower couplings additional CMB constraints come from decays to electron-positron pairs.
{\color[HTML]{C0C0C0}\textbf{Light gray}} regions now represent prior axion constraints from Supernova 1987A~\cite{Ferreira:2022xlw} and red giants (RG) in the globular cluster $\omega$ Centauri~\cite{Capozzi:2020cbu}.
(The decoupling of the RG bound with the axion mass was calculated assuming RG core temperature of $8.6\,\rm keV$~\cite{Raffelt:1990yz}, the actual bound should be slightly stronger, cf. \Refc{Carenza:2020zil}.)
}
\vspace{-0.4cm}
\label{fig:Constraints_gae}
\end{figure}

The remainder of this work is dedicated to extending and formalizing the logic of the above argument, which will result in the constraints shown in \Fig{fig:Constraints} for photophilic axions ($\gae = 0$) and \Fig{fig:Constraints_gae} for the photophobic analogue ($\ga = 0$).
We will first outline the details of how an irreducible contribution to the axion energy density is produced in the early Universe, generalizing \Eq{eq:rhoa10keV}.
Then, we describe how constraints of DM can be recast to constrain these small axion energy densities.
When doing so, we will consider not only late time constraints on DM decay, but also probes from earlier epochs, in particular from not overproducing the cosmic radiation background (CRB) and bounds derived from the cosmic microwave background (CMB).
The need for measurements at earlier times arises for regions of parameter space where $\tau_a \ll \tU$ and the axion density in the present Universe is exponentially depleted.
Details of the analysis and additional/generalized constraints are reserved to the Supplementary Material.
Specifically, we show constraints on axions that couple simultaneously to photons and electrons, the full details of our relic density calculations, and the determination of constraints when there is a misalignment contribution to the axion density.
We also use sterile neutrinos as an example of the generality of the procedure of obtaining robust constraints from irreducible abundances.

%%%%%%%%%%
\vspace{0.2cm}
\noindent {\bf The Irreducible Axion Density.}
%%%%%%%%%%
%
Whenever the SM plasma contains particles that couple to axions and have energy $E \gtrsim m_a$, axions will be produced.
For sufficiently large couplings, the axion will reach thermal equilibrium with the SM, which it will retain until its interaction rate drops below the rate of Hubble expansion and the axion density freezes out; this scenario is well established~\cite{Turner:1986tb,Chang:1993gm,Masso:2002np,Hannestad:2005df,Graf:2010tv}, understood in detail~\cite{Salvio:2013iaa,Ferreira:2018vjj,Arias-Aragon:2020shv,DEramo:2021psx,DEramo:2021lgb}, and not our focus.
Instead, we consider the scenario where axions are produced without reaching thermal equilibrium; the residual freeze-in abundance will form a relic contribution to $\rho_\DM$.
Here we establish the \textit{irreducible} contribution, leaving the observable consequences to the next section.

The higher the initial temperature of the Universe, the longer axions have to freeze-in off the thermal bath, independent of whether the specific production is IR or UV dominated.
We define the \textit{reheating temperature}, $\TRH$, as the temperature of the Universe when it entered its last phase of radiation domination~\footnote{The importance of $\TRH$ in axion cosmology is well appreciated, see e.g. \Refs{Grin:2007yg,Visinelli:2009kt,Depta:2020wmr,Balazs:2022tjl}}.
(We remain completely agnostic as to the state of the Universe before this---it could be dominated by another energy source, or not even exist.)
To obtain the minimal axion abundance, we take the smallest allowed $\TRH$, assume that at the instant of reheating $\rho_a=0$, and compute the abundance accumulated after this time. 
This is \textit{irreducible} since any assumption about the specific state of the Universe before reheating can only ever increase this abundance.
The minimal reheat temperature is chosen to ensure we preserve the successes of BBN, which requires $\TRH^{\rm min} = 5\,{\rm MeV}$~\cite{Hasegawa:2019jsa,deSalas:2015glj,Ichikawa:2006vm,Ichikawa:2005vw,Hannestad:2004px,Kawasaki:2000en,Kawasaki:1999na}.
An immediate implication is that any constraints obtained using this irreducible abundance will only have strength for masses up to ${\cal O}(100\,{\rm MeV})$ since the production of heavier axions is greatly Boltzmann suppressed.

An additional corollary of $\TRH=5\,{\rm MeV}$ is that axions are produced from an epoch when only electrons, positrons, photons, and neutrinos were in equilibrium in the SM plasma.
Disregarding the neutrino coupling (cf. \Refc{Huang:2018cwo}), then there are four relevant processes to consider: photon or electron-positron inverse decay ($\gamma\gamma\to a$ or $e^- e^+\to a$), electron-positron annihilation ($e^-e^+\to\gamma a$), and photon conversions ($e^{\pm}\gamma\to e^{\pm}a$).
(Cf. \Refc{Balazs:2022tjl} where the photophilic irreducible abundance was considered without the inclusion of inverse decays.)
Having specified the processes, and a fixed $\TRH$, we can ask what couplings would be required for the axion to achieve thermal equilibrium.
For the photophobic and photophilic scenarios, the answer is furnished by $\ga\gtrsim 10^{-7}\,\rm GeV^{-1}$~\cite{Cadamuro:2011fd,Balazs:2022tjl,Baumholzer:2020hvx,Dror:2021nyr} and $\gae\gtrsim 10^{-9}$, respectively. 
For $m_a \lesssim 100\,{\rm MeV}$ such couplings are in strong tension with a broad range of astrophysical and laboratory constraints \footnote{The ``cosmological triangle'' in the vicinity of $m_a \sim 1~{\rm GeV}$ and $\ga \sim 10^{-5}~{\rm GeV}^{-1}$, see e.g. \Refs{Cadamuro:2011fd,Millea:2015qra,Dolan:2017osp,Brdar:2020dpr,Depta:2020wmr,Dolan:2021rya}, has been closed~\cite{Caputo:2021rux}}.
Accordingly, freeze-in will dictate the irreducible axion density, and we confirmed that the axion never reaches equilibrium in the parameter space we constrain.

Having identified freeze-in as the relevant paradigm, we then compute the mass fraction of DM the axion constitutes as we depart from the epoch of the early Universe.
Specifically, after freeze-in is completed and the axions become non-relativistic, we can define their abundance through $\rho_a = e^{-t/\tau_a} {\cal F}_a \rho_\DM$, where the exponential accounts for depletion via decays, and we define
\be
  {\cal F}_a = e^{\tU/\tau_a} \rho_{a,0}/\rho_{\DM,0},
  \label{eq:def_F_a}
\ee
as the fraction of DM the axions would constitute today were they absolutely stable.
We determine ${\cal F}_a$ from the Boltzmann equation following established techniques, e.g. \Refs{Gondolo:1990dk,Masso:1995tw,Edsjo:1997bg,Hall:2009bx,Cadamuro:2010cz,Cadamuro:2011fd,Blennow:2013jba,DEramo:2017ecx,DEramo:2018vss,Baumholzer:2020hvx,Depta:2020wmr,Balazs:2022tjl,Bolz:2000fu,Dunsky:2022uoq}, yielding the results shown as the blue contours in Figs.~\ref{fig:Constraints} and~\ref{fig:Constraints_gae}.
(These results are made publicly available~\cite{PubCode}.)
Although the calculations are standard, we emphasize several points.
For $m_a\lesssim \TRH$ photon conversion dominates the production, whereas for $m_a\gtrsim \TRH$, inverse decays are most important---annihilations are always subdominant.
Amongst the processes considered, only two are UV dominated, whereby axions are produced predominantly near reheating: electron-positron annihilation and photon conversion, both as mediated by $\ga$ (the $\gae$ analogues are not UV dominated).
In detail, for $\TRH=5\,{\rm MeV}$, we find that freeze-in is UV dominated only when $m_a\lesssim \TRH$ and $\ga/{\rm GeV}^{-1}\gtrsim 350~\gae$.
For UV dominated freeze in, we find that the process is effectively completed shortly after $\TRH$, whereas for IR dominated process appreciable production continues until $T \sim \min(m_e/10,m_a/10)$.
The full details of our calculation will be provided in the Supplementary Material.
Nevertheless, to provide intuition for our results, for smaller masses the density for the photophilic and photophobic scenarios -- identified by a superscript $(\gamma)$ and $(e)$ -- are well approximated by
\bea
  {\cal F}_a^{(\gamma)} &\simeq  0.20\,
  \left(\frac{m_a}{\rm keV}\right)
  \left(\frac{\ga}{10^{-8} \,{\rm GeV}^{-1}}\right)^2
  \left(\frac{\TRH}{5 \,\mathrm{MeV}}\right)\!, \\
  {\cal F}_a^{(e)} &\simeq 2.4\,
  \left(\frac{m_a}{\rm keV}\right)
  \left(\frac{\gae}{10^{-10}}\right)^2\!.
  \label{eq:F_a_approx}
\eea
Both results assume $m_a \lesssim \TRH$, and the second additionally assumes $m_a < 2 m_e$.
We have also set $g_{\star,s} = 10.75$
We note that in accordance with the above discussion, only the photophilic density depends on $\TRH$ as the production is UV dominated, and in fact it remains UV dominated for $\TRH$ even well above the electroweak scale~\cite{Baumholzer:2020hvx}.

%%%%%%%%%%%%
% Figure 2 %
%%%%%%%%%%%%

%%%%%%%%%%
\vspace{0.2cm}
\noindent {\bf Constraints from Axion Decays.}
%%%%%%%%%%
%
So far we have determined the irreducible axion contribution to the DM density.
We now turn to the question of how one could detect (or at present constrain) a potentially small fraction of $\rho_\DM$. 
To do so, observe that the same couplings that produce axions in the early Universe also mediate their decays in the late Universe.
Indeed, for $m_a \ll m_e$,
\be
  \Gamma(a \to \gamma \gamma) = \frac{m_a^3}{64\pi} \left[ \ga + \frac{\alpha \gae m_a^2}{12 m_e^3 \pi} \right]^2\!,
  \label{eq:Gam_a2gamgam}
\ee
where the electron coupling mediates the decay at one loop.
In analogy to \Eq{eq:F_a_approx}, the decay rate in the two scenarios is
\begin{align}
  \tau_{a \to \gamma \gamma}^{(\gamma)} &\simeq 3.0 \times 10^{-6}\,\tU
  \left(\frac{m_a}{\rm keV}\right)^{-3}
  \left(\frac{\ga}{10^{-8} \,{\rm GeV}^{-1}}\right)^{-2}\!, \nonumber\\
  \tau_{a \to \gamma \gamma}^{(e)} &\simeq 1.4 \times 10^{10}\,\tU
  \left(\frac{m_a}{\rm keV}\right)^{-7}
  \left(\frac{\gae}{10^{-10}}\right)^{-2}\!.
\end{align}
While the photophobic rate is slow for the chosen couplings, the significant $m_a$ scaling implies it becomes relevant at higher masses (although as $m_a$ approaches $2m_e$ the expression in \Eq{eq:Gam_a2gamgam} is modified~\cite{Bauer:2017ris}).
See the Supplementary Material for further discussion.

A sufficiently fast decay rate can compensate for a small ${\cal F}_a$, and render the irreducible axion detectable.
Formalizing that logic, in this section we repurpose various searches for DM decay to justify the constraints shown in Figs.~\ref{fig:Constraints} and~\ref{fig:Constraints_gae}.
To do so, we will assume the axions behave exactly like cold DM (CDM).
There are three effects that will violate this assumption.
Firstly, the lifetime of the axion could be comparable to or much shorter than the age of the Universe, i.e. $\tau_a \ll \tU$.
Here $\tau_a$ refers to the lifetime when accounting for all decay modes, for the masses we consider $a \to \gamma \gamma$ and, if it is open, $a \to e^- e^+$.
To account for this we have weighted ${\cal F}_a$ in \Eq{eq:def_F_a} by the appropriate factor to account for exponential depletion.
The second effect relates to the use of DM searches based on observations of compact objects such as the Milky Way: lighter axions can free stream and cluster less than CDM; cf. scenarios with a small warm DM (WDM) component in addition to CDM (e.g. \Refs{Boyarsky:2008mt,Boyarsky:2008xj,Boyarsky:2006fg})~\footnote{The constraints that exist on light axion DM being too hot if it was produced through freeze-in (see e.g. \Refs{DEramo:2020gpr,Ballesteros:2020adh,Baumholzer:2020hvx}) decouple for ${\cal F}_a \lesssim 0.1$~\cite{DEramo:2020gpr}.}.
In the context of Milky Way like halos, simulations have confirmed that indeed the WDM component will develop a core, taking on a different spatial distribution to CDM~\cite{Anderhalden:2012qt}.
In particular, \Refc{Anderhalden:2012qt} quantified that a WDM component in the Milky Way will have a position dependent fraction well described by
\be
  {\cal F}_a^{\scriptscriptstyle \textrm{MW}}(r) \simeq {\cal F}_a\times
  \left[1+0.008\left(\frac{\mathrm{keV}}{m_a}\right)^{2}\left(\frac{{\rm kpc}}{r}\right)\right]^{-1}\!,
  \label{eq:WDM_Fa}
\ee
which will reduce the decay flux near the Galactic Center.
We found that including this effect impacted our results by less than $5\%$ for current constraints, but did affect the future prospects for $m_a \ll {\rm keV}$.
At present there has not been a similar study for satellite galaxies of the Milky Way, and for this reason we chose to neglect constraints from Leo T~\cite{Wadekar:2021qae}.
We discuss WDM effects further in the Supplementary Material.
Finally, as we consider axions generated from freeze-in, they will not have the same momentum distribution as a canonical thermal relic.
Nevertheless, we confirmed that the axions are always non-relativistic at the time of the decays we consider.

Turning to existing searches, we first consider constraints on DM decaying to two photons locally, for instance, in the Milky Way.
These analyses search for the following differential photon flux arising from DM decays,
\be
  \frac{d\Phi}{dE}=\frac{D}{2\pi\, m_\DM \tau_\DM} \delta(E-m_\DM/2).
  \label{eq:DMdecayFlux}
\ee
Here $m_\DM$ and $\tau_\DM$ are the DM mass and inverse decay rate to two photons, respectively.
The term on the numerator is the $D$-factor, and is given by the DM mass column density of the observed object, integrated over the field of view for the search ($D = \int ds d\Omega\, \rho_\DM(s,\Omega)$, see e.g. \Refc{Lisanti:2017qoz}).
In the absence of a detection, instruments searching for the flux in \Eq{eq:DMdecayFlux} interpret their results as a constraint on the minimal allowed lifetime at that mass, which we denote $\tau_\DM^{\rm min}$.

The irreducible axion relic density can also generate a flux as in \Eq{eq:DMdecayFlux}, however, with $m_\DM \to m_a$, $\tau_\DM \to \tau_{a\to\gamma\gamma}$, and $D \to {\cal F}_a e^{-\tU/\tau_a} D$.
We can therefore recast the DM constraints as requiring
\be
  {\cal F}_a \leq {\cal F}_{\rm local}^{\rm max} \equiv 
  \frac{\tau_{a\to \gamma \gamma}}{\tau_\DM^{\rm min}}
  \exp \left(\frac{\tU}{\tau_a} \right)\!.
  \label{eq:F_local}
\ee
We exploit this criterion to convert results for $\tau_\DM^{\rm min}$ into constraints on ${\cal F}_a$, which we then convert to bounds on the axion couplings.
The constraints we draw on are derived using XMM-Newton~\cite{Boyarsky:2006fg,Boyarsky:2006ag,Boyarsky:2007ay,Foster:2021ngm}, NuSTAR~\cite{Ng:2019gch,Roach:2019ctw,Roach:2022lgo}, and INTEGRAL~\cite{Calore:2022pks} observations of the Milky Way, M31, and the Large Magellanic Cloud.
Taken together, these results are collectively labelled X-ray in Figs.~\ref{fig:Constraints} and~\ref{fig:Constraints_gae}.
The breakdown of specific contributions to these results is provided in the Supplementary Material.
The lower boundary of the constraints are set by $\tau_{a\to \gamma \gamma}/{\cal F}_a \sim \tau_\DM^{\rm min}$, whereas the upper boundary occurs when $\tau_a \ll \tU$, and the local density becomes exponentially suppressed below ${\cal F}_a$.
A dramatic example of the latter effect occurs at $m_a = 2m_e$ in \Fig{fig:Constraints_gae}, the opening up of the $a \to e^- e^+$ channel significantly decreases $\tau_a$.

In order to interpret how future local DM decay searches could improve these constraints, in both figures we show the sensitivity that would be attainable if we had $\tau_\DM^{\rm min}$ at the level of $10^{29}$, $10^{30}$, and $10^{31}~{\rm s}$ at all masses (for a discussion of future local searches, see e.g. \Refs{Thorpe-Morgan:2020rwc,Dekker:2021bos,Boddy:2022knd}).
These projections incorporate the effects encapsulated in \Eq{eq:WDM_Fa}.
Broadly, the improvement occurs at smaller couplings---again the cutoff at larger couplings is primarily set by the exponential depletion.
We can probe earlier epochs by considering constraints on decays outside the Milky Way.
After accounting for the optical depth~\cite{Cirelli:2010xx}, such decays must not overproduce the observed CRB~\cite{Hill:2018trh}, and this allows a region of lower masses and higher couplings to be excluded in \Fig{fig:Constraints}.
(In more detail, these bounds probe redshifts of $0 \lesssim z \lesssim 20$ and energies $50~{\rm eV} \lesssim E_{\gamma} \lesssim 3~{\rm keV}$, corresponding to the ultraviolet and X-ray backgrounds.)

Observations of CMB anisotropies allow us to probe decays at earlier times still.
We require,
\be
  {\cal F}_a\leq {\cal F}_{\CMB,\gamma\gamma/e^-e^+}^{\rm max}\equiv\frac{\tau_{a\to \gamma \gamma/e^-e^+}}{\tau_{\CMB,\gamma \gamma/e^-e^+}^{\rm min}}
  \exp\!\left(\frac{t_\CMB}{\tau_{a}}\right)^{2/3}\!,
  \label{eq:F_a_CMB}
\ee
where we distinguish between CMB constraints set on the $\gamma \gamma$ and $e^- e^+$ final states, and define $t_\CMB=8.7\times 10^{13}\,{\rm s} \simeq 2 \times 10^{-4}\, \tU$ ($z \sim 320$).
The form of \Eq{eq:F_a_CMB} and the value of $t_\CMB$ were derived by fitting the conservative `on-the-spot' approximation of \Refc{Poulin:2016anj}, and were confirmed to be consistent with \Refs{Balazs:2022tjl,Slatyer:2016qyl}.
The constraints which inform $\tau_{\CMB,\gamma \gamma/e^-e^+}^{\rm min}$ were taken from \Refc{Cang:2020exa}, which updated \Refs{Slatyer:2016qyl,Poulin:2016anj} to the Planck 2018 data, and \Refc{Bolliet:2020ofj}.
Our results label the CMB anisotropy constraints as CMB.
Nevertheless, the CMB can probe energy injections at even earlier epochs through spectral distortions.
We make use of the results \Refc{Balazs:2022tjl}, that took advantage of the full spectral shape of the CMB, instead of the commonly used $\mu$ and $y$ distortions alone, specifically for the irreducible photophilic axion.

Finally, let us comment on how increasing $\TRH$ impacts our results.
For a fixed $m_a \ll \TRH$ freeze-in of the photophilic axion is UV dominated, increasing $\TRH$ strengthens the bounds in \Fig{fig:Constraints}, but only weakly.
Recall the lower edge of these constraints is set by comparing $\tau_{a \to \gamma \gamma}/{\cal F}_a$ to current lifetime bounds.
This quantity scales as $\ga^{-4} \TRH^{-1}$, implying the bounds improve as $g_{a\gamma\gamma}\propto \TRH^{-1/4}$.
The improvement at the upper edge is weaker, as this is dictated by the onset of exponential depletion.
For reference, the $\TRH=100~{\rm MeV}$ analogue of \Fig{fig:Constraints} is included in Supplementary Material.
The scaling will be modified once $\TRH \gtrsim 100~{\rm MeV}$, when muons and pions enter the SM plasma.
However, including additional particles can only increase the abundance; therefore, this scaling is conservative. 
The effects on the photophobic axion constraints and photophilic case for $m_a\gtrsim \TRH$ are less pronounced as they are not UV dominated.

%%%%%%%%%%
\newpage
\noindent {\bf Conclusions.}
%%%%%%%%%%
%
A remarkable consequence of our present understanding of the Universe is that we cannot arbitrarily decouple its different epochs.
For example, states that are sufficiently strongly coupled such that they would modify the present evolution of stars will invariably have been produced in the early Universe.
The present work has formalized this argument for axions coupled to photons and electrons, showing that in the $100~{\rm eV} - 100~{\rm MeV}$ range axions will be produced sufficiently and form a detectable fraction of the DM density.

We have sought to cast our results as irreducible in the sense that we make the most conservative assumptions about the early Universe, taking the minimal $\TRH$ consistent with BBN.
This does not imply our results are free of assumptions.
For one, we have adopted a conventional cosmology between BBN and today.
We have assumed a minimal scenario of the axion, coupling only to photons or electrons (although we show results when both couplings are active in the Supplementary Material).
If the axion decays faster to neutrinos or dark sector states then the enhanced $\tau_a$ will exponentially deplete $\rho_a$, cutting our constraints off from above.
While such decays could potentially remove our constraints, they will also remove constraints from probes such as the Solar Basin~\cite{VanTilburg:2020jvl,DeRocco:2022jyq}.

Let us end with several comments on extensions to the ideas we have presented.
We have ignored any contribution from the misalignment mechanism.
This is because any misalignment contribution would only add to the irreducible axion abundance we adopted, and so it was conservative to exclude this possibility.
We discuss how one can include misalignment in the Supplementary Material.
Furthermore, our analysis is not restricted to axions; it can constrain any light particle which couples to the SM.
Within the dark sector framework, for instance, many of the mediators commonly considered could be constrained.
Two specific examples are the dark photon and sterile neutrino.
In the Supplementary Material we show that constraints on the latter are particularly strong.
Here, the irreducible freeze-in contribution arises from the Dodelson-Widrow mechanism~\cite{Dodelson:1993je}, beginning with $\rho_s = 0$ at $\TRH = 5~ \rm{MeV}$~\cite{Gelmini:2008fq,Gelmini:2019wfp}. 
The sterile neutrino is then detectable through its decay to an active neutrino and a photon, $\nu_s \to \nu_a \gamma$.

%%%%%%%%%%%%%%%%%%%%%%%%%%%%%%%%%%%%%%%%
\vspace{0.5 cm}
\noindent {\it Acknowledgements.} 
The present work greatly benefited from conversations with Boris Bolliet, Andrea Caputo, Joshua Foster, Fred Hiskens, Kevin Kelly, Soubhik Kumar, Hongwan Liu, Toby Opferkuch, Brandon Roach, Benjamin Safdi, Tracy Slatyer, Ken Van Tilburg, and Tomer Volansky. 
We particularly thank Lawrence Hall and Simon Knapen for many useful comments, and discussions that inspired this work in its initial stages.
In addition, we further thank Boris Bolliet, Andrea Caputo, Francesco D'Eramo, Simon Knapen, Eric Kuflik, Hongwan Liu, Tracy Slatyer, and Ken Van Tilburg for important feedback on a draft version of our work.
Our work made use of the resources provided by \textcite{AxionLimits}.
The work of NJO was supported in part by the Zuckerman STEM Leadership Program and by the National Science Foundation (NSF) under the grant No.~PHY-1915314.
%%%%%%%%%%%%%%%%%%%%%%%%%%%%%%%%%%%%%%%%

\bibliographystyle{apsrev4-1}
\bibliography{alp.bib}

\begin{thebibliography}{118}%
\makeatletter
\providecommand \@ifxundefined [1]{%
 \@ifx{#1\undefined}
}%
\providecommand \@ifnum [1]{%
 \ifnum #1\expandafter \@firstoftwo
 \else \expandafter \@secondoftwo
 \fi
}%
\providecommand \@ifx [1]{%
 \ifx #1\expandafter \@firstoftwo
 \else \expandafter \@secondoftwo
 \fi
}%
\providecommand \natexlab [1]{#1}%
\providecommand \enquote  [1]{``#1''}%
\providecommand \bibnamefont  [1]{#1}%
\providecommand \bibfnamefont [1]{#1}%
\providecommand \citenamefont [1]{#1}%
\providecommand \href@noop [0]{\@secondoftwo}%
\providecommand \href [0]{\begingroup \@sanitize@url \@href}%
\providecommand \@href[1]{\@@startlink{#1}\@@href}%
\providecommand \@@href[1]{\endgroup#1\@@endlink}%
\providecommand \@sanitize@url [0]{\catcode `\\12\catcode `\$12\catcode
  `\&12\catcode `\#12\catcode `\^12\catcode `\_12\catcode `\%12\relax}%
\providecommand \@@startlink[1]{}%
\providecommand \@@endlink[0]{}%
\providecommand \url  [0]{\begingroup\@sanitize@url \@url }%
\providecommand \@url [1]{\endgroup\@href {#1}{\urlprefix }}%
\providecommand \urlprefix  [0]{URL }%
\providecommand \Eprint [0]{\href }%
\providecommand \doibase [0]{http://dx.doi.org/}%
\providecommand \selectlanguage [0]{\@gobble}%
\providecommand \bibinfo  [0]{\@secondoftwo}%
\providecommand \bibfield  [0]{\@secondoftwo}%
\providecommand \translation [1]{[#1]}%
\providecommand \BibitemOpen [0]{}%
\providecommand \bibitemStop [0]{}%
\providecommand \bibitemNoStop [0]{.\EOS\space}%
\providecommand \EOS [0]{\spacefactor3000\relax}%
\providecommand \BibitemShut  [1]{\csname bibitem#1\endcsname}%
\let\auto@bib@innerbib\@empty
%</preamble>
\bibitem [{\citenamefont {Peccei}\ and\ \citenamefont
  {Quinn}(1977{\natexlab{a}})}]{Peccei:1977hh}%
  \BibitemOpen
  \bibfield  {author} {\bibinfo {author} {\bibfnamefont {R.~D.}\ \bibnamefont
  {Peccei}}\ and\ \bibinfo {author} {\bibfnamefont {H.~R.}\ \bibnamefont
  {Quinn}},\ }\href {\doibase 10.1103/PhysRevLett.38.1440} {\bibfield
  {journal} {\bibinfo  {journal} {Phys. Rev. Lett.}\ }\textbf {\bibinfo
  {volume} {38}},\ \bibinfo {pages} {1440} (\bibinfo {year}
  {1977}{\natexlab{a}})}\BibitemShut {NoStop}%
\bibitem [{\citenamefont {Peccei}\ and\ \citenamefont
  {Quinn}(1977{\natexlab{b}})}]{Peccei:1977ur}%
  \BibitemOpen
  \bibfield  {author} {\bibinfo {author} {\bibfnamefont {R.~D.}\ \bibnamefont
  {Peccei}}\ and\ \bibinfo {author} {\bibfnamefont {H.~R.}\ \bibnamefont
  {Quinn}},\ }\href {\doibase 10.1103/PhysRevD.16.1791} {\bibfield  {journal}
  {\bibinfo  {journal} {Phys. Rev. D}\ }\textbf {\bibinfo {volume} {16}},\
  \bibinfo {pages} {1791} (\bibinfo {year} {1977}{\natexlab{b}})}\BibitemShut
  {NoStop}%
\bibitem [{\citenamefont {Weinberg}(1978)}]{Weinberg:1977ma}%
  \BibitemOpen
  \bibfield  {author} {\bibinfo {author} {\bibfnamefont {S.}~\bibnamefont
  {Weinberg}},\ }\href {\doibase 10.1103/PhysRevLett.40.223} {\bibfield
  {journal} {\bibinfo  {journal} {Phys. Rev. Lett.}\ }\textbf {\bibinfo
  {volume} {40}},\ \bibinfo {pages} {223} (\bibinfo {year} {1978})}\BibitemShut
  {NoStop}%
\bibitem [{\citenamefont {Wilczek}(1978)}]{Wilczek:1977pj}%
  \BibitemOpen
  \bibfield  {author} {\bibinfo {author} {\bibfnamefont {F.}~\bibnamefont
  {Wilczek}},\ }\href {\doibase 10.1103/PhysRevLett.40.279} {\bibfield
  {journal} {\bibinfo  {journal} {Phys. Rev. Lett.}\ }\textbf {\bibinfo
  {volume} {40}},\ \bibinfo {pages} {279} (\bibinfo {year} {1978})}\BibitemShut
  {NoStop}%
\bibitem [{\citenamefont {Preskill}\ \emph {et~al.}(1983)\citenamefont
  {Preskill}, \citenamefont {Wise},\ and\ \citenamefont
  {Wilczek}}]{Preskill:1982cy}%
  \BibitemOpen
  \bibfield  {author} {\bibinfo {author} {\bibfnamefont {J.}~\bibnamefont
  {Preskill}}, \bibinfo {author} {\bibfnamefont {M.~B.}\ \bibnamefont {Wise}},
  \ and\ \bibinfo {author} {\bibfnamefont {F.}~\bibnamefont {Wilczek}},\ }\href
  {\doibase 10.1016/0370-2693(83)90637-8} {\bibfield  {journal} {\bibinfo
  {journal} {Phys. Lett. B}\ }\textbf {\bibinfo {volume} {120}},\ \bibinfo
  {pages} {127} (\bibinfo {year} {1983})}\BibitemShut {NoStop}%
\bibitem [{\citenamefont {Abbott}\ and\ \citenamefont
  {Sikivie}(1983)}]{Abbott:1982af}%
  \BibitemOpen
  \bibfield  {author} {\bibinfo {author} {\bibfnamefont {L.~F.}\ \bibnamefont
  {Abbott}}\ and\ \bibinfo {author} {\bibfnamefont {P.}~\bibnamefont
  {Sikivie}},\ }\href {\doibase 10.1016/0370-2693(83)90638-X} {\bibfield
  {journal} {\bibinfo  {journal} {Phys. Lett. B}\ }\textbf {\bibinfo {volume}
  {120}},\ \bibinfo {pages} {133} (\bibinfo {year} {1983})}\BibitemShut
  {NoStop}%
\bibitem [{\citenamefont {Dine}\ and\ \citenamefont
  {Fischler}(1983)}]{Dine:1982ah}%
  \BibitemOpen
  \bibfield  {author} {\bibinfo {author} {\bibfnamefont {M.}~\bibnamefont
  {Dine}}\ and\ \bibinfo {author} {\bibfnamefont {W.}~\bibnamefont
  {Fischler}},\ }\href {\doibase 10.1016/0370-2693(83)90639-1} {\bibfield
  {journal} {\bibinfo  {journal} {Phys. Lett. B}\ }\textbf {\bibinfo {volume}
  {120}},\ \bibinfo {pages} {137} (\bibinfo {year} {1983})}\BibitemShut
  {NoStop}%
\bibitem [{\citenamefont {Sikivie}(1983)}]{Sikivie:1983ip}%
  \BibitemOpen
  \bibfield  {author} {\bibinfo {author} {\bibfnamefont {P.}~\bibnamefont
  {Sikivie}},\ }\href {\doibase 10.1103/PhysRevLett.51.1415} {\bibfield
  {journal} {\bibinfo  {journal} {Phys. Rev. Lett.}\ }\textbf {\bibinfo
  {volume} {51}},\ \bibinfo {pages} {1415} (\bibinfo {year} {1983})},\ \bibinfo
  {note} {[Erratum: Phys.Rev.Lett. 52, 695 (1984)]}\BibitemShut {NoStop}%
\bibitem [{Note1()}]{Note1}%
  \BibitemOpen
  \bibinfo {note} {For $m_a \lesssim 10~{\protect \rm eV}$ these limits were
  recently strengthened by a factor of $\sim $1.4~\cite {Dolan:2022kul}, and
  the constraint in our mass range can likely be similarly
  improved}\BibitemShut {NoStop}%
\bibitem [{\citenamefont {Ayala}\ \emph {et~al.}(2014)\citenamefont {Ayala},
  \citenamefont {Dom\'\i{}nguez}, \citenamefont {Giannotti}, \citenamefont
  {Mirizzi},\ and\ \citenamefont {Straniero}}]{Ayala:2014pea}%
  \BibitemOpen
  \bibfield  {author} {\bibinfo {author} {\bibfnamefont {A.}~\bibnamefont
  {Ayala}}, \bibinfo {author} {\bibfnamefont {I.}~\bibnamefont
  {Dom\'\i{}nguez}}, \bibinfo {author} {\bibfnamefont {M.}~\bibnamefont
  {Giannotti}}, \bibinfo {author} {\bibfnamefont {A.}~\bibnamefont {Mirizzi}},
  \ and\ \bibinfo {author} {\bibfnamefont {O.}~\bibnamefont {Straniero}},\
  }\href {\doibase 10.1103/PhysRevLett.113.191302} {\bibfield  {journal}
  {\bibinfo  {journal} {Phys. Rev. Lett.}\ }\textbf {\bibinfo {volume} {113}},\
  \bibinfo {pages} {191302} (\bibinfo {year} {2014})},\ \Eprint
  {http://arxiv.org/abs/1406.6053} {arXiv:1406.6053 [astro-ph.SR]} \BibitemShut
  {NoStop}%
\bibitem [{\citenamefont {Carenza}\ \emph {et~al.}(2020)\citenamefont
  {Carenza}, \citenamefont {Straniero}, \citenamefont {D\"obrich},
  \citenamefont {Giannotti}, \citenamefont {Lucente},\ and\ \citenamefont
  {Mirizzi}}]{Carenza:2020zil}%
  \BibitemOpen
  \bibfield  {author} {\bibinfo {author} {\bibfnamefont {P.}~\bibnamefont
  {Carenza}}, \bibinfo {author} {\bibfnamefont {O.}~\bibnamefont {Straniero}},
  \bibinfo {author} {\bibfnamefont {B.}~\bibnamefont {D\"obrich}}, \bibinfo
  {author} {\bibfnamefont {M.}~\bibnamefont {Giannotti}}, \bibinfo {author}
  {\bibfnamefont {G.}~\bibnamefont {Lucente}}, \ and\ \bibinfo {author}
  {\bibfnamefont {A.}~\bibnamefont {Mirizzi}},\ }\href {\doibase
  10.1016/j.physletb.2020.135709} {\bibfield  {journal} {\bibinfo  {journal}
  {Phys. Lett. B}\ }\textbf {\bibinfo {volume} {809}},\ \bibinfo {pages}
  {135709} (\bibinfo {year} {2020})},\ \Eprint
  {http://arxiv.org/abs/2004.08399} {arXiv:2004.08399 [hep-ph]} \BibitemShut
  {NoStop}%
\bibitem [{\citenamefont {Foster}\ \emph {et~al.}(2021)\citenamefont {Foster},
  \citenamefont {Kongsore}, \citenamefont {Dessert}, \citenamefont {Park},
  \citenamefont {Rodd}, \citenamefont {Cranmer},\ and\ \citenamefont
  {Safdi}}]{Foster:2021ngm}%
  \BibitemOpen
  \bibfield  {author} {\bibinfo {author} {\bibfnamefont {J.~W.}\ \bibnamefont
  {Foster}}, \bibinfo {author} {\bibfnamefont {M.}~\bibnamefont {Kongsore}},
  \bibinfo {author} {\bibfnamefont {C.}~\bibnamefont {Dessert}}, \bibinfo
  {author} {\bibfnamefont {Y.}~\bibnamefont {Park}}, \bibinfo {author}
  {\bibfnamefont {N.~L.}\ \bibnamefont {Rodd}}, \bibinfo {author}
  {\bibfnamefont {K.}~\bibnamefont {Cranmer}}, \ and\ \bibinfo {author}
  {\bibfnamefont {B.~R.}\ \bibnamefont {Safdi}},\ }\href {\doibase
  10.1103/PhysRevLett.127.051101} {\bibfield  {journal} {\bibinfo  {journal}
  {Phys. Rev. Lett.}\ }\textbf {\bibinfo {volume} {127}},\ \bibinfo {pages}
  {051101} (\bibinfo {year} {2021})},\ \Eprint
  {http://arxiv.org/abs/2102.02207} {arXiv:2102.02207 [astro-ph.CO]}
  \BibitemShut {NoStop}%
\bibitem [{\citenamefont {Boyarsky}\ \emph {et~al.}(2006)\citenamefont
  {Boyarsky}, \citenamefont {Neronov}, \citenamefont {Ruchayskiy},
  \citenamefont {Shaposhnikov},\ and\ \citenamefont
  {Tkachev}}]{Boyarsky:2006fg}%
  \BibitemOpen
  \bibfield  {author} {\bibinfo {author} {\bibfnamefont {A.}~\bibnamefont
  {Boyarsky}}, \bibinfo {author} {\bibfnamefont {A.}~\bibnamefont {Neronov}},
  \bibinfo {author} {\bibfnamefont {O.}~\bibnamefont {Ruchayskiy}}, \bibinfo
  {author} {\bibfnamefont {M.}~\bibnamefont {Shaposhnikov}}, \ and\ \bibinfo
  {author} {\bibfnamefont {I.}~\bibnamefont {Tkachev}},\ }\href {\doibase
  10.1103/PhysRevLett.97.261302} {\bibfield  {journal} {\bibinfo  {journal}
  {Phys. Rev. Lett.}\ }\textbf {\bibinfo {volume} {97}},\ \bibinfo {pages}
  {261302} (\bibinfo {year} {2006})},\ \Eprint
  {http://arxiv.org/abs/astro-ph/0603660} {arXiv:astro-ph/0603660} \BibitemShut
  {NoStop}%
\bibitem [{\citenamefont {Boyarsky}\ \emph {et~al.}(2007)\citenamefont
  {Boyarsky}, \citenamefont {Nevalainen},\ and\ \citenamefont
  {Ruchayskiy}}]{Boyarsky:2006ag}%
  \BibitemOpen
  \bibfield  {author} {\bibinfo {author} {\bibfnamefont {A.}~\bibnamefont
  {Boyarsky}}, \bibinfo {author} {\bibfnamefont {J.}~\bibnamefont
  {Nevalainen}}, \ and\ \bibinfo {author} {\bibfnamefont {O.}~\bibnamefont
  {Ruchayskiy}},\ }\href {\doibase 10.1051/0004-6361:20066774} {\bibfield
  {journal} {\bibinfo  {journal} {Astron. Astrophys.}\ }\textbf {\bibinfo
  {volume} {471}},\ \bibinfo {pages} {51} (\bibinfo {year} {2007})},\ \Eprint
  {http://arxiv.org/abs/astro-ph/0610961} {arXiv:astro-ph/0610961} \BibitemShut
  {NoStop}%
\bibitem [{\citenamefont {Boyarsky}\ \emph
  {et~al.}(2008{\natexlab{a}})\citenamefont {Boyarsky}, \citenamefont
  {Iakubovskyi}, \citenamefont {Ruchayskiy},\ and\ \citenamefont
  {Savchenko}}]{Boyarsky:2007ay}%
  \BibitemOpen
  \bibfield  {author} {\bibinfo {author} {\bibfnamefont {A.}~\bibnamefont
  {Boyarsky}}, \bibinfo {author} {\bibfnamefont {D.}~\bibnamefont
  {Iakubovskyi}}, \bibinfo {author} {\bibfnamefont {O.}~\bibnamefont
  {Ruchayskiy}}, \ and\ \bibinfo {author} {\bibfnamefont {V.}~\bibnamefont
  {Savchenko}},\ }\href {\doibase 10.1111/j.1365-2966.2008.13266.x} {\bibfield
  {journal} {\bibinfo  {journal} {Mon. Not. Roy. Astron. Soc.}\ }\textbf
  {\bibinfo {volume} {387}},\ \bibinfo {pages} {1361} (\bibinfo {year}
  {2008}{\natexlab{a}})},\ \Eprint {http://arxiv.org/abs/0709.2301}
  {arXiv:0709.2301 [astro-ph]} \BibitemShut {NoStop}%
\bibitem [{\citenamefont {Ng}\ \emph {et~al.}(2019)\citenamefont {Ng},
  \citenamefont {Roach}, \citenamefont {Perez}, \citenamefont {Beacom},
  \citenamefont {Horiuchi}, \citenamefont {Krivonos},\ and\ \citenamefont
  {Wik}}]{Ng:2019gch}%
  \BibitemOpen
  \bibfield  {author} {\bibinfo {author} {\bibfnamefont {K.~C.~Y.}\
  \bibnamefont {Ng}}, \bibinfo {author} {\bibfnamefont {B.~M.}\ \bibnamefont
  {Roach}}, \bibinfo {author} {\bibfnamefont {K.}~\bibnamefont {Perez}},
  \bibinfo {author} {\bibfnamefont {J.~F.}\ \bibnamefont {Beacom}}, \bibinfo
  {author} {\bibfnamefont {S.}~\bibnamefont {Horiuchi}}, \bibinfo {author}
  {\bibfnamefont {R.}~\bibnamefont {Krivonos}}, \ and\ \bibinfo {author}
  {\bibfnamefont {D.~R.}\ \bibnamefont {Wik}},\ }\href {\doibase
  10.1103/PhysRevD.99.083005} {\bibfield  {journal} {\bibinfo  {journal} {Phys.
  Rev. D}\ }\textbf {\bibinfo {volume} {99}},\ \bibinfo {pages} {083005}
  (\bibinfo {year} {2019})},\ \Eprint {http://arxiv.org/abs/1901.01262}
  {arXiv:1901.01262 [astro-ph.HE]} \BibitemShut {NoStop}%
\bibitem [{\citenamefont {Roach}\ \emph {et~al.}(2020)\citenamefont {Roach},
  \citenamefont {Ng}, \citenamefont {Perez}, \citenamefont {Beacom},
  \citenamefont {Horiuchi}, \citenamefont {Krivonos},\ and\ \citenamefont
  {Wik}}]{Roach:2019ctw}%
  \BibitemOpen
  \bibfield  {author} {\bibinfo {author} {\bibfnamefont {B.~M.}\ \bibnamefont
  {Roach}}, \bibinfo {author} {\bibfnamefont {K.~C.~Y.}\ \bibnamefont {Ng}},
  \bibinfo {author} {\bibfnamefont {K.}~\bibnamefont {Perez}}, \bibinfo
  {author} {\bibfnamefont {J.~F.}\ \bibnamefont {Beacom}}, \bibinfo {author}
  {\bibfnamefont {S.}~\bibnamefont {Horiuchi}}, \bibinfo {author}
  {\bibfnamefont {R.}~\bibnamefont {Krivonos}}, \ and\ \bibinfo {author}
  {\bibfnamefont {D.~R.}\ \bibnamefont {Wik}},\ }\href {\doibase
  10.1103/PhysRevD.101.103011} {\bibfield  {journal} {\bibinfo  {journal}
  {Phys. Rev. D}\ }\textbf {\bibinfo {volume} {101}},\ \bibinfo {pages}
  {103011} (\bibinfo {year} {2020})},\ \Eprint
  {http://arxiv.org/abs/1908.09037} {arXiv:1908.09037 [astro-ph.HE]}
  \BibitemShut {NoStop}%
\bibitem [{\citenamefont {Roach}\ \emph {et~al.}(2022)\citenamefont {Roach},
  \citenamefont {Rossland}, \citenamefont {Ng}, \citenamefont {Perez},
  \citenamefont {Beacom}, \citenamefont {Grefenstette}, \citenamefont
  {Horiuchi}, \citenamefont {Krivonos},\ and\ \citenamefont
  {Wik}}]{Roach:2022lgo}%
  \BibitemOpen
  \bibfield  {author} {\bibinfo {author} {\bibfnamefont {B.~M.}\ \bibnamefont
  {Roach}}, \bibinfo {author} {\bibfnamefont {S.}~\bibnamefont {Rossland}},
  \bibinfo {author} {\bibfnamefont {K.~C.~Y.}\ \bibnamefont {Ng}}, \bibinfo
  {author} {\bibfnamefont {K.}~\bibnamefont {Perez}}, \bibinfo {author}
  {\bibfnamefont {J.~F.}\ \bibnamefont {Beacom}}, \bibinfo {author}
  {\bibfnamefont {B.~W.}\ \bibnamefont {Grefenstette}}, \bibinfo {author}
  {\bibfnamefont {S.}~\bibnamefont {Horiuchi}}, \bibinfo {author}
  {\bibfnamefont {R.}~\bibnamefont {Krivonos}}, \ and\ \bibinfo {author}
  {\bibfnamefont {D.~R.}\ \bibnamefont {Wik}},\ }\href@noop {} {\  (\bibinfo
  {year} {2022})},\ \Eprint {http://arxiv.org/abs/2207.04572} {arXiv:2207.04572
  [astro-ph.HE]} \BibitemShut {NoStop}%
\bibitem [{\citenamefont {Calore}\ \emph {et~al.}(2022)\citenamefont {Calore},
  \citenamefont {Dekker}, \citenamefont {Serpico},\ and\ \citenamefont
  {Siegert}}]{Calore:2022pks}%
  \BibitemOpen
  \bibfield  {author} {\bibinfo {author} {\bibfnamefont {F.}~\bibnamefont
  {Calore}}, \bibinfo {author} {\bibfnamefont {A.}~\bibnamefont {Dekker}},
  \bibinfo {author} {\bibfnamefont {P.~D.}\ \bibnamefont {Serpico}}, \ and\
  \bibinfo {author} {\bibfnamefont {T.}~\bibnamefont {Siegert}},\ }\href@noop
  {} {\  (\bibinfo {year} {2022})},\ \Eprint {http://arxiv.org/abs/2209.06299}
  {arXiv:2209.06299 [hep-ph]} \BibitemShut {NoStop}%
\bibitem [{\citenamefont {Hill}\ \emph {et~al.}(2018)\citenamefont {Hill},
  \citenamefont {Masui},\ and\ \citenamefont {Scott}}]{Hill:2018trh}%
  \BibitemOpen
  \bibfield  {author} {\bibinfo {author} {\bibfnamefont {R.}~\bibnamefont
  {Hill}}, \bibinfo {author} {\bibfnamefont {K.~W.}\ \bibnamefont {Masui}}, \
  and\ \bibinfo {author} {\bibfnamefont {D.}~\bibnamefont {Scott}},\ }\href
  {\doibase 10.1177/0003702818767133} {\bibfield  {journal} {\bibinfo
  {journal} {Appl. Spectrosc.}\ }\textbf {\bibinfo {volume} {72}},\ \bibinfo
  {pages} {663} (\bibinfo {year} {2018})},\ \Eprint
  {http://arxiv.org/abs/1802.03694} {arXiv:1802.03694 [astro-ph.CO]}
  \BibitemShut {NoStop}%
\bibitem [{\citenamefont {Slatyer}\ and\ \citenamefont
  {Wu}(2017)}]{Slatyer:2016qyl}%
  \BibitemOpen
  \bibfield  {author} {\bibinfo {author} {\bibfnamefont {T.~R.}\ \bibnamefont
  {Slatyer}}\ and\ \bibinfo {author} {\bibfnamefont {C.-L.}\ \bibnamefont
  {Wu}},\ }\href {\doibase 10.1103/PhysRevD.95.023010} {\bibfield  {journal}
  {\bibinfo  {journal} {Phys. Rev. D}\ }\textbf {\bibinfo {volume} {95}},\
  \bibinfo {pages} {023010} (\bibinfo {year} {2017})},\ \Eprint
  {http://arxiv.org/abs/1610.06933} {arXiv:1610.06933 [astro-ph.CO]}
  \BibitemShut {NoStop}%
\bibitem [{\citenamefont {Poulin}\ \emph {et~al.}(2017)\citenamefont {Poulin},
  \citenamefont {Lesgourgues},\ and\ \citenamefont {Serpico}}]{Poulin:2016anj}%
  \BibitemOpen
  \bibfield  {author} {\bibinfo {author} {\bibfnamefont {V.}~\bibnamefont
  {Poulin}}, \bibinfo {author} {\bibfnamefont {J.}~\bibnamefont {Lesgourgues}},
  \ and\ \bibinfo {author} {\bibfnamefont {P.~D.}\ \bibnamefont {Serpico}},\
  }\href {\doibase 10.1088/1475-7516/2017/03/043} {\bibfield  {journal}
  {\bibinfo  {journal} {JCAP}\ }\textbf {\bibinfo {volume} {03}},\ \bibinfo
  {pages} {043} (\bibinfo {year} {2017})},\ \Eprint
  {http://arxiv.org/abs/1610.10051} {arXiv:1610.10051 [astro-ph.CO]}
  \BibitemShut {NoStop}%
\bibitem [{\citenamefont {Cang}\ \emph {et~al.}(2020)\citenamefont {Cang},
  \citenamefont {Gao},\ and\ \citenamefont {Ma}}]{Cang:2020exa}%
  \BibitemOpen
  \bibfield  {author} {\bibinfo {author} {\bibfnamefont {J.}~\bibnamefont
  {Cang}}, \bibinfo {author} {\bibfnamefont {Y.}~\bibnamefont {Gao}}, \ and\
  \bibinfo {author} {\bibfnamefont {Y.-Z.}\ \bibnamefont {Ma}},\ }\href
  {\doibase 10.1103/PhysRevD.102.103005} {\bibfield  {journal} {\bibinfo
  {journal} {Phys. Rev. D}\ }\textbf {\bibinfo {volume} {102}},\ \bibinfo
  {pages} {103005} (\bibinfo {year} {2020})},\ \Eprint
  {http://arxiv.org/abs/2002.03380} {arXiv:2002.03380 [astro-ph.CO]}
  \BibitemShut {NoStop}%
\bibitem [{\citenamefont {Bolliet}\ \emph {et~al.}(2021)\citenamefont
  {Bolliet}, \citenamefont {Chluba},\ and\ \citenamefont
  {Battye}}]{Bolliet:2020ofj}%
  \BibitemOpen
  \bibfield  {author} {\bibinfo {author} {\bibfnamefont {B.}~\bibnamefont
  {Bolliet}}, \bibinfo {author} {\bibfnamefont {J.}~\bibnamefont {Chluba}}, \
  and\ \bibinfo {author} {\bibfnamefont {R.}~\bibnamefont {Battye}},\ }\href
  {\doibase 10.1093/mnras/stab1997} {\bibfield  {journal} {\bibinfo  {journal}
  {Mon. Not. Roy. Astron. Soc.}\ }\textbf {\bibinfo {volume} {507}},\ \bibinfo
  {pages} {3148} (\bibinfo {year} {2021})},\ \Eprint
  {http://arxiv.org/abs/2012.07292} {arXiv:2012.07292 [astro-ph.CO]}
  \BibitemShut {NoStop}%
\bibitem [{\citenamefont {Bal\'azs}\ \emph {et~al.}(2022)\citenamefont
  {Bal\'azs} \emph {et~al.}}]{Balazs:2022tjl}%
  \BibitemOpen
  \bibfield  {author} {\bibinfo {author} {\bibfnamefont {C.}~\bibnamefont
  {Bal\'azs}} \emph {et~al.},\ }\href@noop {} {\  (\bibinfo {year} {2022})},\
  \Eprint {http://arxiv.org/abs/2205.13549} {arXiv:2205.13549 [astro-ph.CO]}
  \BibitemShut {NoStop}%
\bibitem [{\citenamefont {Baumholzer}\ \emph {et~al.}(2021)\citenamefont
  {Baumholzer}, \citenamefont {Brdar},\ and\ \citenamefont
  {Morgante}}]{Baumholzer:2020hvx}%
  \BibitemOpen
  \bibfield  {author} {\bibinfo {author} {\bibfnamefont {S.}~\bibnamefont
  {Baumholzer}}, \bibinfo {author} {\bibfnamefont {V.}~\bibnamefont {Brdar}}, \
  and\ \bibinfo {author} {\bibfnamefont {E.}~\bibnamefont {Morgante}},\ }\href
  {\doibase 10.1088/1475-7516/2021/05/004} {\bibfield  {journal} {\bibinfo
  {journal} {JCAP}\ }\textbf {\bibinfo {volume} {05}},\ \bibinfo {pages} {004}
  (\bibinfo {year} {2021})},\ \Eprint {http://arxiv.org/abs/2012.09181}
  {arXiv:2012.09181 [hep-ph]} \BibitemShut {NoStop}%
\bibitem [{\citenamefont {Dolan}\ \emph {et~al.}(2022)\citenamefont {Dolan},
  \citenamefont {Hiskens},\ and\ \citenamefont {Volkas}}]{Dolan:2022kul}%
  \BibitemOpen
  \bibfield  {author} {\bibinfo {author} {\bibfnamefont {M.~J.}\ \bibnamefont
  {Dolan}}, \bibinfo {author} {\bibfnamefont {F.~J.}\ \bibnamefont {Hiskens}},
  \ and\ \bibinfo {author} {\bibfnamefont {R.~R.}\ \bibnamefont {Volkas}},\
  }\href@noop {} {\  (\bibinfo {year} {2022})},\ \Eprint
  {http://arxiv.org/abs/2207.03102} {arXiv:2207.03102 [hep-ph]} \BibitemShut
  {NoStop}%
\bibitem [{\citenamefont {Jaeckel}\ \emph {et~al.}(2018)\citenamefont
  {Jaeckel}, \citenamefont {Malta},\ and\ \citenamefont
  {Redondo}}]{Jaeckel:2017tud}%
  \BibitemOpen
  \bibfield  {author} {\bibinfo {author} {\bibfnamefont {J.}~\bibnamefont
  {Jaeckel}}, \bibinfo {author} {\bibfnamefont {P.~C.}\ \bibnamefont {Malta}},
  \ and\ \bibinfo {author} {\bibfnamefont {J.}~\bibnamefont {Redondo}},\ }\href
  {\doibase 10.1103/PhysRevD.98.055032} {\bibfield  {journal} {\bibinfo
  {journal} {Phys. Rev. D}\ }\textbf {\bibinfo {volume} {98}},\ \bibinfo
  {pages} {055032} (\bibinfo {year} {2018})},\ \Eprint
  {http://arxiv.org/abs/1702.02964} {arXiv:1702.02964 [hep-ph]} \BibitemShut
  {NoStop}%
\bibitem [{\citenamefont {Masso}\ and\ \citenamefont
  {Toldra}(1995)}]{Masso:1995tw}%
  \BibitemOpen
  \bibfield  {author} {\bibinfo {author} {\bibfnamefont {E.}~\bibnamefont
  {Masso}}\ and\ \bibinfo {author} {\bibfnamefont {R.}~\bibnamefont {Toldra}},\
  }\href {\doibase 10.1103/PhysRevD.52.1755} {\bibfield  {journal} {\bibinfo
  {journal} {Phys. Rev. D}\ }\textbf {\bibinfo {volume} {52}},\ \bibinfo
  {pages} {1755} (\bibinfo {year} {1995})},\ \Eprint
  {http://arxiv.org/abs/hep-ph/9503293} {arXiv:hep-ph/9503293} \BibitemShut
  {NoStop}%
\bibitem [{\citenamefont {Lucente}\ \emph {et~al.}(2020)\citenamefont
  {Lucente}, \citenamefont {Carenza}, \citenamefont {Fischer}, \citenamefont
  {Giannotti},\ and\ \citenamefont {Mirizzi}}]{Lucente:2020whw}%
  \BibitemOpen
  \bibfield  {author} {\bibinfo {author} {\bibfnamefont {G.}~\bibnamefont
  {Lucente}}, \bibinfo {author} {\bibfnamefont {P.}~\bibnamefont {Carenza}},
  \bibinfo {author} {\bibfnamefont {T.}~\bibnamefont {Fischer}}, \bibinfo
  {author} {\bibfnamefont {M.}~\bibnamefont {Giannotti}}, \ and\ \bibinfo
  {author} {\bibfnamefont {A.}~\bibnamefont {Mirizzi}},\ }\href {\doibase
  10.1088/1475-7516/2020/12/008} {\bibfield  {journal} {\bibinfo  {journal}
  {JCAP}\ }\textbf {\bibinfo {volume} {12}},\ \bibinfo {pages} {008} (\bibinfo
  {year} {2020})},\ \Eprint {http://arxiv.org/abs/2008.04918} {arXiv:2008.04918
  [hep-ph]} \BibitemShut {NoStop}%
\bibitem [{\citenamefont {Caputo}\ \emph
  {et~al.}(2022{\natexlab{a}})\citenamefont {Caputo}, \citenamefont {Janka},
  \citenamefont {Raffelt},\ and\ \citenamefont {Vitagliano}}]{Caputo:2022mah}%
  \BibitemOpen
  \bibfield  {author} {\bibinfo {author} {\bibfnamefont {A.}~\bibnamefont
  {Caputo}}, \bibinfo {author} {\bibfnamefont {H.-T.}\ \bibnamefont {Janka}},
  \bibinfo {author} {\bibfnamefont {G.}~\bibnamefont {Raffelt}}, \ and\
  \bibinfo {author} {\bibfnamefont {E.}~\bibnamefont {Vitagliano}},\ }\href
  {\doibase 10.1103/PhysRevLett.128.221103} {\bibfield  {journal} {\bibinfo
  {journal} {Phys. Rev. Lett.}\ }\textbf {\bibinfo {volume} {128}},\ \bibinfo
  {pages} {221103} (\bibinfo {year} {2022}{\natexlab{a}})},\ \Eprint
  {http://arxiv.org/abs/2201.09890} {arXiv:2201.09890 [astro-ph.HE]}
  \BibitemShut {NoStop}%
\bibitem [{\citenamefont {DeRocco}\ \emph {et~al.}(2022)\citenamefont
  {DeRocco}, \citenamefont {Wegsman}, \citenamefont {Grefenstette},
  \citenamefont {Huang},\ and\ \citenamefont {Van~Tilburg}}]{DeRocco:2022jyq}%
  \BibitemOpen
  \bibfield  {author} {\bibinfo {author} {\bibfnamefont {W.}~\bibnamefont
  {DeRocco}}, \bibinfo {author} {\bibfnamefont {S.}~\bibnamefont {Wegsman}},
  \bibinfo {author} {\bibfnamefont {B.}~\bibnamefont {Grefenstette}}, \bibinfo
  {author} {\bibfnamefont {J.}~\bibnamefont {Huang}}, \ and\ \bibinfo {author}
  {\bibfnamefont {K.}~\bibnamefont {Van~Tilburg}},\ }\href {\doibase
  10.1103/PhysRevLett.129.101101} {\bibfield  {journal} {\bibinfo  {journal}
  {Phys. Rev. Lett.}\ }\textbf {\bibinfo {volume} {129}},\ \bibinfo {pages}
  {101101} (\bibinfo {year} {2022})},\ \Eprint
  {http://arxiv.org/abs/2205.05700} {arXiv:2205.05700 [hep-ph]} \BibitemShut
  {NoStop}%
\bibitem [{\citenamefont {Turner}(1987)}]{Turner:1986tb}%
  \BibitemOpen
  \bibfield  {author} {\bibinfo {author} {\bibfnamefont {M.~S.}\ \bibnamefont
  {Turner}},\ }\href {\doibase 10.1103/PhysRevLett.59.2489} {\bibfield
  {journal} {\bibinfo  {journal} {Phys. Rev. Lett.}\ }\textbf {\bibinfo
  {volume} {59}},\ \bibinfo {pages} {2489} (\bibinfo {year} {1987})},\ \bibinfo
  {note} {[Erratum: Phys.Rev.Lett. 60, 1101 (1988)]}\BibitemShut {NoStop}%
\bibitem [{\citenamefont {Chang}\ and\ \citenamefont
  {Choi}(1993)}]{Chang:1993gm}%
  \BibitemOpen
  \bibfield  {author} {\bibinfo {author} {\bibfnamefont {S.}~\bibnamefont
  {Chang}}\ and\ \bibinfo {author} {\bibfnamefont {K.}~\bibnamefont {Choi}},\
  }\href {\doibase 10.1016/0370-2693(93)90656-3} {\bibfield  {journal}
  {\bibinfo  {journal} {Phys. Lett. B}\ }\textbf {\bibinfo {volume} {316}},\
  \bibinfo {pages} {51} (\bibinfo {year} {1993})},\ \Eprint
  {http://arxiv.org/abs/hep-ph/9306216} {arXiv:hep-ph/9306216} \BibitemShut
  {NoStop}%
\bibitem [{\citenamefont {Masso}\ \emph {et~al.}(2002)\citenamefont {Masso},
  \citenamefont {Rota},\ and\ \citenamefont {Zsembinszki}}]{Masso:2002np}%
  \BibitemOpen
  \bibfield  {author} {\bibinfo {author} {\bibfnamefont {E.}~\bibnamefont
  {Masso}}, \bibinfo {author} {\bibfnamefont {F.}~\bibnamefont {Rota}}, \ and\
  \bibinfo {author} {\bibfnamefont {G.}~\bibnamefont {Zsembinszki}},\ }\href
  {\doibase 10.1103/PhysRevD.66.023004} {\bibfield  {journal} {\bibinfo
  {journal} {Phys. Rev. D}\ }\textbf {\bibinfo {volume} {66}},\ \bibinfo
  {pages} {023004} (\bibinfo {year} {2002})},\ \Eprint
  {http://arxiv.org/abs/hep-ph/0203221} {arXiv:hep-ph/0203221} \BibitemShut
  {NoStop}%
\bibitem [{\citenamefont {Hannestad}\ \emph {et~al.}(2005)\citenamefont
  {Hannestad}, \citenamefont {Mirizzi},\ and\ \citenamefont
  {Raffelt}}]{Hannestad:2005df}%
  \BibitemOpen
  \bibfield  {author} {\bibinfo {author} {\bibfnamefont {S.}~\bibnamefont
  {Hannestad}}, \bibinfo {author} {\bibfnamefont {A.}~\bibnamefont {Mirizzi}},
  \ and\ \bibinfo {author} {\bibfnamefont {G.}~\bibnamefont {Raffelt}},\ }\href
  {\doibase 10.1088/1475-7516/2005/07/002} {\bibfield  {journal} {\bibinfo
  {journal} {JCAP}\ }\textbf {\bibinfo {volume} {07}},\ \bibinfo {pages} {002}
  (\bibinfo {year} {2005})},\ \Eprint {http://arxiv.org/abs/hep-ph/0504059}
  {arXiv:hep-ph/0504059} \BibitemShut {NoStop}%
\bibitem [{\citenamefont {Graf}\ and\ \citenamefont
  {Steffen}(2011)}]{Graf:2010tv}%
  \BibitemOpen
  \bibfield  {author} {\bibinfo {author} {\bibfnamefont {P.}~\bibnamefont
  {Graf}}\ and\ \bibinfo {author} {\bibfnamefont {F.~D.}\ \bibnamefont
  {Steffen}},\ }\href {\doibase 10.1103/PhysRevD.83.075011} {\bibfield
  {journal} {\bibinfo  {journal} {Phys. Rev. D}\ }\textbf {\bibinfo {volume}
  {83}},\ \bibinfo {pages} {075011} (\bibinfo {year} {2011})},\ \Eprint
  {http://arxiv.org/abs/1008.4528} {arXiv:1008.4528 [hep-ph]} \BibitemShut
  {NoStop}%
\bibitem [{\citenamefont {Salvio}\ \emph {et~al.}(2014)\citenamefont {Salvio},
  \citenamefont {Strumia},\ and\ \citenamefont {Xue}}]{Salvio:2013iaa}%
  \BibitemOpen
  \bibfield  {author} {\bibinfo {author} {\bibfnamefont {A.}~\bibnamefont
  {Salvio}}, \bibinfo {author} {\bibfnamefont {A.}~\bibnamefont {Strumia}}, \
  and\ \bibinfo {author} {\bibfnamefont {W.}~\bibnamefont {Xue}},\ }\href
  {\doibase 10.1088/1475-7516/2014/01/011} {\bibfield  {journal} {\bibinfo
  {journal} {JCAP}\ }\textbf {\bibinfo {volume} {01}},\ \bibinfo {pages} {011}
  (\bibinfo {year} {2014})},\ \Eprint {http://arxiv.org/abs/1310.6982}
  {arXiv:1310.6982 [hep-ph]} \BibitemShut {NoStop}%
\bibitem [{\citenamefont {Ferreira}\ and\ \citenamefont
  {Notari}(2018)}]{Ferreira:2018vjj}%
  \BibitemOpen
  \bibfield  {author} {\bibinfo {author} {\bibfnamefont {R.~Z.}\ \bibnamefont
  {Ferreira}}\ and\ \bibinfo {author} {\bibfnamefont {A.}~\bibnamefont
  {Notari}},\ }\href {\doibase 10.1103/PhysRevLett.120.191301} {\bibfield
  {journal} {\bibinfo  {journal} {Phys. Rev. Lett.}\ }\textbf {\bibinfo
  {volume} {120}},\ \bibinfo {pages} {191301} (\bibinfo {year} {2018})},\
  \Eprint {http://arxiv.org/abs/1801.06090} {arXiv:1801.06090 [hep-ph]}
  \BibitemShut {NoStop}%
\bibitem [{\citenamefont {Arias-Arag\'on}\ \emph {et~al.}(2021)\citenamefont
  {Arias-Arag\'on}, \citenamefont {D'Eramo}, \citenamefont {Ferreira},
  \citenamefont {Merlo},\ and\ \citenamefont {Notari}}]{Arias-Aragon:2020shv}%
  \BibitemOpen
  \bibfield  {author} {\bibinfo {author} {\bibfnamefont {F.}~\bibnamefont
  {Arias-Arag\'on}}, \bibinfo {author} {\bibfnamefont {F.}~\bibnamefont
  {D'Eramo}}, \bibinfo {author} {\bibfnamefont {R.~Z.}\ \bibnamefont
  {Ferreira}}, \bibinfo {author} {\bibfnamefont {L.}~\bibnamefont {Merlo}}, \
  and\ \bibinfo {author} {\bibfnamefont {A.}~\bibnamefont {Notari}},\ }\href
  {\doibase 10.1088/1475-7516/2021/03/090} {\bibfield  {journal} {\bibinfo
  {journal} {JCAP}\ }\textbf {\bibinfo {volume} {03}},\ \bibinfo {pages} {090}
  (\bibinfo {year} {2021})},\ \Eprint {http://arxiv.org/abs/2012.04736}
  {arXiv:2012.04736 [hep-ph]} \BibitemShut {NoStop}%
\bibitem [{\citenamefont {D'Eramo}\ \emph {et~al.}(2022)\citenamefont
  {D'Eramo}, \citenamefont {Hajkarim},\ and\ \citenamefont
  {Yun}}]{DEramo:2021psx}%
  \BibitemOpen
  \bibfield  {author} {\bibinfo {author} {\bibfnamefont {F.}~\bibnamefont
  {D'Eramo}}, \bibinfo {author} {\bibfnamefont {F.}~\bibnamefont {Hajkarim}}, \
  and\ \bibinfo {author} {\bibfnamefont {S.}~\bibnamefont {Yun}},\ }\href
  {\doibase 10.1103/PhysRevLett.128.152001} {\bibfield  {journal} {\bibinfo
  {journal} {Phys. Rev. Lett.}\ }\textbf {\bibinfo {volume} {128}},\ \bibinfo
  {pages} {152001} (\bibinfo {year} {2022})},\ \Eprint
  {http://arxiv.org/abs/2108.04259} {arXiv:2108.04259 [hep-ph]} \BibitemShut
  {NoStop}%
\bibitem [{\citenamefont {D'Eramo}\ \emph {et~al.}(2021)\citenamefont
  {D'Eramo}, \citenamefont {Hajkarim},\ and\ \citenamefont
  {Yun}}]{DEramo:2021lgb}%
  \BibitemOpen
  \bibfield  {author} {\bibinfo {author} {\bibfnamefont {F.}~\bibnamefont
  {D'Eramo}}, \bibinfo {author} {\bibfnamefont {F.}~\bibnamefont {Hajkarim}}, \
  and\ \bibinfo {author} {\bibfnamefont {S.}~\bibnamefont {Yun}},\ }\href
  {\doibase 10.1007/JHEP10(2021)224} {\bibfield  {journal} {\bibinfo  {journal}
  {JHEP}\ }\textbf {\bibinfo {volume} {10}},\ \bibinfo {pages} {224} (\bibinfo
  {year} {2021})},\ \Eprint {http://arxiv.org/abs/2108.05371} {arXiv:2108.05371
  [hep-ph]} \BibitemShut {NoStop}%
\bibitem [{Note2()}]{Note2}%
  \BibitemOpen
  \bibinfo {note} {The importance of $T_{\scriptscriptstyle \protect \textrm
  {RH}}$ in axion cosmology is well appreciated, see e.g. Refs.~\cite
  {Grin:2007yg,Visinelli:2009kt,Depta:2020wmr,Balazs:2022tjl}}\BibitemShut
  {NoStop}%
\bibitem [{\citenamefont {Hasegawa}\ \emph {et~al.}(2019)\citenamefont
  {Hasegawa}, \citenamefont {Hiroshima}, \citenamefont {Kohri}, \citenamefont
  {Hansen}, \citenamefont {Tram},\ and\ \citenamefont
  {Hannestad}}]{Hasegawa:2019jsa}%
  \BibitemOpen
  \bibfield  {author} {\bibinfo {author} {\bibfnamefont {T.}~\bibnamefont
  {Hasegawa}}, \bibinfo {author} {\bibfnamefont {N.}~\bibnamefont {Hiroshima}},
  \bibinfo {author} {\bibfnamefont {K.}~\bibnamefont {Kohri}}, \bibinfo
  {author} {\bibfnamefont {R.~S.~L.}\ \bibnamefont {Hansen}}, \bibinfo {author}
  {\bibfnamefont {T.}~\bibnamefont {Tram}}, \ and\ \bibinfo {author}
  {\bibfnamefont {S.}~\bibnamefont {Hannestad}},\ }\href {\doibase
  10.1088/1475-7516/2019/12/012} {\bibfield  {journal} {\bibinfo  {journal}
  {JCAP}\ }\textbf {\bibinfo {volume} {12}},\ \bibinfo {pages} {012} (\bibinfo
  {year} {2019})},\ \Eprint {http://arxiv.org/abs/1908.10189} {arXiv:1908.10189
  [hep-ph]} \BibitemShut {NoStop}%
\bibitem [{\citenamefont {de~Salas}\ \emph {et~al.}(2015)\citenamefont
  {de~Salas}, \citenamefont {Lattanzi}, \citenamefont {Mangano}, \citenamefont
  {Miele}, \citenamefont {Pastor},\ and\ \citenamefont
  {Pisanti}}]{deSalas:2015glj}%
  \BibitemOpen
  \bibfield  {author} {\bibinfo {author} {\bibfnamefont {P.~F.}\ \bibnamefont
  {de~Salas}}, \bibinfo {author} {\bibfnamefont {M.}~\bibnamefont {Lattanzi}},
  \bibinfo {author} {\bibfnamefont {G.}~\bibnamefont {Mangano}}, \bibinfo
  {author} {\bibfnamefont {G.}~\bibnamefont {Miele}}, \bibinfo {author}
  {\bibfnamefont {S.}~\bibnamefont {Pastor}}, \ and\ \bibinfo {author}
  {\bibfnamefont {O.}~\bibnamefont {Pisanti}},\ }\href {\doibase
  10.1103/PhysRevD.92.123534} {\bibfield  {journal} {\bibinfo  {journal} {Phys.
  Rev. D}\ }\textbf {\bibinfo {volume} {92}},\ \bibinfo {pages} {123534}
  (\bibinfo {year} {2015})},\ \Eprint {http://arxiv.org/abs/1511.00672}
  {arXiv:1511.00672 [astro-ph.CO]} \BibitemShut {NoStop}%
\bibitem [{\citenamefont {Ichikawa}\ \emph {et~al.}(2007)\citenamefont
  {Ichikawa}, \citenamefont {Kawasaki},\ and\ \citenamefont
  {Takahashi}}]{Ichikawa:2006vm}%
  \BibitemOpen
  \bibfield  {author} {\bibinfo {author} {\bibfnamefont {K.}~\bibnamefont
  {Ichikawa}}, \bibinfo {author} {\bibfnamefont {M.}~\bibnamefont {Kawasaki}},
  \ and\ \bibinfo {author} {\bibfnamefont {F.}~\bibnamefont {Takahashi}},\
  }\href {\doibase 10.1088/1475-7516/2007/05/007} {\bibfield  {journal}
  {\bibinfo  {journal} {JCAP}\ }\textbf {\bibinfo {volume} {05}},\ \bibinfo
  {pages} {007} (\bibinfo {year} {2007})},\ \Eprint
  {http://arxiv.org/abs/astro-ph/0611784} {arXiv:astro-ph/0611784} \BibitemShut
  {NoStop}%
\bibitem [{\citenamefont {Ichikawa}\ \emph {et~al.}(2005)\citenamefont
  {Ichikawa}, \citenamefont {Kawasaki},\ and\ \citenamefont
  {Takahashi}}]{Ichikawa:2005vw}%
  \BibitemOpen
  \bibfield  {author} {\bibinfo {author} {\bibfnamefont {K.}~\bibnamefont
  {Ichikawa}}, \bibinfo {author} {\bibfnamefont {M.}~\bibnamefont {Kawasaki}},
  \ and\ \bibinfo {author} {\bibfnamefont {F.}~\bibnamefont {Takahashi}},\
  }\href {\doibase 10.1103/PhysRevD.72.043522} {\bibfield  {journal} {\bibinfo
  {journal} {Phys. Rev. D}\ }\textbf {\bibinfo {volume} {72}},\ \bibinfo
  {pages} {043522} (\bibinfo {year} {2005})},\ \Eprint
  {http://arxiv.org/abs/astro-ph/0505395} {arXiv:astro-ph/0505395} \BibitemShut
  {NoStop}%
\bibitem [{\citenamefont {Hannestad}(2004)}]{Hannestad:2004px}%
  \BibitemOpen
  \bibfield  {author} {\bibinfo {author} {\bibfnamefont {S.}~\bibnamefont
  {Hannestad}},\ }\href {\doibase 10.1103/PhysRevD.70.043506} {\bibfield
  {journal} {\bibinfo  {journal} {Phys. Rev. D}\ }\textbf {\bibinfo {volume}
  {70}},\ \bibinfo {pages} {043506} (\bibinfo {year} {2004})},\ \Eprint
  {http://arxiv.org/abs/astro-ph/0403291} {arXiv:astro-ph/0403291} \BibitemShut
  {NoStop}%
\bibitem [{\citenamefont {Kawasaki}\ \emph {et~al.}(2000)\citenamefont
  {Kawasaki}, \citenamefont {Kohri},\ and\ \citenamefont
  {Sugiyama}}]{Kawasaki:2000en}%
  \BibitemOpen
  \bibfield  {author} {\bibinfo {author} {\bibfnamefont {M.}~\bibnamefont
  {Kawasaki}}, \bibinfo {author} {\bibfnamefont {K.}~\bibnamefont {Kohri}}, \
  and\ \bibinfo {author} {\bibfnamefont {N.}~\bibnamefont {Sugiyama}},\ }\href
  {\doibase 10.1103/PhysRevD.62.023506} {\bibfield  {journal} {\bibinfo
  {journal} {Phys. Rev. D}\ }\textbf {\bibinfo {volume} {62}},\ \bibinfo
  {pages} {023506} (\bibinfo {year} {2000})},\ \Eprint
  {http://arxiv.org/abs/astro-ph/0002127} {arXiv:astro-ph/0002127} \BibitemShut
  {NoStop}%
\bibitem [{\citenamefont {Kawasaki}\ \emph {et~al.}(1999)\citenamefont
  {Kawasaki}, \citenamefont {Kohri},\ and\ \citenamefont
  {Sugiyama}}]{Kawasaki:1999na}%
  \BibitemOpen
  \bibfield  {author} {\bibinfo {author} {\bibfnamefont {M.}~\bibnamefont
  {Kawasaki}}, \bibinfo {author} {\bibfnamefont {K.}~\bibnamefont {Kohri}}, \
  and\ \bibinfo {author} {\bibfnamefont {N.}~\bibnamefont {Sugiyama}},\ }\href
  {\doibase 10.1103/PhysRevLett.82.4168} {\bibfield  {journal} {\bibinfo
  {journal} {Phys. Rev. Lett.}\ }\textbf {\bibinfo {volume} {82}},\ \bibinfo
  {pages} {4168} (\bibinfo {year} {1999})},\ \Eprint
  {http://arxiv.org/abs/astro-ph/9811437} {arXiv:astro-ph/9811437} \BibitemShut
  {NoStop}%
\bibitem [{\citenamefont {Huang}\ and\ \citenamefont
  {Nath}(2018)}]{Huang:2018cwo}%
  \BibitemOpen
  \bibfield  {author} {\bibinfo {author} {\bibfnamefont {G.-Y.}\ \bibnamefont
  {Huang}}\ and\ \bibinfo {author} {\bibfnamefont {N.}~\bibnamefont {Nath}},\
  }\href {\doibase 10.1140/epjc/s10052-018-6391-y} {\bibfield  {journal}
  {\bibinfo  {journal} {Eur. Phys. J. C}\ }\textbf {\bibinfo {volume} {78}},\
  \bibinfo {pages} {922} (\bibinfo {year} {2018})},\ \Eprint
  {http://arxiv.org/abs/1809.01111} {arXiv:1809.01111 [hep-ph]} \BibitemShut
  {NoStop}%
\bibitem [{\citenamefont {Cadamuro}\ and\ \citenamefont
  {Redondo}(2012)}]{Cadamuro:2011fd}%
  \BibitemOpen
  \bibfield  {author} {\bibinfo {author} {\bibfnamefont {D.}~\bibnamefont
  {Cadamuro}}\ and\ \bibinfo {author} {\bibfnamefont {J.}~\bibnamefont
  {Redondo}},\ }\href {\doibase 10.1088/1475-7516/2012/02/032} {\bibfield
  {journal} {\bibinfo  {journal} {JCAP}\ }\textbf {\bibinfo {volume} {02}},\
  \bibinfo {pages} {032} (\bibinfo {year} {2012})},\ \Eprint
  {http://arxiv.org/abs/1110.2895} {arXiv:1110.2895 [hep-ph]} \BibitemShut
  {NoStop}%
\bibitem [{\citenamefont {Dror}\ \emph {et~al.}(2021)\citenamefont {Dror},
  \citenamefont {Murayama},\ and\ \citenamefont {Rodd}}]{Dror:2021nyr}%
  \BibitemOpen
  \bibfield  {author} {\bibinfo {author} {\bibfnamefont {J.~A.}\ \bibnamefont
  {Dror}}, \bibinfo {author} {\bibfnamefont {H.}~\bibnamefont {Murayama}}, \
  and\ \bibinfo {author} {\bibfnamefont {N.~L.}\ \bibnamefont {Rodd}},\ }\href
  {\doibase 10.1103/PhysRevD.103.115004} {\bibfield  {journal} {\bibinfo
  {journal} {Phys. Rev. D}\ }\textbf {\bibinfo {volume} {103}},\ \bibinfo
  {pages} {115004} (\bibinfo {year} {2021})},\ \Eprint
  {http://arxiv.org/abs/2101.09287} {arXiv:2101.09287 [hep-ph]} \BibitemShut
  {NoStop}%
\bibitem [{Note3()}]{Note3}%
  \BibitemOpen
  \bibinfo {note} {The ``cosmological triangle'' in the vicinity of $m_a \sim
  1~{\protect \rm GeV}$ and $g_{a\gamma \gamma }\sim 10^{-5}~{\protect \rm
  GeV}^{-1}$, see e.g. Refs.~\cite
  {Cadamuro:2011fd,Millea:2015qra,Dolan:2017osp,Brdar:2020dpr,Depta:2020wmr,Dolan:2021rya},
  has been closed~\cite {Caputo:2021rux}}\BibitemShut {NoStop}%
\bibitem [{\citenamefont {Gondolo}\ and\ \citenamefont
  {Gelmini}(1991)}]{Gondolo:1990dk}%
  \BibitemOpen
  \bibfield  {author} {\bibinfo {author} {\bibfnamefont {P.}~\bibnamefont
  {Gondolo}}\ and\ \bibinfo {author} {\bibfnamefont {G.}~\bibnamefont
  {Gelmini}},\ }\href {\doibase 10.1016/0550-3213(91)90438-4} {\bibfield
  {journal} {\bibinfo  {journal} {Nucl. Phys. B}\ }\textbf {\bibinfo {volume}
  {360}},\ \bibinfo {pages} {145} (\bibinfo {year} {1991})}\BibitemShut
  {NoStop}%
\bibitem [{\citenamefont {Edsjo}\ and\ \citenamefont
  {Gondolo}(1997)}]{Edsjo:1997bg}%
  \BibitemOpen
  \bibfield  {author} {\bibinfo {author} {\bibfnamefont {J.}~\bibnamefont
  {Edsjo}}\ and\ \bibinfo {author} {\bibfnamefont {P.}~\bibnamefont
  {Gondolo}},\ }\href {\doibase 10.1103/PhysRevD.56.1879} {\bibfield  {journal}
  {\bibinfo  {journal} {Phys. Rev. D}\ }\textbf {\bibinfo {volume} {56}},\
  \bibinfo {pages} {1879} (\bibinfo {year} {1997})},\ \Eprint
  {http://arxiv.org/abs/hep-ph/9704361} {arXiv:hep-ph/9704361} \BibitemShut
  {NoStop}%
\bibitem [{\citenamefont {Hall}\ \emph {et~al.}(2010)\citenamefont {Hall},
  \citenamefont {Jedamzik}, \citenamefont {March-Russell},\ and\ \citenamefont
  {West}}]{Hall:2009bx}%
  \BibitemOpen
  \bibfield  {author} {\bibinfo {author} {\bibfnamefont {L.~J.}\ \bibnamefont
  {Hall}}, \bibinfo {author} {\bibfnamefont {K.}~\bibnamefont {Jedamzik}},
  \bibinfo {author} {\bibfnamefont {J.}~\bibnamefont {March-Russell}}, \ and\
  \bibinfo {author} {\bibfnamefont {S.~M.}\ \bibnamefont {West}},\ }\href
  {\doibase 10.1007/JHEP03(2010)080} {\bibfield  {journal} {\bibinfo  {journal}
  {JHEP}\ }\textbf {\bibinfo {volume} {03}},\ \bibinfo {pages} {080} (\bibinfo
  {year} {2010})},\ \Eprint {http://arxiv.org/abs/0911.1120} {arXiv:0911.1120
  [hep-ph]} \BibitemShut {NoStop}%
\bibitem [{\citenamefont {Cadamuro}\ \emph {et~al.}(2011)\citenamefont
  {Cadamuro}, \citenamefont {Hannestad}, \citenamefont {Raffelt},\ and\
  \citenamefont {Redondo}}]{Cadamuro:2010cz}%
  \BibitemOpen
  \bibfield  {author} {\bibinfo {author} {\bibfnamefont {D.}~\bibnamefont
  {Cadamuro}}, \bibinfo {author} {\bibfnamefont {S.}~\bibnamefont {Hannestad}},
  \bibinfo {author} {\bibfnamefont {G.}~\bibnamefont {Raffelt}}, \ and\
  \bibinfo {author} {\bibfnamefont {J.}~\bibnamefont {Redondo}},\ }\href
  {\doibase 10.1088/1475-7516/2011/02/003} {\bibfield  {journal} {\bibinfo
  {journal} {JCAP}\ }\textbf {\bibinfo {volume} {02}},\ \bibinfo {pages} {003}
  (\bibinfo {year} {2011})},\ \Eprint {http://arxiv.org/abs/1011.3694}
  {arXiv:1011.3694 [hep-ph]} \BibitemShut {NoStop}%
\bibitem [{\citenamefont {Blennow}\ \emph {et~al.}(2014)\citenamefont
  {Blennow}, \citenamefont {Fernandez-Mart{\'\i ne}z},\ and\ \citenamefont
  {Zaldivar}}]{Blennow:2013jba}%
  \BibitemOpen
  \bibfield  {author} {\bibinfo {author} {\bibfnamefont {M.}~\bibnamefont
  {Blennow}}, \bibinfo {author} {\bibfnamefont {E.}~\bibnamefont
  {Fernandez-Mart{\'\i ne}z}}, \ and\ \bibinfo {author} {\bibfnamefont
  {B.}~\bibnamefont {Zaldivar}},\ }\href {\doibase
  10.1088/1475-7516/2014/01/003} {\bibfield  {journal} {\bibinfo  {journal}
  {JCAP}\ }\textbf {\bibinfo {volume} {01}},\ \bibinfo {pages} {003} (\bibinfo
  {year} {2014})},\ \Eprint {http://arxiv.org/abs/1309.7348} {arXiv:1309.7348
  [hep-ph]} \BibitemShut {NoStop}%
\bibitem [{\citenamefont {D'Eramo}\ \emph
  {et~al.}(2018{\natexlab{a}})\citenamefont {D'Eramo}, \citenamefont
  {Fernandez},\ and\ \citenamefont {Profumo}}]{DEramo:2017ecx}%
  \BibitemOpen
  \bibfield  {author} {\bibinfo {author} {\bibfnamefont {F.}~\bibnamefont
  {D'Eramo}}, \bibinfo {author} {\bibfnamefont {N.}~\bibnamefont {Fernandez}},
  \ and\ \bibinfo {author} {\bibfnamefont {S.}~\bibnamefont {Profumo}},\ }\href
  {\doibase 10.1088/1475-7516/2018/02/046} {\bibfield  {journal} {\bibinfo
  {journal} {JCAP}\ }\textbf {\bibinfo {volume} {02}},\ \bibinfo {pages} {046}
  (\bibinfo {year} {2018}{\natexlab{a}})},\ \Eprint
  {http://arxiv.org/abs/1712.07453} {arXiv:1712.07453 [hep-ph]} \BibitemShut
  {NoStop}%
\bibitem [{\citenamefont {D'Eramo}\ \emph
  {et~al.}(2018{\natexlab{b}})\citenamefont {D'Eramo}, \citenamefont
  {Ferreira}, \citenamefont {Notari},\ and\ \citenamefont
  {Bernal}}]{DEramo:2018vss}%
  \BibitemOpen
  \bibfield  {author} {\bibinfo {author} {\bibfnamefont {F.}~\bibnamefont
  {D'Eramo}}, \bibinfo {author} {\bibfnamefont {R.~Z.}\ \bibnamefont
  {Ferreira}}, \bibinfo {author} {\bibfnamefont {A.}~\bibnamefont {Notari}}, \
  and\ \bibinfo {author} {\bibfnamefont {J.~L.}\ \bibnamefont {Bernal}},\
  }\href {\doibase 10.1088/1475-7516/2018/11/014} {\bibfield  {journal}
  {\bibinfo  {journal} {JCAP}\ }\textbf {\bibinfo {volume} {11}},\ \bibinfo
  {pages} {014} (\bibinfo {year} {2018}{\natexlab{b}})},\ \Eprint
  {http://arxiv.org/abs/1808.07430} {arXiv:1808.07430 [hep-ph]} \BibitemShut
  {NoStop}%
\bibitem [{\citenamefont {Depta}\ \emph {et~al.}(2020)\citenamefont {Depta},
  \citenamefont {Hufnagel},\ and\ \citenamefont
  {Schmidt-Hoberg}}]{Depta:2020wmr}%
  \BibitemOpen
  \bibfield  {author} {\bibinfo {author} {\bibfnamefont {P.~F.}\ \bibnamefont
  {Depta}}, \bibinfo {author} {\bibfnamefont {M.}~\bibnamefont {Hufnagel}}, \
  and\ \bibinfo {author} {\bibfnamefont {K.}~\bibnamefont {Schmidt-Hoberg}},\
  }\href {\doibase 10.1088/1475-7516/2020/05/009} {\bibfield  {journal}
  {\bibinfo  {journal} {JCAP}\ }\textbf {\bibinfo {volume} {05}},\ \bibinfo
  {pages} {009} (\bibinfo {year} {2020})},\ \Eprint
  {http://arxiv.org/abs/2002.08370} {arXiv:2002.08370 [hep-ph]} \BibitemShut
  {NoStop}%
\bibitem [{\citenamefont {Bolz}\ \emph {et~al.}(2001)\citenamefont {Bolz},
  \citenamefont {Brandenburg},\ and\ \citenamefont {Buchmuller}}]{Bolz:2000fu}%
  \BibitemOpen
  \bibfield  {author} {\bibinfo {author} {\bibfnamefont {M.}~\bibnamefont
  {Bolz}}, \bibinfo {author} {\bibfnamefont {A.}~\bibnamefont {Brandenburg}}, \
  and\ \bibinfo {author} {\bibfnamefont {W.}~\bibnamefont {Buchmuller}},\
  }\href {\doibase 10.1016/S0550-3213(01)00132-8} {\bibfield  {journal}
  {\bibinfo  {journal} {Nucl. Phys. B}\ }\textbf {\bibinfo {volume} {606}},\
  \bibinfo {pages} {518} (\bibinfo {year} {2001})},\ \bibinfo {note} {[Erratum:
  Nucl.Phys.B 790, 336--337 (2008)]},\ \Eprint
  {http://arxiv.org/abs/hep-ph/0012052} {arXiv:hep-ph/0012052} \BibitemShut
  {NoStop}%
\bibitem [{\citenamefont {Dunsky}\ \emph {et~al.}(2022)\citenamefont {Dunsky},
  \citenamefont {Hall},\ and\ \citenamefont {Harigaya}}]{Dunsky:2022uoq}%
  \BibitemOpen
  \bibfield  {author} {\bibinfo {author} {\bibfnamefont {D.~I.}\ \bibnamefont
  {Dunsky}}, \bibinfo {author} {\bibfnamefont {L.~J.}\ \bibnamefont {Hall}}, \
  and\ \bibinfo {author} {\bibfnamefont {K.}~\bibnamefont {Harigaya}},\
  }\href@noop {} {\  (\bibinfo {year} {2022})},\ \Eprint
  {http://arxiv.org/abs/2205.11540} {arXiv:2205.11540 [hep-ph]} \BibitemShut
  {NoStop}%
\bibitem [{Pub()}]{PubCode}%
  \BibitemOpen
  \href@noop {} {}\bibinfo {howpublished}
  {\url{https://github.com/nickrodd/IrreducibleAxionAbundance}}\BibitemShut
  {NoStop}%
\bibitem [{\citenamefont {Ferreira}\ \emph {et~al.}(2022)\citenamefont
  {Ferreira}, \citenamefont {Marsh},\ and\ \citenamefont
  {M\"uller}}]{Ferreira:2022xlw}%
  \BibitemOpen
  \bibfield  {author} {\bibinfo {author} {\bibfnamefont {R.~Z.}\ \bibnamefont
  {Ferreira}}, \bibinfo {author} {\bibfnamefont {M.~C.~D.}\ \bibnamefont
  {Marsh}}, \ and\ \bibinfo {author} {\bibfnamefont {E.}~\bibnamefont
  {M\"uller}},\ }\href@noop {} {\  (\bibinfo {year} {2022})},\ \Eprint
  {http://arxiv.org/abs/2205.07896} {arXiv:2205.07896 [hep-ph]} \BibitemShut
  {NoStop}%
\bibitem [{\citenamefont {Capozzi}\ and\ \citenamefont
  {Raffelt}(2020)}]{Capozzi:2020cbu}%
  \BibitemOpen
  \bibfield  {author} {\bibinfo {author} {\bibfnamefont {F.}~\bibnamefont
  {Capozzi}}\ and\ \bibinfo {author} {\bibfnamefont {G.}~\bibnamefont
  {Raffelt}},\ }\href {\doibase 10.1103/PhysRevD.102.083007} {\bibfield
  {journal} {\bibinfo  {journal} {Phys. Rev. D}\ }\textbf {\bibinfo {volume}
  {102}},\ \bibinfo {pages} {083007} (\bibinfo {year} {2020})},\ \Eprint
  {http://arxiv.org/abs/2007.03694} {arXiv:2007.03694 [astro-ph.SR]}
  \BibitemShut {NoStop}%
\bibitem [{\citenamefont {Raffelt}(1990)}]{Raffelt:1990yz}%
  \BibitemOpen
  \bibfield  {author} {\bibinfo {author} {\bibfnamefont {G.~G.}\ \bibnamefont
  {Raffelt}},\ }\href {\doibase 10.1016/0370-1573(90)90054-6} {\bibfield
  {journal} {\bibinfo  {journal} {Phys. Rept.}\ }\textbf {\bibinfo {volume}
  {198}},\ \bibinfo {pages} {1} (\bibinfo {year} {1990})}\BibitemShut {NoStop}%
\bibitem [{\citenamefont {Bauer}\ \emph {et~al.}(2017)\citenamefont {Bauer},
  \citenamefont {Neubert},\ and\ \citenamefont {Thamm}}]{Bauer:2017ris}%
  \BibitemOpen
  \bibfield  {author} {\bibinfo {author} {\bibfnamefont {M.}~\bibnamefont
  {Bauer}}, \bibinfo {author} {\bibfnamefont {M.}~\bibnamefont {Neubert}}, \
  and\ \bibinfo {author} {\bibfnamefont {A.}~\bibnamefont {Thamm}},\ }\href
  {\doibase 10.1007/JHEP12(2017)044} {\bibfield  {journal} {\bibinfo  {journal}
  {JHEP}\ }\textbf {\bibinfo {volume} {12}},\ \bibinfo {pages} {044} (\bibinfo
  {year} {2017})},\ \Eprint {http://arxiv.org/abs/1708.00443} {arXiv:1708.00443
  [hep-ph]} \BibitemShut {NoStop}%
\bibitem [{\citenamefont {Boyarsky}\ \emph
  {et~al.}(2009{\natexlab{a}})\citenamefont {Boyarsky}, \citenamefont
  {Lesgourgues}, \citenamefont {Ruchayskiy},\ and\ \citenamefont
  {Viel}}]{Boyarsky:2008mt}%
  \BibitemOpen
  \bibfield  {author} {\bibinfo {author} {\bibfnamefont {A.}~\bibnamefont
  {Boyarsky}}, \bibinfo {author} {\bibfnamefont {J.}~\bibnamefont
  {Lesgourgues}}, \bibinfo {author} {\bibfnamefont {O.}~\bibnamefont
  {Ruchayskiy}}, \ and\ \bibinfo {author} {\bibfnamefont {M.}~\bibnamefont
  {Viel}},\ }\href {\doibase 10.1103/PhysRevLett.102.201304} {\bibfield
  {journal} {\bibinfo  {journal} {Phys. Rev. Lett.}\ }\textbf {\bibinfo
  {volume} {102}},\ \bibinfo {pages} {201304} (\bibinfo {year}
  {2009}{\natexlab{a}})},\ \Eprint {http://arxiv.org/abs/0812.3256}
  {arXiv:0812.3256 [hep-ph]} \BibitemShut {NoStop}%
\bibitem [{\citenamefont {Boyarsky}\ \emph
  {et~al.}(2009{\natexlab{b}})\citenamefont {Boyarsky}, \citenamefont
  {Lesgourgues}, \citenamefont {Ruchayskiy},\ and\ \citenamefont
  {Viel}}]{Boyarsky:2008xj}%
  \BibitemOpen
  \bibfield  {author} {\bibinfo {author} {\bibfnamefont {A.}~\bibnamefont
  {Boyarsky}}, \bibinfo {author} {\bibfnamefont {J.}~\bibnamefont
  {Lesgourgues}}, \bibinfo {author} {\bibfnamefont {O.}~\bibnamefont
  {Ruchayskiy}}, \ and\ \bibinfo {author} {\bibfnamefont {M.}~\bibnamefont
  {Viel}},\ }\href {\doibase 10.1088/1475-7516/2009/05/012} {\bibfield
  {journal} {\bibinfo  {journal} {JCAP}\ }\textbf {\bibinfo {volume} {05}},\
  \bibinfo {pages} {012} (\bibinfo {year} {2009}{\natexlab{b}})},\ \Eprint
  {http://arxiv.org/abs/0812.0010} {arXiv:0812.0010 [astro-ph]} \BibitemShut
  {NoStop}%
\bibitem [{Note4()}]{Note4}%
  \BibitemOpen
  \bibinfo {note} {The constraints that exist on light axion DM being too hot
  if it was produced through freeze-in (see e.g. Refs.~\cite
  {DEramo:2020gpr,Ballesteros:2020adh,Baumholzer:2020hvx}) decouple for
  ${\protect \cal F}_a \lesssim 0.1$~\cite {DEramo:2020gpr}.}\BibitemShut
  {Stop}%
\bibitem [{\citenamefont {Anderhalden}\ \emph {et~al.}(2012)\citenamefont
  {Anderhalden}, \citenamefont {Diemand}, \citenamefont {Bertone},
  \citenamefont {Maccio},\ and\ \citenamefont
  {Schneider}}]{Anderhalden:2012qt}%
  \BibitemOpen
  \bibfield  {author} {\bibinfo {author} {\bibfnamefont {D.}~\bibnamefont
  {Anderhalden}}, \bibinfo {author} {\bibfnamefont {J.}~\bibnamefont
  {Diemand}}, \bibinfo {author} {\bibfnamefont {G.}~\bibnamefont {Bertone}},
  \bibinfo {author} {\bibfnamefont {A.~V.}\ \bibnamefont {Maccio}}, \ and\
  \bibinfo {author} {\bibfnamefont {A.}~\bibnamefont {Schneider}},\ }\href
  {\doibase 10.1088/1475-7516/2012/10/047} {\bibfield  {journal} {\bibinfo
  {journal} {JCAP}\ }\textbf {\bibinfo {volume} {10}},\ \bibinfo {pages} {047}
  (\bibinfo {year} {2012})},\ \Eprint {http://arxiv.org/abs/1206.3788}
  {arXiv:1206.3788 [astro-ph.CO]} \BibitemShut {NoStop}%
\bibitem [{\citenamefont {Wadekar}\ and\ \citenamefont
  {Wang}(2021)}]{Wadekar:2021qae}%
  \BibitemOpen
  \bibfield  {author} {\bibinfo {author} {\bibfnamefont {D.}~\bibnamefont
  {Wadekar}}\ and\ \bibinfo {author} {\bibfnamefont {Z.}~\bibnamefont {Wang}},\
  }\href@noop {} {\  (\bibinfo {year} {2021})},\ \Eprint
  {http://arxiv.org/abs/2111.08025} {arXiv:2111.08025 [hep-ph]} \BibitemShut
  {NoStop}%
\bibitem [{\citenamefont {Lisanti}\ \emph {et~al.}(2018)\citenamefont
  {Lisanti}, \citenamefont {Mishra-Sharma}, \citenamefont {Rodd}, \citenamefont
  {Safdi},\ and\ \citenamefont {Wechsler}}]{Lisanti:2017qoz}%
  \BibitemOpen
  \bibfield  {author} {\bibinfo {author} {\bibfnamefont {M.}~\bibnamefont
  {Lisanti}}, \bibinfo {author} {\bibfnamefont {S.}~\bibnamefont
  {Mishra-Sharma}}, \bibinfo {author} {\bibfnamefont {N.~L.}\ \bibnamefont
  {Rodd}}, \bibinfo {author} {\bibfnamefont {B.~R.}\ \bibnamefont {Safdi}}, \
  and\ \bibinfo {author} {\bibfnamefont {R.~H.}\ \bibnamefont {Wechsler}},\
  }\href {\doibase 10.1103/PhysRevD.97.063005} {\bibfield  {journal} {\bibinfo
  {journal} {Phys. Rev. D}\ }\textbf {\bibinfo {volume} {97}},\ \bibinfo
  {pages} {063005} (\bibinfo {year} {2018})},\ \Eprint
  {http://arxiv.org/abs/1709.00416} {arXiv:1709.00416 [astro-ph.CO]}
  \BibitemShut {NoStop}%
\bibitem [{\citenamefont {Thorpe-Morgan}\ \emph {et~al.}(2020)\citenamefont
  {Thorpe-Morgan}, \citenamefont {Malyshev}, \citenamefont {Santangelo},
  \citenamefont {Jochum}, \citenamefont {J\"ager}, \citenamefont {Sasaki},\
  and\ \citenamefont {Saeedi}}]{Thorpe-Morgan:2020rwc}%
  \BibitemOpen
  \bibfield  {author} {\bibinfo {author} {\bibfnamefont {C.}~\bibnamefont
  {Thorpe-Morgan}}, \bibinfo {author} {\bibfnamefont {D.}~\bibnamefont
  {Malyshev}}, \bibinfo {author} {\bibfnamefont {A.}~\bibnamefont
  {Santangelo}}, \bibinfo {author} {\bibfnamefont {J.}~\bibnamefont {Jochum}},
  \bibinfo {author} {\bibfnamefont {B.}~\bibnamefont {J\"ager}}, \bibinfo
  {author} {\bibfnamefont {M.}~\bibnamefont {Sasaki}}, \ and\ \bibinfo {author}
  {\bibfnamefont {S.}~\bibnamefont {Saeedi}},\ }\href {\doibase
  10.1103/PhysRevD.102.123003} {\bibfield  {journal} {\bibinfo  {journal}
  {Phys. Rev. D}\ }\textbf {\bibinfo {volume} {102}},\ \bibinfo {pages}
  {123003} (\bibinfo {year} {2020})},\ \Eprint
  {http://arxiv.org/abs/2008.08306} {arXiv:2008.08306 [astro-ph.HE]}
  \BibitemShut {NoStop}%
\bibitem [{\citenamefont {Dekker}\ \emph {et~al.}(2021)\citenamefont {Dekker},
  \citenamefont {Peerbooms}, \citenamefont {Zimmer}, \citenamefont {Ng},\ and\
  \citenamefont {Ando}}]{Dekker:2021bos}%
  \BibitemOpen
  \bibfield  {author} {\bibinfo {author} {\bibfnamefont {A.}~\bibnamefont
  {Dekker}}, \bibinfo {author} {\bibfnamefont {E.}~\bibnamefont {Peerbooms}},
  \bibinfo {author} {\bibfnamefont {F.}~\bibnamefont {Zimmer}}, \bibinfo
  {author} {\bibfnamefont {K.~C.~Y.}\ \bibnamefont {Ng}}, \ and\ \bibinfo
  {author} {\bibfnamefont {S.}~\bibnamefont {Ando}},\ }\href {\doibase
  10.1103/PhysRevD.104.023021} {\bibfield  {journal} {\bibinfo  {journal}
  {Phys. Rev. D}\ }\textbf {\bibinfo {volume} {104}},\ \bibinfo {pages}
  {023021} (\bibinfo {year} {2021})},\ \Eprint
  {http://arxiv.org/abs/2103.13241} {arXiv:2103.13241 [astro-ph.HE]}
  \BibitemShut {NoStop}%
\bibitem [{\citenamefont {Boddy}\ \emph {et~al.}(2022)\citenamefont {Boddy}
  \emph {et~al.}}]{Boddy:2022knd}%
  \BibitemOpen
  \bibfield  {author} {\bibinfo {author} {\bibfnamefont {K.~K.}\ \bibnamefont
  {Boddy}} \emph {et~al.},\ }in\ \href@noop {} {\emph {\bibinfo {booktitle}
  {{2022 Snowmass Summer Study}}}}\ (\bibinfo {year} {2022})\ \Eprint
  {http://arxiv.org/abs/2203.06380} {arXiv:2203.06380 [hep-ph]} \BibitemShut
  {NoStop}%
\bibitem [{\citenamefont {Cirelli}\ \emph {et~al.}(2011)\citenamefont
  {Cirelli}, \citenamefont {Corcella}, \citenamefont {Hektor}, \citenamefont
  {Hutsi}, \citenamefont {Kadastik}, \citenamefont {Panci}, \citenamefont
  {Raidal}, \citenamefont {Sala},\ and\ \citenamefont
  {Strumia}}]{Cirelli:2010xx}%
  \BibitemOpen
  \bibfield  {author} {\bibinfo {author} {\bibfnamefont {M.}~\bibnamefont
  {Cirelli}}, \bibinfo {author} {\bibfnamefont {G.}~\bibnamefont {Corcella}},
  \bibinfo {author} {\bibfnamefont {A.}~\bibnamefont {Hektor}}, \bibinfo
  {author} {\bibfnamefont {G.}~\bibnamefont {Hutsi}}, \bibinfo {author}
  {\bibfnamefont {M.}~\bibnamefont {Kadastik}}, \bibinfo {author}
  {\bibfnamefont {P.}~\bibnamefont {Panci}}, \bibinfo {author} {\bibfnamefont
  {M.}~\bibnamefont {Raidal}}, \bibinfo {author} {\bibfnamefont
  {F.}~\bibnamefont {Sala}}, \ and\ \bibinfo {author} {\bibfnamefont
  {A.}~\bibnamefont {Strumia}},\ }\href {\doibase
  10.1088/1475-7516/2012/10/E01} {\bibfield  {journal} {\bibinfo  {journal}
  {JCAP}\ }\textbf {\bibinfo {volume} {03}},\ \bibinfo {pages} {051} (\bibinfo
  {year} {2011})},\ \bibinfo {note} {[Erratum: JCAP 10, E01 (2012)]},\ \Eprint
  {http://arxiv.org/abs/1012.4515} {arXiv:1012.4515 [hep-ph]} \BibitemShut
  {NoStop}%
\bibitem [{\citenamefont {Van~Tilburg}(2021)}]{VanTilburg:2020jvl}%
  \BibitemOpen
  \bibfield  {author} {\bibinfo {author} {\bibfnamefont {K.}~\bibnamefont
  {Van~Tilburg}},\ }\href {\doibase 10.1103/PhysRevD.104.023019} {\bibfield
  {journal} {\bibinfo  {journal} {Phys. Rev. D}\ }\textbf {\bibinfo {volume}
  {104}},\ \bibinfo {pages} {023019} (\bibinfo {year} {2021})},\ \Eprint
  {http://arxiv.org/abs/2006.12431} {arXiv:2006.12431 [hep-ph]} \BibitemShut
  {NoStop}%
\bibitem [{\citenamefont {Dodelson}\ and\ \citenamefont
  {Widrow}(1994)}]{Dodelson:1993je}%
  \BibitemOpen
  \bibfield  {author} {\bibinfo {author} {\bibfnamefont {S.}~\bibnamefont
  {Dodelson}}\ and\ \bibinfo {author} {\bibfnamefont {L.~M.}\ \bibnamefont
  {Widrow}},\ }\href {\doibase 10.1103/PhysRevLett.72.17} {\bibfield  {journal}
  {\bibinfo  {journal} {Phys. Rev. Lett.}\ }\textbf {\bibinfo {volume} {72}},\
  \bibinfo {pages} {17} (\bibinfo {year} {1994})},\ \Eprint
  {http://arxiv.org/abs/hep-ph/9303287} {arXiv:hep-ph/9303287} \BibitemShut
  {NoStop}%
\bibitem [{\citenamefont {Gelmini}\ \emph {et~al.}(2008)\citenamefont
  {Gelmini}, \citenamefont {Osoba}, \citenamefont {Palomares-Ruiz},\ and\
  \citenamefont {Pascoli}}]{Gelmini:2008fq}%
  \BibitemOpen
  \bibfield  {author} {\bibinfo {author} {\bibfnamefont {G.}~\bibnamefont
  {Gelmini}}, \bibinfo {author} {\bibfnamefont {E.}~\bibnamefont {Osoba}},
  \bibinfo {author} {\bibfnamefont {S.}~\bibnamefont {Palomares-Ruiz}}, \ and\
  \bibinfo {author} {\bibfnamefont {S.}~\bibnamefont {Pascoli}},\ }\href
  {\doibase 10.1088/1475-7516/2008/10/029} {\bibfield  {journal} {\bibinfo
  {journal} {JCAP}\ }\textbf {\bibinfo {volume} {10}},\ \bibinfo {pages} {029}
  (\bibinfo {year} {2008})},\ \Eprint {http://arxiv.org/abs/0803.2735}
  {arXiv:0803.2735 [astro-ph]} \BibitemShut {NoStop}%
\bibitem [{\citenamefont {Gelmini}\ \emph {et~al.}(2019)\citenamefont
  {Gelmini}, \citenamefont {Lu},\ and\ \citenamefont
  {Takhistov}}]{Gelmini:2019wfp}%
  \BibitemOpen
  \bibfield  {author} {\bibinfo {author} {\bibfnamefont {G.~B.}\ \bibnamefont
  {Gelmini}}, \bibinfo {author} {\bibfnamefont {P.}~\bibnamefont {Lu}}, \ and\
  \bibinfo {author} {\bibfnamefont {V.}~\bibnamefont {Takhistov}},\ }\href
  {\doibase 10.1088/1475-7516/2019/12/047} {\bibfield  {journal} {\bibinfo
  {journal} {JCAP}\ }\textbf {\bibinfo {volume} {12}},\ \bibinfo {pages} {047}
  (\bibinfo {year} {2019})},\ \Eprint {http://arxiv.org/abs/1909.13328}
  {arXiv:1909.13328 [hep-ph]} \BibitemShut {NoStop}%
\bibitem [{\citenamefont {O'Hare}(2020)}]{AxionLimits}%
  \BibitemOpen
  \bibfield  {author} {\bibinfo {author} {\bibfnamefont {C.}~\bibnamefont
  {O'Hare}},\ }\href {\doibase 10.5281/zenodo.3932430} {\enquote {\bibinfo
  {title} {cajohare/axionlimits: Axionlimits},}\ }\bibinfo {howpublished}
  {\url{https://cajohare.github.io/AxionLimits/}} (\bibinfo {year}
  {2020})\BibitemShut {NoStop}%
\bibitem [{\citenamefont {Grin}\ \emph {et~al.}(2008)\citenamefont {Grin},
  \citenamefont {Smith},\ and\ \citenamefont {Kamionkowski}}]{Grin:2007yg}%
  \BibitemOpen
  \bibfield  {author} {\bibinfo {author} {\bibfnamefont {D.}~\bibnamefont
  {Grin}}, \bibinfo {author} {\bibfnamefont {T.~L.}\ \bibnamefont {Smith}}, \
  and\ \bibinfo {author} {\bibfnamefont {M.}~\bibnamefont {Kamionkowski}},\
  }\href {\doibase 10.1103/PhysRevD.77.085020} {\bibfield  {journal} {\bibinfo
  {journal} {Phys. Rev. D}\ }\textbf {\bibinfo {volume} {77}},\ \bibinfo
  {pages} {085020} (\bibinfo {year} {2008})},\ \Eprint
  {http://arxiv.org/abs/0711.1352} {arXiv:0711.1352 [astro-ph]} \BibitemShut
  {NoStop}%
\bibitem [{\citenamefont {Visinelli}\ and\ \citenamefont
  {Gondolo}(2010)}]{Visinelli:2009kt}%
  \BibitemOpen
  \bibfield  {author} {\bibinfo {author} {\bibfnamefont {L.}~\bibnamefont
  {Visinelli}}\ and\ \bibinfo {author} {\bibfnamefont {P.}~\bibnamefont
  {Gondolo}},\ }\href {\doibase 10.1103/PhysRevD.81.063508} {\bibfield
  {journal} {\bibinfo  {journal} {Phys. Rev. D}\ }\textbf {\bibinfo {volume}
  {81}},\ \bibinfo {pages} {063508} (\bibinfo {year} {2010})},\ \Eprint
  {http://arxiv.org/abs/0912.0015} {arXiv:0912.0015 [astro-ph.CO]} \BibitemShut
  {NoStop}%
\bibitem [{\citenamefont {Millea}\ \emph {et~al.}(2015)\citenamefont {Millea},
  \citenamefont {Knox},\ and\ \citenamefont {Fields}}]{Millea:2015qra}%
  \BibitemOpen
  \bibfield  {author} {\bibinfo {author} {\bibfnamefont {M.}~\bibnamefont
  {Millea}}, \bibinfo {author} {\bibfnamefont {L.}~\bibnamefont {Knox}}, \ and\
  \bibinfo {author} {\bibfnamefont {B.}~\bibnamefont {Fields}},\ }\href
  {\doibase 10.1103/PhysRevD.92.023010} {\bibfield  {journal} {\bibinfo
  {journal} {Phys. Rev. D}\ }\textbf {\bibinfo {volume} {92}},\ \bibinfo
  {pages} {023010} (\bibinfo {year} {2015})},\ \Eprint
  {http://arxiv.org/abs/1501.04097} {arXiv:1501.04097 [astro-ph.CO]}
  \BibitemShut {NoStop}%
\bibitem [{\citenamefont {Dolan}\ \emph {et~al.}(2017)\citenamefont {Dolan},
  \citenamefont {Ferber}, \citenamefont {Hearty}, \citenamefont {Kahlhoefer},\
  and\ \citenamefont {Schmidt-Hoberg}}]{Dolan:2017osp}%
  \BibitemOpen
  \bibfield  {author} {\bibinfo {author} {\bibfnamefont {M.~J.}\ \bibnamefont
  {Dolan}}, \bibinfo {author} {\bibfnamefont {T.}~\bibnamefont {Ferber}},
  \bibinfo {author} {\bibfnamefont {C.}~\bibnamefont {Hearty}}, \bibinfo
  {author} {\bibfnamefont {F.}~\bibnamefont {Kahlhoefer}}, \ and\ \bibinfo
  {author} {\bibfnamefont {K.}~\bibnamefont {Schmidt-Hoberg}},\ }\href
  {\doibase 10.1007/JHEP12(2017)094} {\bibfield  {journal} {\bibinfo  {journal}
  {JHEP}\ }\textbf {\bibinfo {volume} {12}},\ \bibinfo {pages} {094} (\bibinfo
  {year} {2017})},\ \bibinfo {note} {[Erratum: JHEP 03, 190 (2021)]},\ \Eprint
  {http://arxiv.org/abs/1709.00009} {arXiv:1709.00009 [hep-ph]} \BibitemShut
  {NoStop}%
\bibitem [{\citenamefont {Brdar}\ \emph {et~al.}(2021)\citenamefont {Brdar},
  \citenamefont {Dutta}, \citenamefont {Jang}, \citenamefont {Kim},
  \citenamefont {Shoemaker}, \citenamefont {Tabrizi}, \citenamefont
  {Thompson},\ and\ \citenamefont {Yu}}]{Brdar:2020dpr}%
  \BibitemOpen
  \bibfield  {author} {\bibinfo {author} {\bibfnamefont {V.}~\bibnamefont
  {Brdar}}, \bibinfo {author} {\bibfnamefont {B.}~\bibnamefont {Dutta}},
  \bibinfo {author} {\bibfnamefont {W.}~\bibnamefont {Jang}}, \bibinfo {author}
  {\bibfnamefont {D.}~\bibnamefont {Kim}}, \bibinfo {author} {\bibfnamefont
  {I.~M.}\ \bibnamefont {Shoemaker}}, \bibinfo {author} {\bibfnamefont
  {Z.}~\bibnamefont {Tabrizi}}, \bibinfo {author} {\bibfnamefont
  {A.}~\bibnamefont {Thompson}}, \ and\ \bibinfo {author} {\bibfnamefont
  {J.}~\bibnamefont {Yu}},\ }\href {\doibase 10.1103/PhysRevLett.126.201801}
  {\bibfield  {journal} {\bibinfo  {journal} {Phys. Rev. Lett.}\ }\textbf
  {\bibinfo {volume} {126}},\ \bibinfo {pages} {201801} (\bibinfo {year}
  {2021})},\ \Eprint {http://arxiv.org/abs/2011.07054} {arXiv:2011.07054
  [hep-ph]} \BibitemShut {NoStop}%
\bibitem [{\citenamefont {Dolan}\ \emph {et~al.}(2021)\citenamefont {Dolan},
  \citenamefont {Hiskens},\ and\ \citenamefont {Volkas}}]{Dolan:2021rya}%
  \BibitemOpen
  \bibfield  {author} {\bibinfo {author} {\bibfnamefont {M.~J.}\ \bibnamefont
  {Dolan}}, \bibinfo {author} {\bibfnamefont {F.~J.}\ \bibnamefont {Hiskens}},
  \ and\ \bibinfo {author} {\bibfnamefont {R.~R.}\ \bibnamefont {Volkas}},\
  }\href {\doibase 10.1088/1475-7516/2021/09/010} {\bibfield  {journal}
  {\bibinfo  {journal} {JCAP}\ }\textbf {\bibinfo {volume} {09}},\ \bibinfo
  {pages} {010} (\bibinfo {year} {2021})},\ \Eprint
  {http://arxiv.org/abs/2102.00379} {arXiv:2102.00379 [hep-ph]} \BibitemShut
  {NoStop}%
\bibitem [{\citenamefont {Caputo}\ \emph
  {et~al.}(2022{\natexlab{b}})\citenamefont {Caputo}, \citenamefont {Raffelt},\
  and\ \citenamefont {Vitagliano}}]{Caputo:2021rux}%
  \BibitemOpen
  \bibfield  {author} {\bibinfo {author} {\bibfnamefont {A.}~\bibnamefont
  {Caputo}}, \bibinfo {author} {\bibfnamefont {G.}~\bibnamefont {Raffelt}}, \
  and\ \bibinfo {author} {\bibfnamefont {E.}~\bibnamefont {Vitagliano}},\
  }\href {\doibase 10.1103/PhysRevD.105.035022} {\bibfield  {journal} {\bibinfo
   {journal} {Phys. Rev. D}\ }\textbf {\bibinfo {volume} {105}},\ \bibinfo
  {pages} {035022} (\bibinfo {year} {2022}{\natexlab{b}})},\ \Eprint
  {http://arxiv.org/abs/2109.03244} {arXiv:2109.03244 [hep-ph]} \BibitemShut
  {NoStop}%
\bibitem [{\citenamefont {D'Eramo}\ and\ \citenamefont
  {Lenoci}(2021)}]{DEramo:2020gpr}%
  \BibitemOpen
  \bibfield  {author} {\bibinfo {author} {\bibfnamefont {F.}~\bibnamefont
  {D'Eramo}}\ and\ \bibinfo {author} {\bibfnamefont {A.}~\bibnamefont
  {Lenoci}},\ }\href {\doibase 10.1088/1475-7516/2021/10/045} {\bibfield
  {journal} {\bibinfo  {journal} {JCAP}\ }\textbf {\bibinfo {volume} {10}},\
  \bibinfo {pages} {045} (\bibinfo {year} {2021})},\ \Eprint
  {http://arxiv.org/abs/2012.01446} {arXiv:2012.01446 [hep-ph]} \BibitemShut
  {NoStop}%
\bibitem [{\citenamefont {Ballesteros}\ \emph {et~al.}(2021)\citenamefont
  {Ballesteros}, \citenamefont {Garcia},\ and\ \citenamefont
  {Pierre}}]{Ballesteros:2020adh}%
  \BibitemOpen
  \bibfield  {author} {\bibinfo {author} {\bibfnamefont {G.}~\bibnamefont
  {Ballesteros}}, \bibinfo {author} {\bibfnamefont {M.~A.~G.}\ \bibnamefont
  {Garcia}}, \ and\ \bibinfo {author} {\bibfnamefont {M.}~\bibnamefont
  {Pierre}},\ }\href {\doibase 10.1088/1475-7516/2021/03/101} {\bibfield
  {journal} {\bibinfo  {journal} {JCAP}\ }\textbf {\bibinfo {volume} {03}},\
  \bibinfo {pages} {101} (\bibinfo {year} {2021})},\ \Eprint
  {http://arxiv.org/abs/2011.13458} {arXiv:2011.13458 [hep-ph]} \BibitemShut
  {NoStop}%
\bibitem [{\citenamefont {Calore}\ \emph {et~al.}(2020)\citenamefont {Calore},
  \citenamefont {Carenza}, \citenamefont {Giannotti}, \citenamefont {Jaeckel},\
  and\ \citenamefont {Mirizzi}}]{Calore:2020tjw}%
  \BibitemOpen
  \bibfield  {author} {\bibinfo {author} {\bibfnamefont {F.}~\bibnamefont
  {Calore}}, \bibinfo {author} {\bibfnamefont {P.}~\bibnamefont {Carenza}},
  \bibinfo {author} {\bibfnamefont {M.}~\bibnamefont {Giannotti}}, \bibinfo
  {author} {\bibfnamefont {J.}~\bibnamefont {Jaeckel}}, \ and\ \bibinfo
  {author} {\bibfnamefont {A.}~\bibnamefont {Mirizzi}},\ }\href {\doibase
  10.1103/PhysRevD.102.123005} {\bibfield  {journal} {\bibinfo  {journal}
  {Phys. Rev. D}\ }\textbf {\bibinfo {volume} {102}},\ \bibinfo {pages}
  {123005} (\bibinfo {year} {2020})},\ \Eprint
  {http://arxiv.org/abs/2008.11741} {arXiv:2008.11741 [hep-ph]} \BibitemShut
  {NoStop}%
\bibitem [{\citenamefont {Shifman}\ \emph {et~al.}(1979)\citenamefont
  {Shifman}, \citenamefont {Vainshtein}, \citenamefont {Voloshin},\ and\
  \citenamefont {Zakharov}}]{Shifman:1979eb}%
  \BibitemOpen
  \bibfield  {author} {\bibinfo {author} {\bibfnamefont {M.~A.}\ \bibnamefont
  {Shifman}}, \bibinfo {author} {\bibfnamefont {A.~I.}\ \bibnamefont
  {Vainshtein}}, \bibinfo {author} {\bibfnamefont {M.~B.}\ \bibnamefont
  {Voloshin}}, \ and\ \bibinfo {author} {\bibfnamefont {V.~I.}\ \bibnamefont
  {Zakharov}},\ }\href@noop {} {\bibfield  {journal} {\bibinfo  {journal} {Sov.
  J. Nucl. Phys.}\ }\textbf {\bibinfo {volume} {30}},\ \bibinfo {pages} {711}
  (\bibinfo {year} {1979})}\BibitemShut {NoStop}%
\bibitem [{\citenamefont {Essig}\ \emph {et~al.}(2013)\citenamefont {Essig},
  \citenamefont {Kuflik}, \citenamefont {McDermott}, \citenamefont {Volansky},\
  and\ \citenamefont {Zurek}}]{Essig:2013goa}%
  \BibitemOpen
  \bibfield  {author} {\bibinfo {author} {\bibfnamefont {R.}~\bibnamefont
  {Essig}}, \bibinfo {author} {\bibfnamefont {E.}~\bibnamefont {Kuflik}},
  \bibinfo {author} {\bibfnamefont {S.~D.}\ \bibnamefont {McDermott}}, \bibinfo
  {author} {\bibfnamefont {T.}~\bibnamefont {Volansky}}, \ and\ \bibinfo
  {author} {\bibfnamefont {K.~M.}\ \bibnamefont {Zurek}},\ }\href {\doibase
  10.1007/JHEP11(2013)193} {\bibfield  {journal} {\bibinfo  {journal} {JHEP}\
  }\textbf {\bibinfo {volume} {11}},\ \bibinfo {pages} {193} (\bibinfo {year}
  {2013})},\ \Eprint {http://arxiv.org/abs/1309.4091} {arXiv:1309.4091
  [hep-ph]} \BibitemShut {NoStop}%
\bibitem [{\citenamefont {Laha}\ \emph {et~al.}(2020)\citenamefont {Laha},
  \citenamefont {Mu\~noz},\ and\ \citenamefont {Slatyer}}]{Laha:2020ivk}%
  \BibitemOpen
  \bibfield  {author} {\bibinfo {author} {\bibfnamefont {R.}~\bibnamefont
  {Laha}}, \bibinfo {author} {\bibfnamefont {J.~B.}\ \bibnamefont {Mu\~noz}}, \
  and\ \bibinfo {author} {\bibfnamefont {T.~R.}\ \bibnamefont {Slatyer}},\
  }\href {\doibase 10.1103/PhysRevD.101.123514} {\bibfield  {journal} {\bibinfo
   {journal} {Phys. Rev. D}\ }\textbf {\bibinfo {volume} {101}},\ \bibinfo
  {pages} {123514} (\bibinfo {year} {2020})},\ \Eprint
  {http://arxiv.org/abs/2004.00627} {arXiv:2004.00627 [astro-ph.CO]}
  \BibitemShut {NoStop}%
\bibitem [{\citenamefont {Boyarsky}\ \emph
  {et~al.}(2008{\natexlab{b}})\citenamefont {Boyarsky}, \citenamefont
  {Malyshev}, \citenamefont {Neronov},\ and\ \citenamefont
  {Ruchayskiy}}]{Boyarsky:2007ge}%
  \BibitemOpen
  \bibfield  {author} {\bibinfo {author} {\bibfnamefont {A.}~\bibnamefont
  {Boyarsky}}, \bibinfo {author} {\bibfnamefont {D.}~\bibnamefont {Malyshev}},
  \bibinfo {author} {\bibfnamefont {A.}~\bibnamefont {Neronov}}, \ and\
  \bibinfo {author} {\bibfnamefont {O.}~\bibnamefont {Ruchayskiy}},\ }\href
  {\doibase 10.1111/j.1365-2966.2008.13003.x} {\bibfield  {journal} {\bibinfo
  {journal} {Mon. Not. Roy. Astron. Soc.}\ }\textbf {\bibinfo {volume} {387}},\
  \bibinfo {pages} {1345} (\bibinfo {year} {2008}{\natexlab{b}})},\ \Eprint
  {http://arxiv.org/abs/0710.4922} {arXiv:0710.4922 [astro-ph]} \BibitemShut
  {NoStop}%
\bibitem [{\citenamefont {Dessert}\ \emph {et~al.}(2020)\citenamefont
  {Dessert}, \citenamefont {Rodd},\ and\ \citenamefont
  {Safdi}}]{Dessert:2018qih}%
  \BibitemOpen
  \bibfield  {author} {\bibinfo {author} {\bibfnamefont {C.}~\bibnamefont
  {Dessert}}, \bibinfo {author} {\bibfnamefont {N.~L.}\ \bibnamefont {Rodd}}, \
  and\ \bibinfo {author} {\bibfnamefont {B.~R.}\ \bibnamefont {Safdi}},\ }\href
  {\doibase 10.1126/science.aaw3772} {\bibfield  {journal} {\bibinfo  {journal}
  {Science}\ }\textbf {\bibinfo {volume} {367}},\ \bibinfo {pages} {1465}
  (\bibinfo {year} {2020})},\ \Eprint {http://arxiv.org/abs/1812.06976}
  {arXiv:1812.06976 [astro-ph.CO]} \BibitemShut {NoStop}%
\bibitem [{\citenamefont {Ade}\ \emph {et~al.}(2016)\citenamefont {Ade} \emph
  {et~al.}}]{Planck:2015fie}%
  \BibitemOpen
  \bibfield  {author} {\bibinfo {author} {\bibfnamefont {P.~A.~R.}\
  \bibnamefont {Ade}} \emph {et~al.} (\bibinfo {collaboration} {Planck}),\
  }\href {\doibase 10.1051/0004-6361/201525830} {\bibfield  {journal} {\bibinfo
   {journal} {Astron. Astrophys.}\ }\textbf {\bibinfo {volume} {594}},\
  \bibinfo {pages} {A13} (\bibinfo {year} {2016})},\ \Eprint
  {http://arxiv.org/abs/1502.01589} {arXiv:1502.01589 [astro-ph.CO]}
  \BibitemShut {NoStop}%
\bibitem [{\citenamefont {Gorghetto}\ \emph {et~al.}(2018)\citenamefont
  {Gorghetto}, \citenamefont {Hardy},\ and\ \citenamefont
  {Villadoro}}]{Gorghetto:2018myk}%
  \BibitemOpen
  \bibfield  {author} {\bibinfo {author} {\bibfnamefont {M.}~\bibnamefont
  {Gorghetto}}, \bibinfo {author} {\bibfnamefont {E.}~\bibnamefont {Hardy}}, \
  and\ \bibinfo {author} {\bibfnamefont {G.}~\bibnamefont {Villadoro}},\ }\href
  {\doibase 10.1007/JHEP07(2018)151} {\bibfield  {journal} {\bibinfo  {journal}
  {JHEP}\ }\textbf {\bibinfo {volume} {07}},\ \bibinfo {pages} {151} (\bibinfo
  {year} {2018})},\ \Eprint {http://arxiv.org/abs/1806.04677} {arXiv:1806.04677
  [hep-ph]} \BibitemShut {NoStop}%
\bibitem [{\citenamefont {Gorghetto}\ \emph {et~al.}(2021)\citenamefont
  {Gorghetto}, \citenamefont {Hardy},\ and\ \citenamefont
  {Villadoro}}]{Gorghetto:2020qws}%
  \BibitemOpen
  \bibfield  {author} {\bibinfo {author} {\bibfnamefont {M.}~\bibnamefont
  {Gorghetto}}, \bibinfo {author} {\bibfnamefont {E.}~\bibnamefont {Hardy}}, \
  and\ \bibinfo {author} {\bibfnamefont {G.}~\bibnamefont {Villadoro}},\ }\href
  {\doibase 10.21468/SciPostPhys.10.2.050} {\bibfield  {journal} {\bibinfo
  {journal} {SciPost Phys.}\ }\textbf {\bibinfo {volume} {10}},\ \bibinfo
  {pages} {050} (\bibinfo {year} {2021})},\ \Eprint
  {http://arxiv.org/abs/2007.04990} {arXiv:2007.04990 [hep-ph]} \BibitemShut
  {NoStop}%
\bibitem [{\citenamefont {Buschmann}\ \emph {et~al.}(2020)\citenamefont
  {Buschmann}, \citenamefont {Foster},\ and\ \citenamefont
  {Safdi}}]{Buschmann:2019icd}%
  \BibitemOpen
  \bibfield  {author} {\bibinfo {author} {\bibfnamefont {M.}~\bibnamefont
  {Buschmann}}, \bibinfo {author} {\bibfnamefont {J.~W.}\ \bibnamefont
  {Foster}}, \ and\ \bibinfo {author} {\bibfnamefont {B.~R.}\ \bibnamefont
  {Safdi}},\ }\href {\doibase 10.1103/PhysRevLett.124.161103} {\bibfield
  {journal} {\bibinfo  {journal} {Phys. Rev. Lett.}\ }\textbf {\bibinfo
  {volume} {124}},\ \bibinfo {pages} {161103} (\bibinfo {year} {2020})},\
  \Eprint {http://arxiv.org/abs/1906.00967} {arXiv:1906.00967 [astro-ph.CO]}
  \BibitemShut {NoStop}%
\bibitem [{\citenamefont {Dine}\ \emph {et~al.}(2021)\citenamefont {Dine},
  \citenamefont {Fernandez}, \citenamefont {Ghalsasi},\ and\ \citenamefont
  {Patel}}]{Dine:2020pds}%
  \BibitemOpen
  \bibfield  {author} {\bibinfo {author} {\bibfnamefont {M.}~\bibnamefont
  {Dine}}, \bibinfo {author} {\bibfnamefont {N.}~\bibnamefont {Fernandez}},
  \bibinfo {author} {\bibfnamefont {A.}~\bibnamefont {Ghalsasi}}, \ and\
  \bibinfo {author} {\bibfnamefont {H.~H.}\ \bibnamefont {Patel}},\ }\href
  {\doibase 10.1088/1475-7516/2021/11/041} {\bibfield  {journal} {\bibinfo
  {journal} {JCAP}\ }\textbf {\bibinfo {volume} {11}},\ \bibinfo {pages} {041}
  (\bibinfo {year} {2021})},\ \Eprint {http://arxiv.org/abs/2012.13065}
  {arXiv:2012.13065 [hep-ph]} \BibitemShut {NoStop}%
\bibitem [{\citenamefont {Buschmann}\ \emph {et~al.}(2022)\citenamefont
  {Buschmann}, \citenamefont {Foster}, \citenamefont {Hook}, \citenamefont
  {Peterson}, \citenamefont {Willcox}, \citenamefont {Zhang},\ and\
  \citenamefont {Safdi}}]{Buschmann:2021sdq}%
  \BibitemOpen
  \bibfield  {author} {\bibinfo {author} {\bibfnamefont {M.}~\bibnamefont
  {Buschmann}}, \bibinfo {author} {\bibfnamefont {J.~W.}\ \bibnamefont
  {Foster}}, \bibinfo {author} {\bibfnamefont {A.}~\bibnamefont {Hook}},
  \bibinfo {author} {\bibfnamefont {A.}~\bibnamefont {Peterson}}, \bibinfo
  {author} {\bibfnamefont {D.~E.}\ \bibnamefont {Willcox}}, \bibinfo {author}
  {\bibfnamefont {W.}~\bibnamefont {Zhang}}, \ and\ \bibinfo {author}
  {\bibfnamefont {B.~R.}\ \bibnamefont {Safdi}},\ }\href {\doibase
  10.1038/s41467-022-28669-y} {\bibfield  {journal} {\bibinfo  {journal}
  {Nature Commun.}\ }\textbf {\bibinfo {volume} {13}},\ \bibinfo {pages} {1049}
  (\bibinfo {year} {2022})},\ \Eprint {http://arxiv.org/abs/2108.05368}
  {arXiv:2108.05368 [hep-ph]} \BibitemShut {NoStop}%
\bibitem [{\citenamefont {Blinov}\ \emph {et~al.}(2019)\citenamefont {Blinov},
  \citenamefont {Dolan}, \citenamefont {Draper},\ and\ \citenamefont
  {Kozaczuk}}]{Blinov:2019rhb}%
  \BibitemOpen
  \bibfield  {author} {\bibinfo {author} {\bibfnamefont {N.}~\bibnamefont
  {Blinov}}, \bibinfo {author} {\bibfnamefont {M.~J.}\ \bibnamefont {Dolan}},
  \bibinfo {author} {\bibfnamefont {P.}~\bibnamefont {Draper}}, \ and\ \bibinfo
  {author} {\bibfnamefont {J.}~\bibnamefont {Kozaczuk}},\ }\href {\doibase
  10.1103/PhysRevD.100.015049} {\bibfield  {journal} {\bibinfo  {journal}
  {Phys. Rev. D}\ }\textbf {\bibinfo {volume} {100}},\ \bibinfo {pages}
  {015049} (\bibinfo {year} {2019})},\ \Eprint
  {http://arxiv.org/abs/1905.06952} {arXiv:1905.06952 [hep-ph]} \BibitemShut
  {NoStop}%
\bibitem [{\citenamefont {Foster}\ \emph {et~al.}(2022)\citenamefont {Foster},
  \citenamefont {Kumar}, \citenamefont {Safdi},\ and\ \citenamefont
  {Soreq}}]{Foster:2022ajl}%
  \BibitemOpen
  \bibfield  {author} {\bibinfo {author} {\bibfnamefont {J.~W.}\ \bibnamefont
  {Foster}}, \bibinfo {author} {\bibfnamefont {S.}~\bibnamefont {Kumar}},
  \bibinfo {author} {\bibfnamefont {B.~R.}\ \bibnamefont {Safdi}}, \ and\
  \bibinfo {author} {\bibfnamefont {Y.}~\bibnamefont {Soreq}},\ }\href@noop {}
  {\  (\bibinfo {year} {2022})},\ \Eprint {http://arxiv.org/abs/2208.10504}
  {arXiv:2208.10504 [hep-ph]} \BibitemShut {NoStop}%
\bibitem [{\citenamefont {Panci}\ \emph {et~al.}(2022)\citenamefont {Panci},
  \citenamefont {Redigolo}, \citenamefont {Schwetz},\ and\ \citenamefont
  {Ziegler}}]{Panci:2022wlc}%
  \BibitemOpen
  \bibfield  {author} {\bibinfo {author} {\bibfnamefont {P.}~\bibnamefont
  {Panci}}, \bibinfo {author} {\bibfnamefont {D.}~\bibnamefont {Redigolo}},
  \bibinfo {author} {\bibfnamefont {T.}~\bibnamefont {Schwetz}}, \ and\
  \bibinfo {author} {\bibfnamefont {R.}~\bibnamefont {Ziegler}},\ }\href@noop
  {} {\  (\bibinfo {year} {2022})},\ \Eprint {http://arxiv.org/abs/2209.03371}
  {arXiv:2209.03371 [hep-ph]} \BibitemShut {NoStop}%
\bibitem [{\citenamefont {Pal}\ and\ \citenamefont
  {Wolfenstein}(1982)}]{Pal:1981rm}%
  \BibitemOpen
  \bibfield  {author} {\bibinfo {author} {\bibfnamefont {P.~B.}\ \bibnamefont
  {Pal}}\ and\ \bibinfo {author} {\bibfnamefont {L.}~\bibnamefont
  {Wolfenstein}},\ }\href {\doibase 10.1103/PhysRevD.25.766} {\bibfield
  {journal} {\bibinfo  {journal} {Phys. Rev. D}\ }\textbf {\bibinfo {volume}
  {25}},\ \bibinfo {pages} {766} (\bibinfo {year} {1982})}\BibitemShut
  {NoStop}%
\bibitem [{\citenamefont {Shi}\ and\ \citenamefont
  {Fuller}(1999)}]{Shi:1998km}%
  \BibitemOpen
  \bibfield  {author} {\bibinfo {author} {\bibfnamefont {X.-D.}\ \bibnamefont
  {Shi}}\ and\ \bibinfo {author} {\bibfnamefont {G.~M.}\ \bibnamefont
  {Fuller}},\ }\href {\doibase 10.1103/PhysRevLett.82.2832} {\bibfield
  {journal} {\bibinfo  {journal} {Phys. Rev. Lett.}\ }\textbf {\bibinfo
  {volume} {82}},\ \bibinfo {pages} {2832} (\bibinfo {year} {1999})},\ \Eprint
  {http://arxiv.org/abs/astro-ph/9810076} {arXiv:astro-ph/9810076} \BibitemShut
  {NoStop}%
\bibitem [{\citenamefont {Bolton}\ \emph {et~al.}(2020)\citenamefont {Bolton},
  \citenamefont {Deppisch},\ and\ \citenamefont {Bhupal~Dev}}]{Bolton:2019pcu}%
  \BibitemOpen
  \bibfield  {author} {\bibinfo {author} {\bibfnamefont {P.~D.}\ \bibnamefont
  {Bolton}}, \bibinfo {author} {\bibfnamefont {F.~F.}\ \bibnamefont
  {Deppisch}}, \ and\ \bibinfo {author} {\bibfnamefont {P.~S.}\ \bibnamefont
  {Bhupal~Dev}},\ }\href {\doibase 10.1007/JHEP03(2020)170} {\bibfield
  {journal} {\bibinfo  {journal} {JHEP}\ }\textbf {\bibinfo {volume} {03}},\
  \bibinfo {pages} {170} (\bibinfo {year} {2020})},\ \Eprint
  {http://arxiv.org/abs/1912.03058} {arXiv:1912.03058 [hep-ph]} \BibitemShut
  {NoStop}%
\bibitem [{\citenamefont {Abdurashitov}\ \emph {et~al.}(2017)\citenamefont
  {Abdurashitov} \emph {et~al.}}]{Abdurashitov:2017kka}%
  \BibitemOpen
  \bibfield  {author} {\bibinfo {author} {\bibfnamefont {J.~N.}\ \bibnamefont
  {Abdurashitov}} \emph {et~al.},\ }\href {\doibase 10.1134/S0021364017120013}
  {\bibfield  {journal} {\bibinfo  {journal} {Pisma Zh. Eksp. Teor. Fiz.}\
  }\textbf {\bibinfo {volume} {105}},\ \bibinfo {pages} {723} (\bibinfo {year}
  {2017})},\ \Eprint {http://arxiv.org/abs/1703.10779} {arXiv:1703.10779
  [hep-ex]} \BibitemShut {NoStop}%
\bibitem [{\citenamefont {Holzschuh}\ \emph {et~al.}(1999)\citenamefont
  {Holzschuh}, \citenamefont {Kundig}, \citenamefont {Palermo}, \citenamefont
  {Stussi},\ and\ \citenamefont {Wenk}}]{Holzschuh:1999vy}%
  \BibitemOpen
  \bibfield  {author} {\bibinfo {author} {\bibfnamefont {E.}~\bibnamefont
  {Holzschuh}}, \bibinfo {author} {\bibfnamefont {W.}~\bibnamefont {Kundig}},
  \bibinfo {author} {\bibfnamefont {L.}~\bibnamefont {Palermo}}, \bibinfo
  {author} {\bibfnamefont {H.}~\bibnamefont {Stussi}}, \ and\ \bibinfo {author}
  {\bibfnamefont {P.}~\bibnamefont {Wenk}},\ }\href {\doibase
  10.1016/S0370-2693(99)00200-2} {\bibfield  {journal} {\bibinfo  {journal}
  {Phys. Lett. B}\ }\textbf {\bibinfo {volume} {451}},\ \bibinfo {pages} {247}
  (\bibinfo {year} {1999})}\BibitemShut {NoStop}%
\bibitem [{\citenamefont {Holzschuh}\ \emph {et~al.}(2000)\citenamefont
  {Holzschuh}, \citenamefont {Palermo}, \citenamefont {Stussi},\ and\
  \citenamefont {Wenk}}]{Holzschuh:2000nj}%
  \BibitemOpen
  \bibfield  {author} {\bibinfo {author} {\bibfnamefont {E.}~\bibnamefont
  {Holzschuh}}, \bibinfo {author} {\bibfnamefont {L.}~\bibnamefont {Palermo}},
  \bibinfo {author} {\bibfnamefont {H.}~\bibnamefont {Stussi}}, \ and\ \bibinfo
  {author} {\bibfnamefont {P.}~\bibnamefont {Wenk}},\ }\href {\doibase
  10.1016/S0370-2693(00)00476-7} {\bibfield  {journal} {\bibinfo  {journal}
  {Phys. Lett. B}\ }\textbf {\bibinfo {volume} {482}},\ \bibinfo {pages} {1}
  (\bibinfo {year} {2000})}\BibitemShut {NoStop}%
\bibitem [{\citenamefont {Friedrich}\ \emph {et~al.}(2021)\citenamefont
  {Friedrich} \emph {et~al.}}]{Friedrich:2020nze}%
  \BibitemOpen
  \bibfield  {author} {\bibinfo {author} {\bibfnamefont {S.}~\bibnamefont
  {Friedrich}} \emph {et~al.},\ }\href {\doibase
  10.1103/PhysRevLett.126.021803} {\bibfield  {journal} {\bibinfo  {journal}
  {Phys. Rev. Lett.}\ }\textbf {\bibinfo {volume} {126}},\ \bibinfo {pages}
  {021803} (\bibinfo {year} {2021})},\ \Eprint
  {http://arxiv.org/abs/2010.09603} {arXiv:2010.09603 [nucl-ex]} \BibitemShut
  {NoStop}%
\bibitem [{\citenamefont {Bellini}\ \emph {et~al.}(2013)\citenamefont {Bellini}
  \emph {et~al.}}]{Borexino:2013bot}%
  \BibitemOpen
  \bibfield  {author} {\bibinfo {author} {\bibfnamefont {G.}~\bibnamefont
  {Bellini}} \emph {et~al.} (\bibinfo {collaboration} {Borexino}),\ }\href
  {\doibase 10.1103/PhysRevD.88.072010} {\bibfield  {journal} {\bibinfo
  {journal} {Phys. Rev. D}\ }\textbf {\bibinfo {volume} {88}},\ \bibinfo
  {pages} {072010} (\bibinfo {year} {2013})},\ \Eprint
  {http://arxiv.org/abs/1311.5347} {arXiv:1311.5347 [hep-ex]} \BibitemShut
  {NoStop}%
\bibitem [{\citenamefont {Aguilar-Arevalo}\ \emph {et~al.}(2018)\citenamefont
  {Aguilar-Arevalo} \emph {et~al.}}]{PIENU:2017wbj}%
  \BibitemOpen
  \bibfield  {author} {\bibinfo {author} {\bibfnamefont {A.}~\bibnamefont
  {Aguilar-Arevalo}} \emph {et~al.} (\bibinfo {collaboration} {PIENU}),\ }\href
  {\doibase 10.1103/PhysRevD.97.072012} {\bibfield  {journal} {\bibinfo
  {journal} {Phys. Rev. D}\ }\textbf {\bibinfo {volume} {97}},\ \bibinfo
  {pages} {072012} (\bibinfo {year} {2018})},\ \Eprint
  {http://arxiv.org/abs/1712.03275} {arXiv:1712.03275 [hep-ex]} \BibitemShut
  {NoStop}%
\bibitem [{\citenamefont {Shi}\ and\ \citenamefont {Sigl}(1994)}]{Shi:1993ee}%
  \BibitemOpen
  \bibfield  {author} {\bibinfo {author} {\bibfnamefont {X.}~\bibnamefont
  {Shi}}\ and\ \bibinfo {author} {\bibfnamefont {G.}~\bibnamefont {Sigl}},\
  }\href {\doibase 10.1016/0370-2693(94)91232-7} {\bibfield  {journal}
  {\bibinfo  {journal} {Phys. Lett. B}\ }\textbf {\bibinfo {volume} {323}},\
  \bibinfo {pages} {360} (\bibinfo {year} {1994})},\ \bibinfo {note} {[Erratum:
  Phys.Lett.B 324, 516--516 (1994)]},\ \Eprint
  {http://arxiv.org/abs/hep-ph/9312247} {arXiv:hep-ph/9312247} \BibitemShut
  {NoStop}%
\end{thebibliography}%

%%%%%%%%%%%%%%%%%%%%%%%%%%
% Supplementary Material %
%%%%%%%%%%%%%%%%%%%%%%%%%%
\clearpage
\onecolumngrid
\begin{center}
   \textbf{\large SUPPLEMENTARY MATERIAL \\[.2cm] ``The Irreducible Axion Background''}\\[.2cm]
  \vspace{0.05in}
  {Kevin Langhoff, Nadav Outmezguine, and Nicholas L. Rodd}
\end{center}

\setcounter{equation}{0}
\setcounter{figure}{0}
\setcounter{table}{0}
\setcounter{section}{0}
\setcounter{page}{1}
\makeatletter
\renewcommand{\theequation}{S\arabic{equation}}
\renewcommand{\thefigure}{S\arabic{figure}}
\renewcommand{\thetable}{S\arabic{table}}

\onecolumngrid
%%%%%%%%%%%%%%%%%%%%%%%%%%

In the main text of this {\it Letter} we established that for previously unconstrained axion couplings, an irreducible DM contribution is produced via freeze-in, which can then be detected via searches for DM decay.
Here we will provide a more explicit description of our calculations and analysis, as well as several simple extensions to our main findings.
Section~\ref{sec:couplings} outlines the IR action and model parameters we consider throughout, and provides the details of the axion decay rate in vacuum.
In Secs.~\ref{sec:abundance} and \ref{sec:constraints} we provide the full details of the two main arguments in the main text: the axion relic density and the constraints we derive from searches for DM decay.
The final three sections then outline minimal extensions of our main argument.
In Sec.~\ref{sec:bothcouplings} we show results for axions that couple to photons and electrons simultaneously, and Secs.~\ref{sec:extension-Mis} and \ref{sec:extension-Nu} demonstrate how constraints on an irreducible relic density can be extended to include axions generation from misalignment, or even to another state such as sterile neutrinos.

%%%%%%%%%
\section{Axion Couplings and Decay Rates}
\label{sec:couplings}
%%%%%%%%%

We begin by detailing the specific axion interactions we consider throughout this work, and reviewing the associated decay rates.
As described in the main text, we focus on a situation where the Universe reheats at 5 MeV.
At such temperatures, the thermal bath contains photons, electrons, and neutrinos.
The axion coupling to neutrinos is expected to be highly suppressed (see e.g. \Refc{Bauer:2017ris}), and so we focus on the coupling to photons and electrons.
Generically, an axion will also couple to protons and neutrons, however, since these will be non-relativistic during the epoch under consideration, production from such interactions is negligible.
(Note, however, that there also exists constraints coming from a diffuse supernovae axion background, in which case the axions are dominantly produced from interactions with nuclei~\cite{Calore:2020tjw}.)
Following these considerations, we will focus on the following Lagrangian:
\be
  \mathcal{L} \supset \frac{1}{2} (\partial_\mu a)^2 - \frac{1}{2}m_a^2 a^2 + \bar{e} \left( i \slashed{D} - m_e  \right) e - \frac{1}{4}F_{\mu \nu}F^{\mu \nu}
  - \frac{1}{4} \ga a F_{\mu \nu}\tilde{F}^{\mu \nu}  + \frac{\gae}{2m_e} (\partial_\mu a) \bar{e}\gamma^\mu \gamma_5 e.
  \label{eq:fullL}
\ee
We remain fully agnostic as to the UV origin of the three parameters $\{m_a,\,\ga,\,\gae\}$.
In the main text we considered two classes of axions: photophilic ($\ga \neq 0$, $\gae = 0$) and photophobic ($\ga = 0$, $\gae \neq 0$).
Extending this, in Sec.~\ref{sec:bothcouplings} we show results for axions that couple to electrons and photons simultaneously.
As a note of caution, given that the axion is a real scalar field, we can always redefine $a \to - a$, which will shift the sign of the interactions in \Eq{eq:fullL} -- equivalent to sending $\ga \to - \ga$ and $\gae \to -\gae$ -- and both choices are in common use.
If we consider the interactions in isolation, there is no physical significance to this choice, the couplings shown in Figs.~\ref{fig:Constraints} and \ref{fig:Constraints_gae} can simply be interpreted as $|\ga|$ and $|\gae|$, respectively.
When both couplings are active, the relative sign matters and we will keep track of the distinction.

The interactions in \Eq{eq:fullL} are what allow the axion to decay to two photons in the late Universe, or for $m_a > 2 m_e$, to an electron-positron pair.
The decay rate to two photons, assuming the photons have no thermal mass, is given by~\cite{Shifman:1979eb,Bauer:2017ris}
\be
  \Gamma(a \to \gamma \gamma) = \frac{m_a^3}{64\pi} \left| \ga - \frac{\alpha \gae}{m_e \pi} [1-\tau f^2(\tau)] \right|^2\!,
  \label{eq:Fulla2gg}
\ee
where $\tau = (2m_e/m_a)^2$ and 
\bea
  f(\tau) = \left\{\begin{array}{lc} \text{arcsin}\tfrac{1}{\sqrt{\tau}} & \tau \geq 1, \\
  \tfrac{\pi}{2} + \tfrac{i}{2} \ln \tfrac{1+\sqrt{1-\tau}}{1-\sqrt{1-\tau}} & \tau < 1.
  \end{array} \right.
\eea
The appearance of $\gae$ in \Eq{eq:Fulla2gg} is due to a one-loop triangle diagram with electrons in the loop.
For $m_a \ll m_e$, the decay rate reduces to \Eq{eq:Gam_a2gamgam}, whereas for $m_a \gg m_e$, we instead have 
\be
  \Gamma(a \to \gamma \gamma) = \frac{m_a^3}{64\pi} \left[ \ga - \frac{\alpha \gae}{m_e \pi} \right]^2\!.
  \label{eq:Gam_a2gg_smallma}
\ee

The second relevant decay channel is to two electrons,
\be
  \Gamma(a \to e^- e^+) = \frac{\gae^2 m_a}{8\pi} \sqrt{1-\frac{4m_e^2}{m_a^2}} \Theta(m_a-2m_e).
\ee
The inclusion of this channel is critical not only as it can significantly impact the total axion decay rate, but also as decays to electron-positron pairs can observably modify the CMB.
At one loop, $\ga$ will also mediate $a \to e^- e^+$, modifying the above.
This correction is, however, never important in the parameter space we consider.

%%%%%%%%%
\section{Calculation of the Irreducible Axion Relic Abundance from Freeze-In}
\label{sec:abundance}
%%%%%%%%%

In this section we will expound on the production of axions in the early Universe via the freeze-in mechanism.
The goal is to determine $\mathcal{F}_a$ as a function of $\{m_a,\,\ga,\,\gae,\,\TRH\}$, which was an essential ingredient in our analysis.
Again, we emphasize that the techniques we use are standard, and similar computations have appeared in, for example,  \Refs{Gondolo:1990dk,Masso:1995tw,Edsjo:1997bg,Hall:2009bx,Cadamuro:2010cz,Cadamuro:2011fd,Blennow:2013jba,DEramo:2017ecx,DEramo:2018vss,Baumholzer:2020hvx,Depta:2020wmr,Balazs:2022tjl,Bolz:2000fu,Dunsky:2022uoq}.
We begin our discussion below with a general overview of the calculation and then individually focus on the two relevant processes: inverse decays and 2-to-2 production.
Finally, we will describe why the axions produced can be well approximated as a contribution to CDM.

\subsection{Overview}

As outlined in the main text, to be maximally conservative, we consider an early Universe cosmology with $\TRH = 5~{\rm MeV}$.
From this starting point, the relevant Lagrangian for describing the interactions between the SM bath and the axions produced from it is given in \Eq{eq:fullL}.
These interactions continue to describe the relevant physics up to $\TRH \sim 100~{\rm MeV}$, beyond which additional degrees of freedom re-enter equilibrium.
Accordingly, without adding additional interactions we can explore the dependence of our results on the reheat temperature within the range $5~{\rm MeV} \leq \TRH \leq 100~{\rm MeV}$.
Given these assumptions, there are four processes that contribute to the freeze-in abundance of axions:

\vspace{0.3cm}
\begin{minipage}{0.475 \textwidth}
\begin{enumerate}[(I)]
    \item Photon Conversion: $e^\pm \gamma \to e^\pm a$
    \item Electron-Positron Annihilation: $ e^- e^+ \to \gamma a$
\end{enumerate}
\end{minipage}
\begin{minipage}{0.475 \textwidth}
\begin{enumerate}[(I)]
    \setcounter{enumi}{2}
    \item Photon Inverse Decay: $\gamma \gamma \to a$
    \item Electron-Positron Inverse Decay: $e^- e^+ \to a$
\end{enumerate}
\end{minipage}
\vspace{0.3cm}

\begin{figure*}[t!]
\centering
\includegraphics[width=0.9\textwidth]{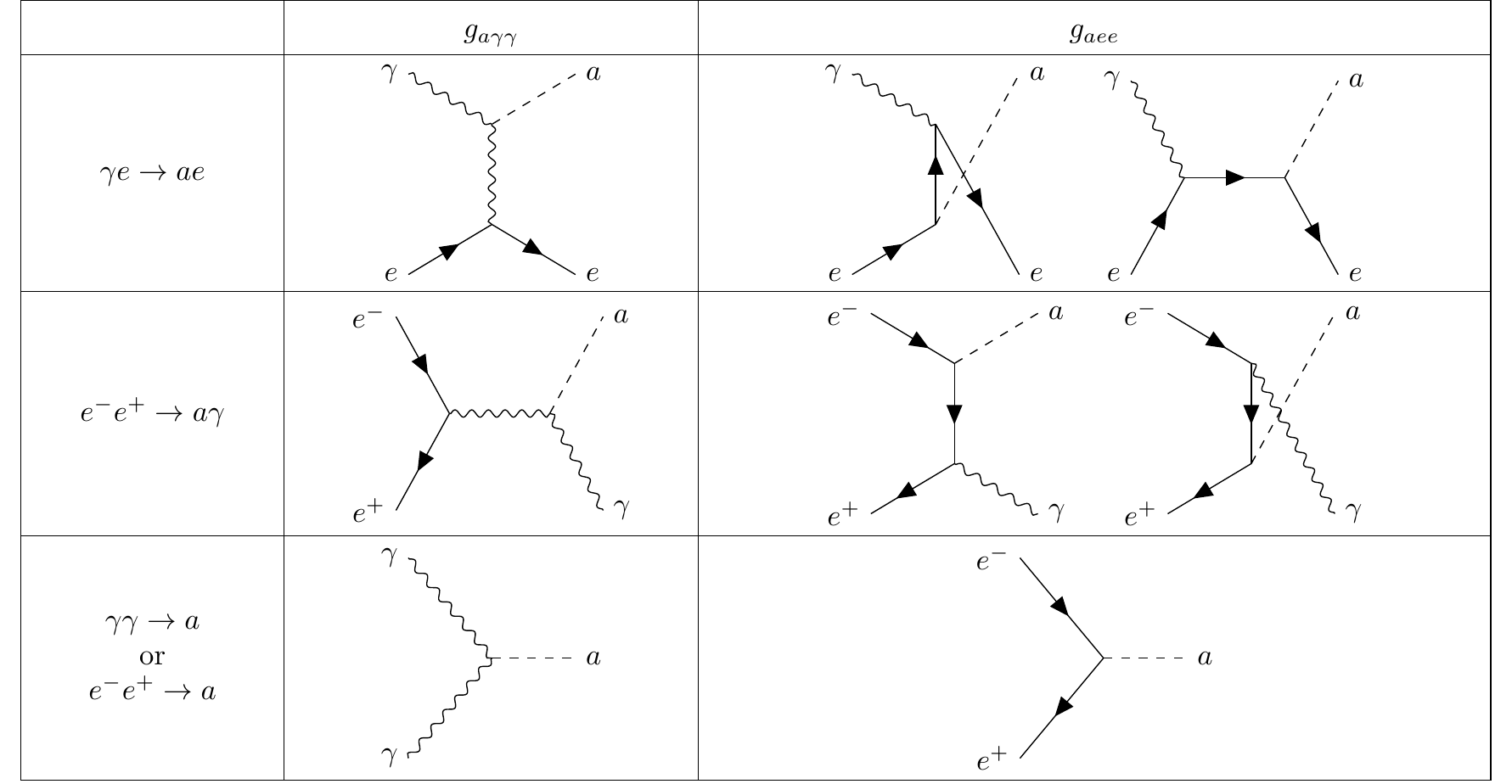}
\caption{The relevant Feynman diagrams contributing to the freeze-in production of axions.
}
\label{fig:FeynmandDiagramTable}
\end{figure*}

At tree level, (I) and (II) can be mediated by either $\ga$ or $\gae$, whereas (III) and (IV) are only supported by a single coupling each.
The relevant Feynman diagrams are given in \Fig{fig:FeynmandDiagramTable}.
At loop level, $\gae$ induces an effective $\ga$ (cf. \Eq{eq:Fulla2gg}), and we include these effects as in \Refc{Ferreira:2022xlw}.
However, for production, these effects are always small and can safely be neglected.

Having specified the relevant interactions, we can now use the Boltzmann equations to determine the evolution of the axion number density, $n_a$.
In a Friedmann–Lemaitre-Robertson–Walker spacetime, we have
\be
  \frac{dn_a}{dt} + 3H n_a = R(t),
  \label{eq:Boltzmann1}
\ee
where the production rate, $R(t)$, is given by
\be
  R(t)=\sum_\text{process} \int 
  \Big(\prod_{i} d\Pi_i \Big) 
  \Big(\prod_{f} d\Pi_f \Big)
  (2\pi)^4\delta^4\Big(\sum_i p_i -\sum_f p_f \Big)
  \left|{\cal M}_{i\to f}\right|^2\times\Phi.
  \label{eq:rate}
\ee
Here, the sum is over the appropriate processes denoted (I-IV) above, we use $i$ and $f$ to index all initial and final state particles, respectively, $d\Pi = d^3\mathbf{p}/[2E\,(2\pi)^3]$, and $\left|\mathcal{M}_{i\to f}\right|^2$ is the squared amplitude for the relevant process, which we sum over the initial and final helicities and polarizations  without averaging.
Lastly, the phase space densities of all states in the process is encoded in $\Phi$ as
\be
  \Phi \equiv \prod_f \Big( 1 \pm f_f\Big) \prod_i f_i
  -  \prod_i \Big(1\pm f_i\Big) \prod_f  f_f,
  \label{eq:Phi_base}
\ee
where the $\pm$ is resolved as a $+$ for bosons, and $-$ for fermions.
For all our production processes, the axion is a final state particle, and all states involved are in thermal equilibrium, except the axion.
Accordingly, we can rewrite the phase space contribution in the following form
\be
  \Phi = (f_a^{\rm eq}-f_a)\times
  \Bigg[\prod_i \Big(1\pm f_i^{\rm eq} \Big) \prod_{f\neq a}  f_f^{\rm eq} - \prod_{f \neq a} \Big( 1 \pm f_f^{\rm eq}\Big) \prod_i f_i^{\rm eq}
  \Bigg].
  \label{eq:Phi_eq}
\ee
In the parameter space relevant for this work, the freeze-in contribution to the relic density is found to be always sufficiently small, such that $f_a\ll f_a^{\rm eq}$.
This allows us to neglect inverse processes, which we do by replacing $f_a^{\rm eq}-f_a\to f_a^{\rm eq}$ in the above expression for $\Phi$.
This guarantees that in our calculations $R(t)$ is a positive function for all processes considered and the (co-moving) axion abundance only grows, the hallmark of freeze-in production.

Rather than evolving the system as a function of increasing time or decreasing temperature, it is instead convenient to use the variable $x = m_a/T$.
Further, in order to factor out the expansion of the Universe, instead of $n_a$ we compute the axion yield, defined as the number density per entropy density, $Y_a = n_a/s$.
With these variables the Boltzmann equation in \Eq{eq:Boltzmann1} takes the form
\be
  \frac{dY_a}{dx} = \frac{\tilde{g}(x)}{x H(x) s(x)}R(x),
  \label{eq:Boltzmann2}
\ee
where $\tilde{g}(x) = (1-\frac{1}{3}d\ln g_{\star,s}/d\ln x)$.
Again neglecting the inverse processes that would deplete axions, ${\cal F}_a$ is given by
\be
  \mathcal{F}_a \simeq \frac{m_a s_0}{\rho^0_\DM}Y_a(\infty)
  = \frac{m_a s_0}{\rho^0_\DM}\int_{x_\RH}^{\infty} dx\,\frac{\tilde{g}(x)}{x H(x) s(x)}R(x),
  \label{eq:F_a}
\ee
where $s_0$ and $\rho^0_\DM$ are the present day entropy and DM density, and $x_\RH = m_a/\TRH$.
Importantly, note that ${\cal F}_a$ is linear in $R(x)$.
This allows us to calculate the axion abundance by simply summing the abundance produced by each of the processes (I-IV) discussed above separately.
In what follows, we determine $R$ for inverse decays, processes (III-IV), and then 2-to-2 interactions, processes (I-II).

\newpage

\begin{figure}[t!]
\centering
\subfloat{\includegraphics[width= 0.47 \textwidth]{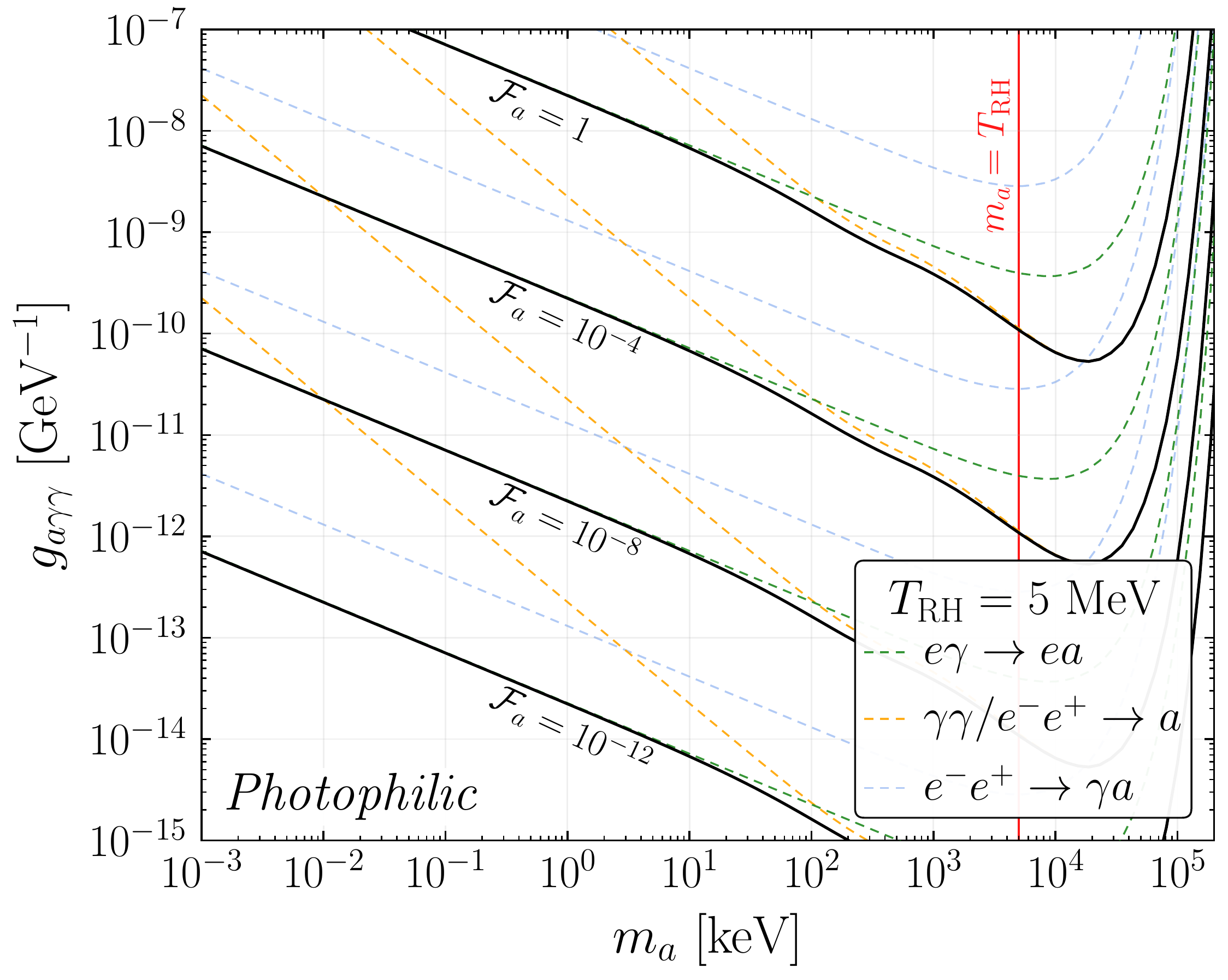}}
\hspace{0.5cm}
\subfloat{\includegraphics[width= 0.47 \textwidth]{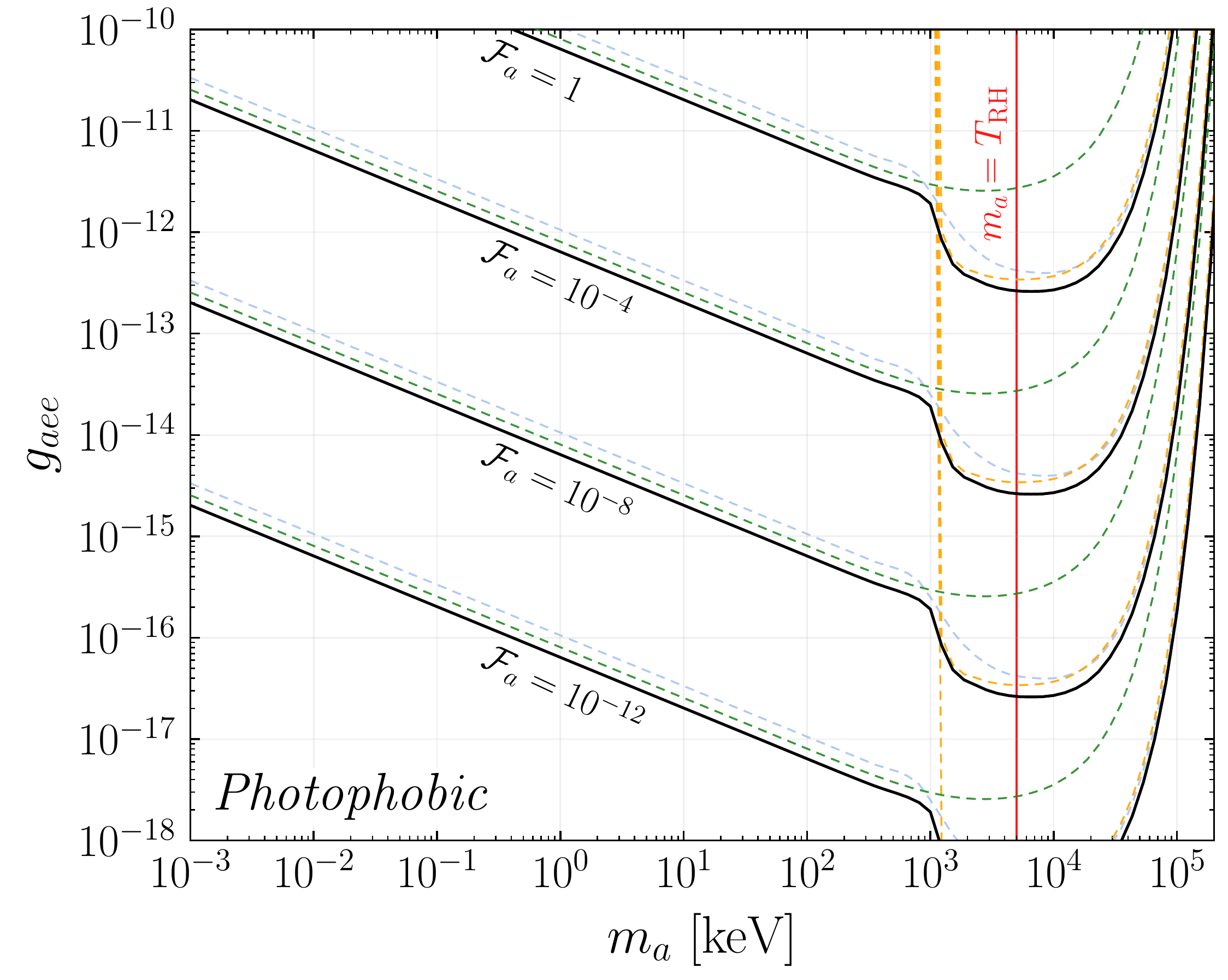}}
\vspace{-0.4cm}
\caption{Contours of constant effective DM fraction, ${\cal F}_a$ for photophilic (left) and photophobic (right) axions, as a function of axion mass and relevant coupling, both for a reheat temperatures $\TRH = 5~{\rm MeV}$.
The solid black line is the sum of all relevant processes that we consider, photon conversion (dashed green), inverse decays (dashed orange), and annihilation (dashed blue).
In both cases, for $m_a \ll \TRH$ we see that photon conversion dominates, whereas for $m_a \gtrsim \TRH$ the dominant contribution transitions to inverse decays.
Fermion annihilation remains subdominant throughout.}
\label{fig:FDM}
\end{figure}

\begin{figure}[h!]
\includegraphics[width=0.49\textwidth]{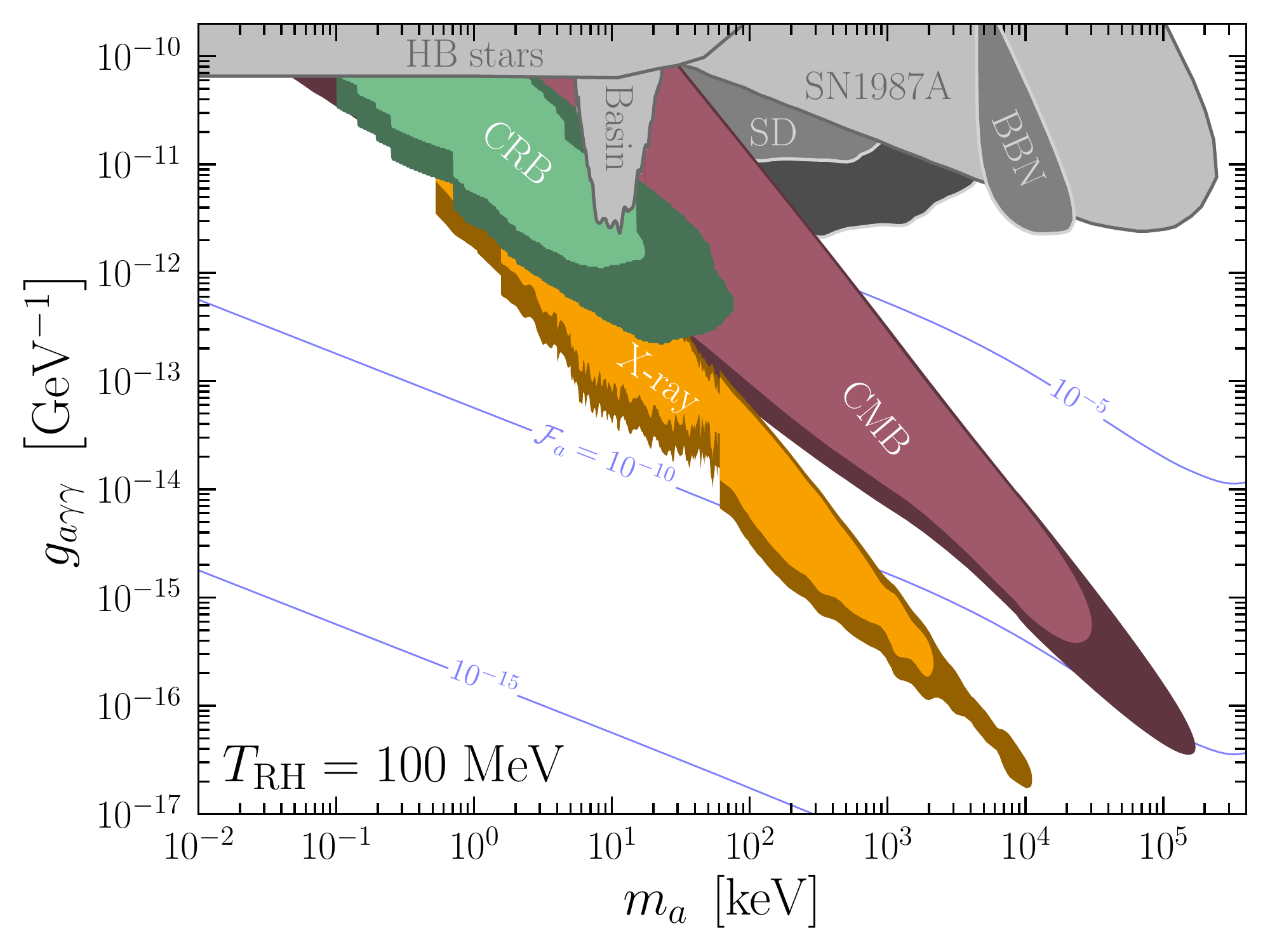}
\vspace{-0.4cm}
\caption{The analogue of \Fig{fig:Constraints}, but for $\TRH = 100~{\rm MeV}$.
The irreducible constraints derived for $\TRH = 5~{\rm MeV}$ are shown in solid colors, whereas the darker shade represent the higher $\TRH$ .
The contour of ${\cal F}_a$ are also shown now for $\TRH = 100~{\rm MeV}$.
We see that the constraints not only move downward due to an increased axion density, but the constraints also extend to larger masses which become less Boltzmann suppressed.
(Note that the enhanced densities at higher masses makes the DM search performed in \Refc{Essig:2013goa} using COMPTEL relevant, and we have added this into our X-ray bound.)
}
\label{fig:Constraints_100MeV}
\end{figure}

The result of solving \Eq{eq:F_a} for all processes is shown in \Fig{fig:FDM} for both the photophilic and photophobic axion, and the total abundances are made publicly available~\cite{PubCode}.
We further show the total number of axions produced, which again, is simply the sum of the individual contributions (so long as the axions never approach equilibrium).
For $\TRH = 5~{\rm MeV}$ we learn that production is dominated by photon conversion when $m_a \ll \TRH$ (and that the abundance is well approximated by \Eq{eq:F_a_approx}), whereas inverse decays take over once the mass is comparable to or larger than $\TRH$.
From the numerical results, we further find that the abundance of the photophilic axion grows roughly linearly with the reheat temperature.
Throughout this work, we almost exclusively focus on $\TRH = 5~{\rm MeV}$ to determine our most robust constraints; however, we can explore how our results are impacted if we extend the period of radiation domination to higher temperatures.
In particular, in \Fig{fig:Constraints_100MeV} we contrast our findings to the scenario where $\TRH = 100~{\rm MeV}$.
Note that at $100~{\rm MeV}$, it is still reasonable to ignore the contribution from more massive particles such as the muon or pion.
If we moved to higher temperatures still, these additional degrees of freedom would contribute to the production of axions.
This, however, would only strengthen the constraints.

\subsection{The Reaction Rate for Inverse Decays}

For an inverse decay of the form $1+2\to a$, the phase space factor \Eq{eq:Phi_eq} becomes
\be
  \Phi_{1+2\to a} \simeq f_a^{\rm eq} [1\pm (f^{\rm eq}_1 + f^{\rm eq}_2)],
\ee
where we have used the fact the initial states must obey the same statistics.
The rate is then given by
\be
  R_{\rm ID}(T) =|\mathcal{M}_{1+2\to a}|^2 \int d\Pi_1 d\Pi_2 d\Pi_a (2\pi)^4\delta^4\left(p_1 + p_2 - p_a \right)\times f_a^{\rm eq} [1\pm (f^{\rm eq}_1 + f^{\rm eq}_2)].
  \label{eq:InverseDecayRate}
\ee
For the processes we consider $f_1^{\rm eq}=f_2^{\rm eq}$ (and $m_1=m_2$), so that we can reduce the result to
\be
  R_{\rm ID}(T) =
  \frac{|\mathcal{M}_{1+2\to a}|^2}{32\pi^3 }\int_{m_a}^\infty dE_a\, f_a^{\rm eq} \left(\beta \, p_a + 2T\ln\left[\frac{1\mp e^{-E_+/T}}{1\mp e^{-E_-/T}}\right]\right)\!,
\ee
with $E_\pm = \left(E_a\pm \beta p_a\right)/2$ and $\beta = \sqrt{1-4m_1^2/m_a^2}$.
The rate is now completely specified up to the amplitude, which we can compute for the two specific processes of interest.

Firstly, for $\gamma \gamma \to a$ we have
\be
  \sum |\mathcal{M}_{\gamma \gamma \to a}|^2 = \frac{1}{2} g_{a\gamma \gamma}^2m_a^2(m_a^2-4m_\gamma^2).
\ee
where the photon's thermal mass is determined from $m_\gamma(T) \simeq eT/3 \simeq T/10$ at the temperatures under consideration.
The initial condition for this contribution to the yield is fixed by $Y(T=\text{min}\{\TRH,T_*\}) = 0$, where $T_*$ is the temperature at which $m_\gamma(T_*) = m_a/2$, since for higher temperatures the inverse decay is kinematically forbidden.
Further, when $T \lesssim m_e$, the electrons will begin to become non-relativistic, which in turn leads the thermal mass of the photon to decrease to zero.
In the calculation we set $m_\gamma = 0$ for $T< m_e/2$; however, our results are very insensitive to the exact temperature at which the photon mass is switched off.

For the $e^- e^+ \to a$ inverse decay, the amplitude is given by
\be
  \sum |\mathcal{M}_{e^- e^+ \to a}|^2 = 2 \gae^2m_a^2.
\ee
For this process we set the initial condition to be $Y(T=\TRH) = 0$.
We note that here it is necessary that $m_a>2m_e$ for the process to be kinematically allowed.
Further, we note that production turns off exponentially fast near $T = m_e$ as the electrons become non-relativistic.

\subsection{The Reaction Rate for $2\to 2$ Processes}

Next we consider $2\to 2$ production, as appropriate for photon conversion and electron-positron annihilation.
For these cases, we make the additional approximation that the distributions can be treated as roughly Boltzmannian and that Bose enhancement or Pauli blocking can be ignored.
(We have estimated that this approximation is accurate to a few percent.)
As a result, the rate for a single $2 \to 2$ process of the form $1+2 \to 3+a$ (i.e. $e^\pm \gamma \to e^\pm a$ or $e^- e^+ \to \gamma a$) can be written as
\be
  R_{2\to 2}(T) 
  =
  \int d\Pi_1 d\Pi_2 d\Pi_3 d\Pi_a (2\pi)^4\,\delta^4(p_1 + p_2 - p_3 - p_a)  |\mathcal{M}_{12\to3a}|^2 e^{-(E_1+E_2)/T}.
\ee
For this result we used the fact that within the Boltzmannian approximation $f_{i,f} \ll 1$, and further as the density of the axions is always subdominant, from \Eq{eq:Phi_base}, $\Phi \simeq f_1 f_2 = e^{-(E_1+E_2)/T}$.

To simplify the rate equation, we insert the identity $1 = \int d^4P\, \delta^4(P - p_3 - p_a)$ which gives
\bea
  R_{2\to 2}(T) 
  =&
  \int d^4P d\Pi_1 d\Pi_2\, \delta^4(P - p_1 - p_2) e^{-P_0/T}\times \left[\int d\Pi_3 d\Pi_a (2\pi)^4\, \delta^4(P-p_3 - p_a) |\mathcal{M}_{12\to3a}|^2\right]\\
  =&\int d^4P d\Pi_1 d\Pi_2\, \delta^4(P - p_1 - p_2) e^{-P_0/T}\times \left[2g_1 g_2\lambda^{1/2}(P^2,m_1^2,m_2^2)\sigma(P^2) \right]\!,
  \label{eq:2to2Integral1}
\eea
where $\lambda(a,b,c) \equiv a^2 + b^2 + c^2 - 2 ab - 2bc - 2ca$, $g_i$ is the number of internal degrees of freedom of particle $i$, and $\sigma(P^2)$ is the spin averaged cross section for $1+2\to 3 +a$ with center-of-mass squared energy $P^2$.
The integral over the incoming particles can be readily evaluated,
\be
  \int d\Pi_1 d\Pi_2 \delta^4(P - p_1 - p_2) = \frac{1}{(2\pi)^6}\frac{4\pi \lambda^{1/2}(P^2,m_1^2, m_2^2)}{8 P^2}.
\ee
Before using this in \Eq{eq:2to2Integral1}, we rewrite the final integration measure as $d^4P = 2\pi \sqrt{P_0^2 - s}~ dP_0 ds$, where we use $s = P^2 \in [s_{\rm min},\infty)$, with $s_{\rm min} = (m_1+m_2)^2$.
The integral over $P_0$ is evaluated in terms of a modified Bessel function of the second kind, and the rate is given by
\be
  R_{2\to 2}(T) 
  =\frac{g_1 g_2 T}{32\pi^4} \int^\infty_{s_{\rm min}} ds\, \lambda(s,m_1^2, m_2^2)  
  \frac{K_1(\sqrt{s}/T)}{\sqrt{s}}\sigma_{12\to 3a}(s).
  \label{eq:intC_P}
\ee

Therefore, the determination of the rate has been reduced to computing the cross sections for each of the two processes.
The majority of the production occurs at temperatures $T\gg m_e,m_\gamma$, which implies that it is an excellent approximation to remove these masses everywhere they are not acting as an IR cutoff.
(This follows as $m_\gamma(T) \simeq T/10$ is generally subdominant, and further as the temperature approaches $m_e$, the electrons and positrons leave equilibrium, greatly depleting their number density.)
Firstly, the fermion annihilation cross section is given by
\bea
  \sigma_{\rm FA}(s) = &
  \frac{\alpha  g_{a\gamma \gamma}^2 }{24\beta  }\left(1-\frac{m_a^2}{s}\right)^3\left(1+\frac{2m_e^2}{s}\right)
  +
  \frac{\alpha  \gae^2 }{ 2s^2\left(s-m_a^2\right)\beta^2}\left[(s^2-4m_e^2 m_a^2+m_a^4)\ln\left(\frac{1+\beta}{1-\beta}\right) - 2 \beta m_a^2 s \right]\\&-
  \frac{\alpha g_{a\gamma \gamma} \gae m_e}{ 2 s \beta^2} \left(1-\frac{m_a^2}{s}\right)^2\ln\left(\frac{1+\beta}{1-\beta}\right)\!,
\eea
where $\beta = \sqrt{1-4m_e^2/s}$.
Although it is not required, as stated above, we have retained the full $m_e$ dependence in this result.
Next, for photon conversion, the cross section is (dropping the electron mass everywhere it is not required),
\bea
  \sigma_{\rm PC}(s) = &\frac{\alpha  g_{a\gamma \gamma}^2}{32 s^2} \left[2(2s^2-2m_a^2 s + m_a^4)\ln\left(\frac{s-m_a^2}{m_\gamma^2}\right) - 7s^2+10 m_a^2 s-5m_a^4\right]\\
  &+\frac{\alpha  \gae^2}{8s^3} \left[2 \left(2 s^2-2 m_a^2 s+m_a^4\right) \ln \left(\frac{s}{m_e^2}\right)-3 s^2+10 m_a^2 s-7 m_a^4\right]\\
  &- \frac{\alpha  g_{a\gamma \gamma} \gae m_e}{8 s^3(s-m_a^2+m_e^2)}\left[  2(s^3 + m_a^6)\ln\left(\frac{(s-m_a^2)^2}{(s+m_a^2)m_e^2}\right) - 3(s+m_a^2)(s-m_a^2)^2\right]\!.
\eea

\subsection{Are the axions non-relativistic at the time of their decay?}

Importantly, the constraints we derived on the irreducible axion density assume that the particles are non-relativistic at the time of their decay, effectively decaying as we would expect cold DM to do so.
Given the axions are produced through freeze-in, they were never in thermal equilibrium with the SM bath, and therefore do not share the same temperature: in fact, they will not have a thermal phase space distribution.
Nonetheless, the axions produced via freeze-in will have an energy of the same order as the temperature at production.
This energy will then approximately redshift proportionally to the temperature of the SM bath.
Therefore, for the axions to decay with a non-relativistic velocity, the temperature at which an order one fraction of axions have decayed should satisfy $T_\text{decay} \ll m_a$, where $T_{\rm decay}$ is defined through $\tau_a H(T_{\rm decay})\simeq1$.
The earliest decays we constraint are due to spectral distortions of the CMB, which probe an epoch when the Universe was still radiation dominated.
Accordingly, the most conservative requirement we can state can be determined in that epoch, and as we require $H(m_a)\tau_a\gg1$, that becomes the following requirement for the photophilic and photophobic axion
\bea
g_{a\gamma \gamma}\ll~& 10^{-7~}\left(\frac{m_a}{\text{MeV}}\right)^{-1/2}~\text{GeV}^{-1}, &&\quad  \text{for all } m_a,\\
\gae\ll~& 10^{-10}\left(\frac{m_a}{\text{MeV}}\right)^{+1/2},&& \quad  \text{if } m_a > 2m_e.
\eea
Both of these are clearly satisfied in all of the parameter space we constrain.

%%%%%%%%%
\section{Further details of the Astrophysical and Cosmological Constraints}
\label{sec:constraints}
%%%%%%%%%

Here we expand on the derivation of the constraints used in this work.
As emphasized many times, our approach is to reappropriate DM constraints to search for the irreducible axion abundance.
Broadly, assuming we have a single axion coupling $g_a$, we can imagine the most general scenario as being described by three parameters: mass, coupling strength to the SM, and a relic density, $\{m_a,\,g_a,\,\rho_a\}$.
DM axion searches probe this space along the axis defined by $\rho_a = \rho_\DM$.
It is also common to consider fractional DM scenarios, with $\rho_a < \rho_\DM$.
Our approach is slightly different, in that we profile the parameter space according to $\rho_a = {\cal F}_a(m_a,g_a) \times \rho_\DM$, where ${\cal F}_a(m_a,g_a)$ is the irreducible density for a given mass and coupling as determined in Sec.~\ref{sec:abundance}.
With this general picture in mind, we now provide an expanded discussion of the various constraints we considered in the main body.

\subsection{Galactic Constraints}

\begin{figure}[t!]
    \centering
    \subfloat{\includegraphics[width= 0.47 \textwidth]{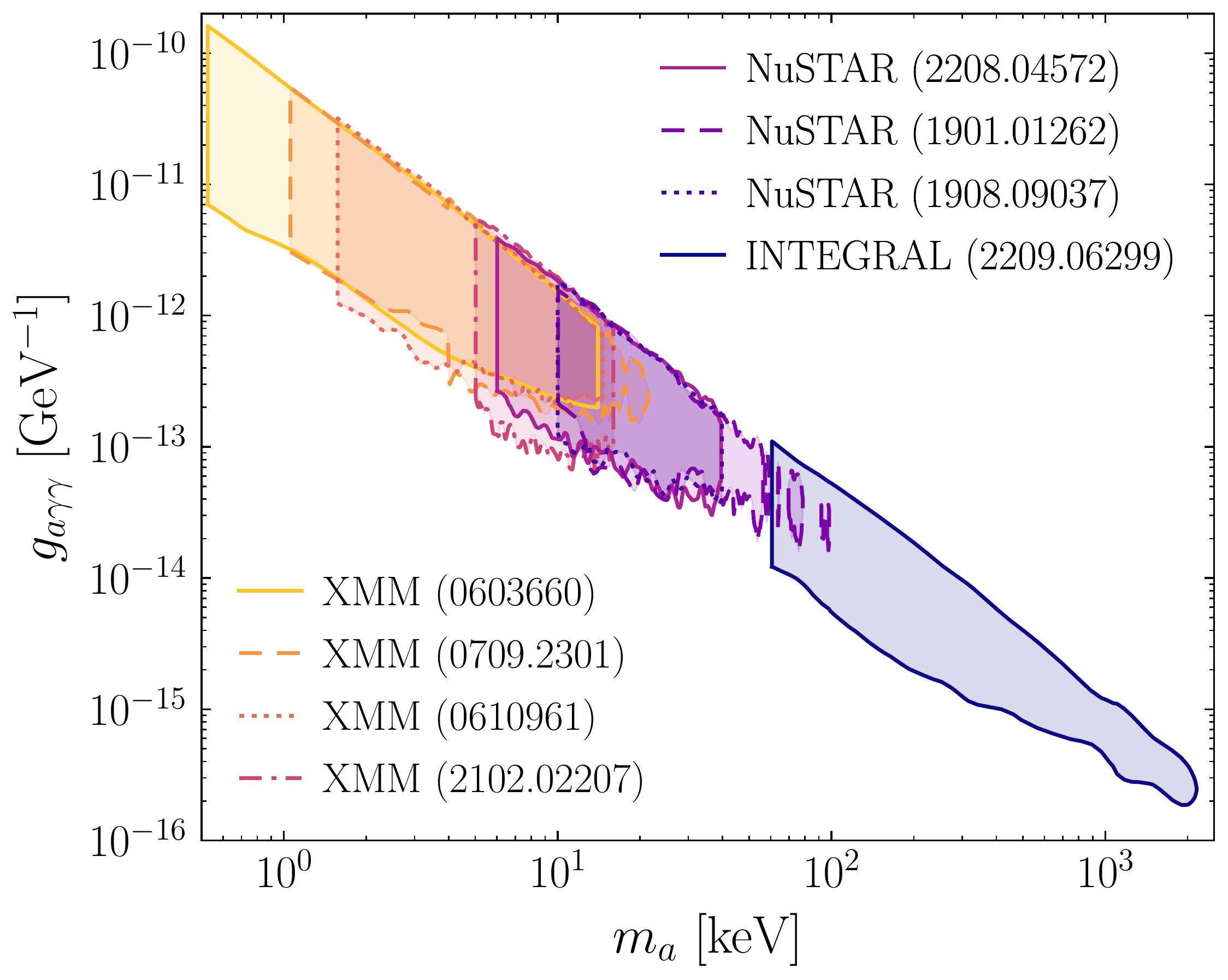}}
    \hspace{0.5cm}
    \subfloat{\includegraphics[width= 0.47 \textwidth]{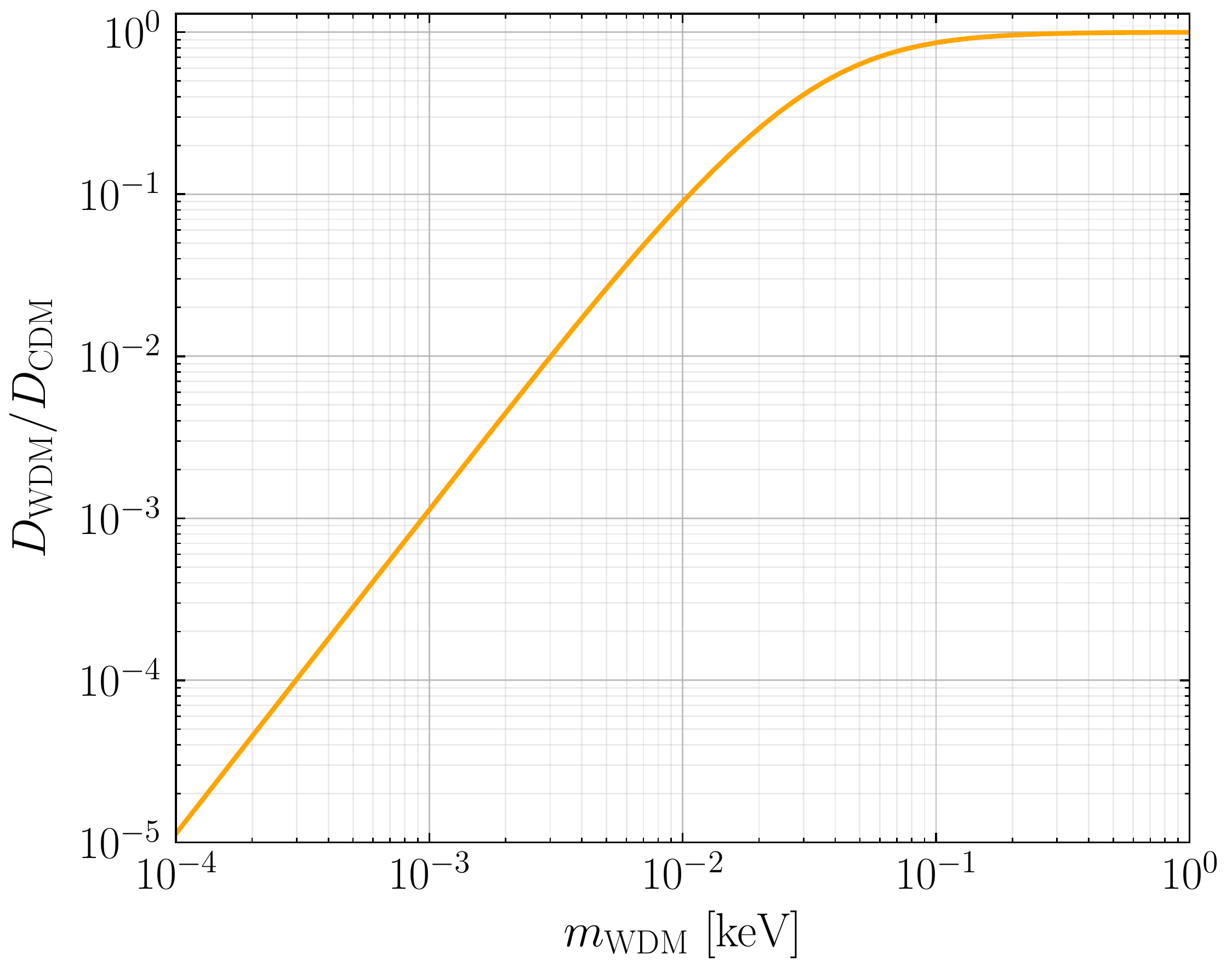}}
    \vspace{-0.4cm}
    \caption{{\it Left:} The breakdown of the X-ray constraint in \Fig{fig:Constraints} into the eight individual analyses that were used, specifically \Refs{Boyarsky:2006fg,Boyarsky:2006ag,Boyarsky:2007ay,Foster:2021ngm,Ng:2019gch,Roach:2019ctw,Roach:2022lgo,Calore:2022pks}.
    In each case, the analysis is identified by the instrument which collected the dataset used, and the \texttt{arXiv} reference in which the analysis appeared.
    {\it Right:} The impact on the DM decay flux expected from the Milky Way as a result of the reduced clustering experienced by WDM.
    We compute the average $D$-factor in the region analyzed in \Refc{Foster:2021ngm} for CDM, and show the relative reduction for WDM when including the effect in \Eq{eq:WDM_Fa}.
    At lower masses, there is a clear suppression in the flux, however, the effect does not arise until masses below those we consider (shown in the left plot).
    }
    \label{fig:Galactic-Additional}
\end{figure}

Across a wide range of masses, the strongest constraints on DM decay originate from searches for those decays within the Milky Way.
Accordingly, we consider these bounds first.
As noted in the main body, the constraints we labeled as X-ray actually originated from eight independent analyses.
In \Fig{fig:Galactic-Additional} we show the breakdown of the full contour that appeared in \Fig{fig:Constraints} by its constituents, where each analysis is labeled by the instruments that collected the data used, and the \texttt{arXiv} reference to the relevant work.
To provide more detail, four of the results used data from XMM-Newton~\cite{Boyarsky:2006fg,Boyarsky:2006ag,Boyarsky:2007ay,Foster:2021ngm}, three from NuSTAR~\cite{Ng:2019gch,Roach:2019ctw,Roach:2022lgo}, and one INTEGRAL~\cite{Calore:2022pks}.
As for the object observed, five of the analyses considered emission from the Milky Way halo~\cite{Boyarsky:2006ag,Foster:2021ngm,Roach:2019ctw,Roach:2022lgo,Calore:2022pks}, two considered M31~\cite{Boyarsky:2007ay,Ng:2019gch}, and a single work drew on emission from the Large Magellanic Cloud~\cite{Boyarsky:2006fg}.
For the mass range probed, we considered each of these to be sufficiently robust against the issue of light axions behaving as WDM, and therefore not clustering in the same way as CDM.

As \Fig{fig:Galactic-Additional} demonstrates, for most masses probed by our X-ray constraints, multiple results give comparable constraints, so that individual analyses could be removed without qualitatively changing our results.
Relatedly, we briefly comment on several results we chose not to include.
\Refc{Essig:2013goa} provides constraints from a number of instruments across our mass range, however, the results there were in all cases subdominant to the other bounds we considered, noting that their INTEGRAL results were updated in \Refc{Laha:2020ivk}.
(For $\TRH = 100~{\rm MeV}$ their COMPTEL constraint is relevant, and we have included it in \Fig{fig:Constraints_100MeV}.)
X-ray constraints on axion decays were also considered in \Refc{Cadamuro:2011fd}, and we have drawn on all the analyses they exploited there, except for the INTEGRAL constraints in \Refc{Boyarsky:2007ge}, as we instead used the recent analysis of \Refc{Calore:2022pks}, and the XMM-Newton analysis of \Refc{Boyarsky:2006ag} which drew on observations of Milky Way dwarf satellite Ursa Minor, given it could be more impacted by the WDM considerations.
Finally, we did not include the strong constraints near $m_a \simeq 7~{\rm keV}$ from \Refc{Dessert:2018qih}, as we instead made use of the extended and more conservative limits in \Refc{Foster:2021ngm}.

As discussed in the main text, as we reduce $m_a$, eventually $\rho_a$ will begin behaving like WDM rather than CDM, and negate a simple recasting of CDM results.
For a Milky Way like halo, the modification was studied with numerical simulations in \Refc{Anderhalden:2012qt}.
The authors of that work quantified that a WDM component would have a modified clustering as described by \Eq{eq:WDM_Fa}.
Importantly, the spatial profile is modified according to $m_a$, and was found to be independent of ${\cal F}_a$.
Intuitively this should be expected: once the energy density of the WDM component becomes sufficiently subdominant, it will have no impact on the global potential, and instead simply behave as a test particle in the potential.
Accordingly, we are confident in extending the findings from \Refc{Anderhalden:2012qt}, which only simulated fractions as small as ${\cal F}_a = 0.05$, down to the very small fractions we considered.

The X-ray bounds we use extend down to 0.5 keV.
From \Eq{eq:WDM_Fa}, we can see that 1 keV is the characteristic mass, below which these effects begin to become important.
However, we can be more quantitative.
Our archetypal X-ray bound is from \Refc{Foster:2021ngm}, which searched for DM decays in the Milky Way in a region extending from 2 to 48 degrees from the Galactic Center.
The expected DM flux from that region is controlled by the associated $D$-factor, as described below \Eq{eq:DMdecayFlux}, and we can compute the average $D$-factor in that region for CDM, and also WDM by including \Eq{eq:WDM_Fa} in the DM distribution.
In \Fig{fig:Galactic-Additional} we show the ratio of the two $D$-factors as a function of the WDM particle mass.
Appreciable suppression does not begin until $m_a \lesssim 0.1$ keV, below our smallest mass.
For much smaller masses, the suppression can be significant.
Ultimately, this effect must be accounted for in any individual analysis, but in our future X-ray projections that we displayed in \Fig{fig:Constraints}, we included the suppression as in \Fig{fig:Galactic-Additional}, which explains the change in slope of the constraints below 0.1 keV.

\subsection{Extragalactic Constraints}

\begin{figure*}
\centering
\includegraphics[width=0.49\textwidth]{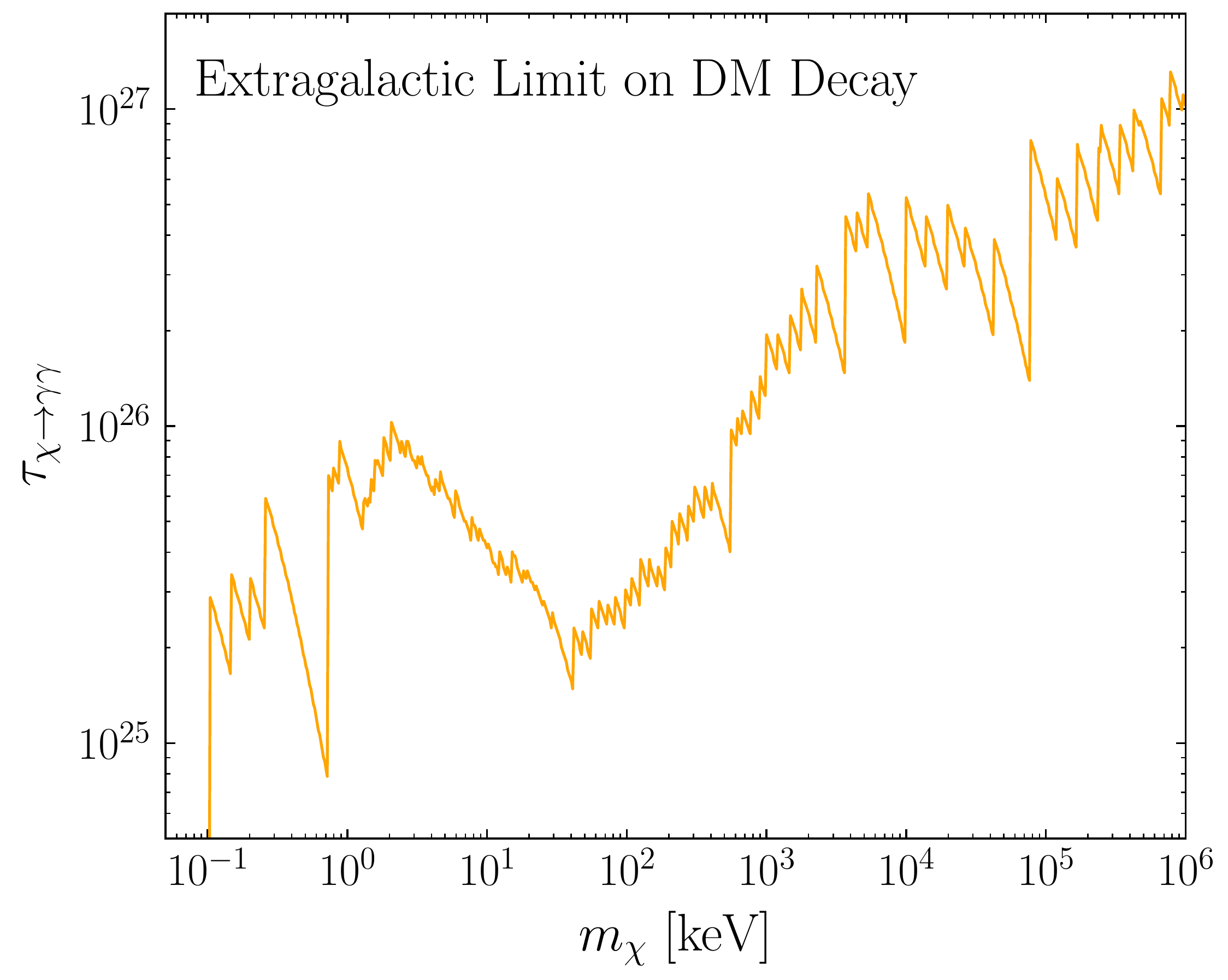}
\vspace{-0.4cm}
\caption{The bound one would obtain on DM, if our procedure for setting limits on decays of the irreducible axion abundance using the CRB, were repeated in the case of ${\cal F}_a = 1$.
For all masses shown, stronger bounds on DM decaying to two photons exist, yet it provides a relevant bound in \Fig{fig:Constraints} as it probes epochs at times before $\tU$.
}
\label{fig:EG-CRB-DM}
\end{figure*}

Galactic constraints probe the present epoch, and as discussed many times, for $\tau_a \ll \tU$ the bounds decouple.
To probe earlier times we can consider the decays of DM throughout the Universe, and not just within the Milky Way.
Focusing on $a \to \gamma \gamma$, these decays will once more produce photons that arrive at Earth, but as this process occurs across a range of cosmological redshifts, a continuum of photons is observed rather than a line.
In detail, integrating over the contributions from all redshifts, the flux (in units of [photons/cm$^2$/s/eV]) is given by
\be
  \frac{d\Phi}{dE} = \frac{2 {\cal F}_a \rho^0_\DM \dH}{m_a\,E} \frac{e^{-t(E)/\tau_a}}{\tau_{a\to \gamma \gamma}} \frac{e^{-\kappa(m_a/2E-1,\,E)}}{\sqrt{\Omega_m (m_a/2E)^3 + \Omega_\Lambda}} \Theta(m_a-2E).
  \label{eq:EGFlux}
\ee
Here $\dH = c/H_0$ with $H_0$ the present Hubble constant, whilst $\Omega_m$ and $\Omega_\Lambda$ are the matter and dark energy densities, where we use the Planck 2015 cosmological parameters~\cite{Planck:2015fie}.
The flux further depends on the optical depth, $\kappa(z,E)$, a function of the emission redshift and photon energy, for which we use the results of \Refc{Cirelli:2010xx}.
Finally, $t(E)$ is the age of the Universe when a photon that is observed to have energy $E$ was emitted,
\be
  t(E) = \frac{2}{3H_0 \sqrt{\Omega_\Lambda}} {\rm ArcSinh} \left( \sqrt{\frac{\Omega_{\Lambda}}{\Omega_m(m_a/2E)^3}} \right)\!. 
\ee

One can search for the flux in \Eq{eq:EGFlux} in a variety of ways.
Here we take the conservative approach of not allowing the flux to exceed the one-sigma upper errors on the measurement of the CRB, or at energies where no measurement has been made, we ensure the flux is below the established upper limits.
For both, we take the compiled results from \Refc{Hill:2018trh}.
(Note, a similar analysis was performed in \Refc{Cadamuro:2011fd}, although we utilize both updated measurements of the optical depth and CRB.)

To provide further intuition for the CRB bound, we can set ${\cal F}_a = 1$ in \Eq{eq:EGFlux}, and use this procedure to set a limit on the lifetime of DM to decay to two photons, $\tau_{\chi \to \gamma \gamma}$.
The results of this analysis are shown in \Fig{fig:EG-CRB-DM}.
Taken as constraints on DM, these are weak (for example, cf. \Refc{Essig:2013goa}).
However, this only reflects their conservative nature.
As the constraints can probe earlier epochs, they can be leading for parameters where the galactic bounds are not.

\begin{figure}
\includegraphics[width=0.49\textwidth]{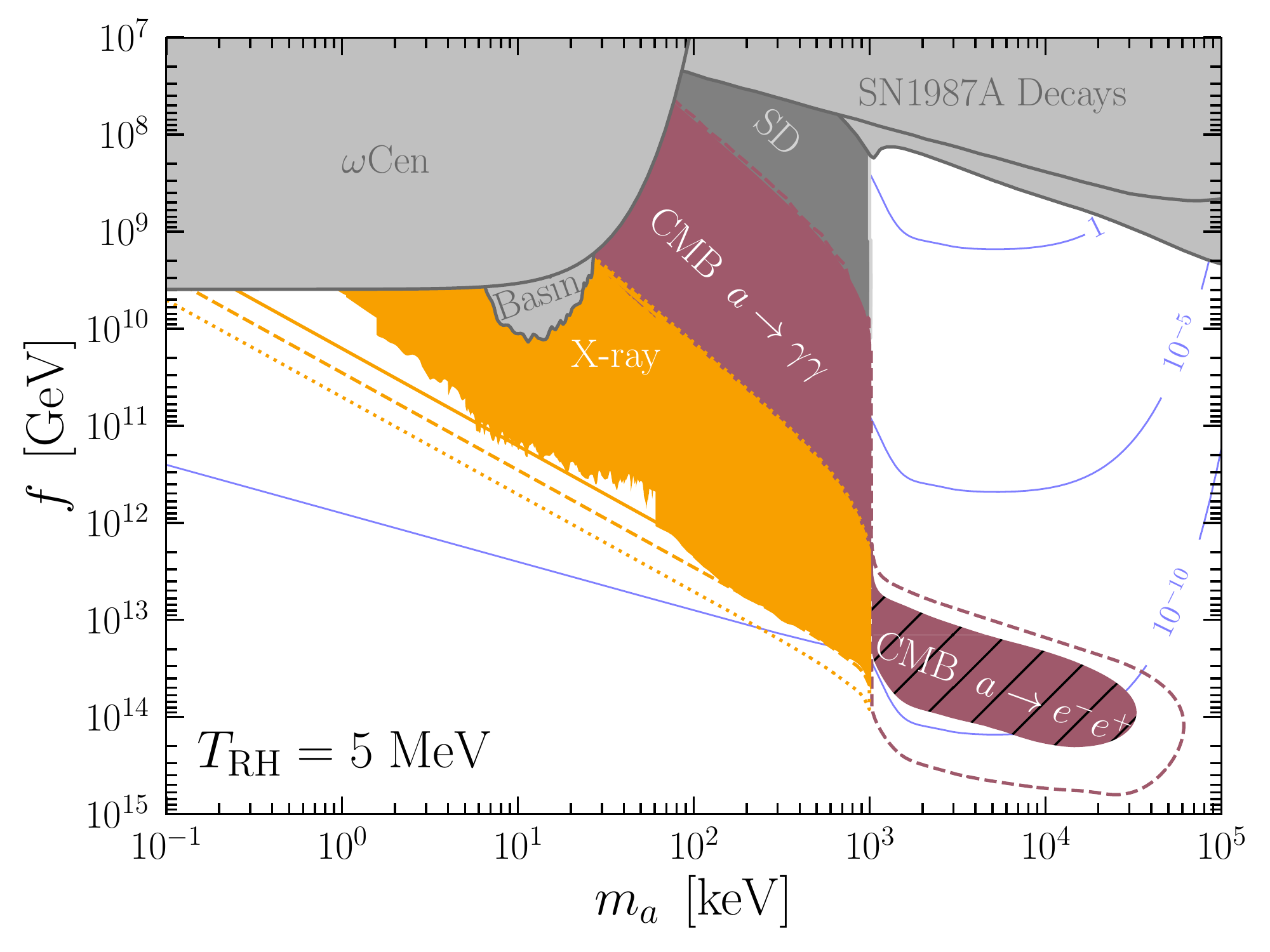}
\vspace{-0.4cm}
\caption{An extension of the results in Figs.~\ref{fig:Constraints} and \ref{fig:Constraints_gae} to a scenario with \textit{universal couplings}: $\ga = \alpha/2\pi f_a$ and $\gae = m_e/f_a$.
We see that these constraints resemble the constraints of the photophobic axion.
This is largely because the production is dominated by the coupling of axions to electrons, as well as the fact that just above $m_a> 2m_e$, the decay to an electron-positron pair is significantly stronger than the decay to photons.}
\label{fig:Constraints_universal}
\vspace{-0.4cm}
\end{figure}

\subsection{CMB Constraints}

The CMB provides a powerful probe of axion decays to both photons and electrons, as its observed properties can be modified by energy injected at extremely early times.
As discussed in the main body, we drew on two CMB observables.
The earliest probe is due to CMB spectral distortions, and is sensitive to energy injections at $z \lesssim 2 \times 10^6$.
Here, we follow the results in \Refc{Balazs:2022tjl}.
Even though the focus of that work was $\ga$, as their constraints are presented as a function of ${\cal F}_a$ and $\tau_{a\to\gamma \gamma}$ we can readily apply them to our scenario, even for models where $\gae \neq 0$ through the use of \Eq{eq:Fulla2gg}.
This approach yielded the constraint we labeled as spectral distortions (SD) in our figures.

CMB anisotropies provide a later probe, and here we recast the bounds determined in \Refs{Slatyer:2016qyl,Poulin:2016anj,Cang:2020exa,Bolliet:2020ofj}.
Below $10~{\rm keV}$ we made use of \Refc{Bolliet:2020ofj} that reported CMB anisotropies constraints for $m_a\lesssim 20\,\rm keV$.
The constraints in that work were also provided as a function of lifetime and DM fraction, so they could be directly implemented.
For $m_a>10~{\rm keV}$ we use the constraints derived in \Refc{Cang:2020exa} by updating \Refs{Slatyer:2016qyl,Poulin:2016anj} to incorporate the Planck 2018 data.
These works primarily focused on constraining the lifetime for all of the DM to decay, rather than just a fraction, and so we developed a procedure to extend this to fractional decays.
For decaying particles in the mass range $10~{\rm keV} - 10~{\rm TeV}$, \Refs{Slatyer:2016qyl,Poulin:2016anj} report the maximal fraction of DM that can decay electromagnetically consistent with CMB observations as a function of the decay lifetime.
As discussed in the text, we found that those results were conservatively reproduced by \Eq{eq:F_a_CMB}.
That the result exhibits a weaker cutoff in time than a pure exponential is to be expected: the CMB probes energy deposited from decays over a wide range of epochs.
We note that \Refc{Balazs:2022tjl} performed a more exhaustive analysis of CMB anisotropy constraints on the photophilic irreducible axion, however, we found that the constraints we derived were qualitatively similar to those found in that reference.

%%%%%%%%%
\section{Probing Axions That Couple to both Photons and Electrons}
\label{sec:bothcouplings}
%%%%%%%%%

\begin{figure}[t!]
\centering
\includegraphics[width=0.49\textwidth]{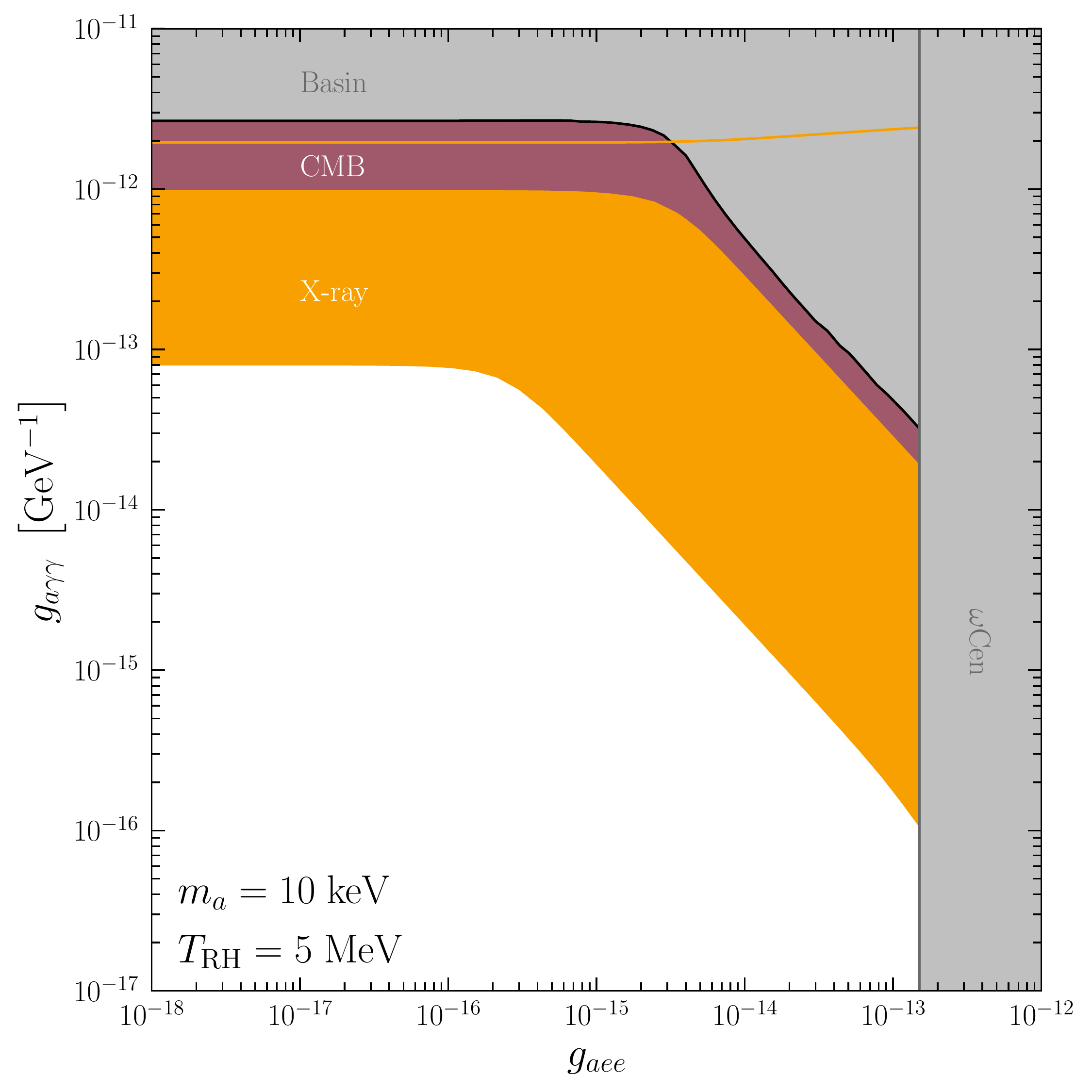}\;\;\includegraphics[width=0.49\textwidth]{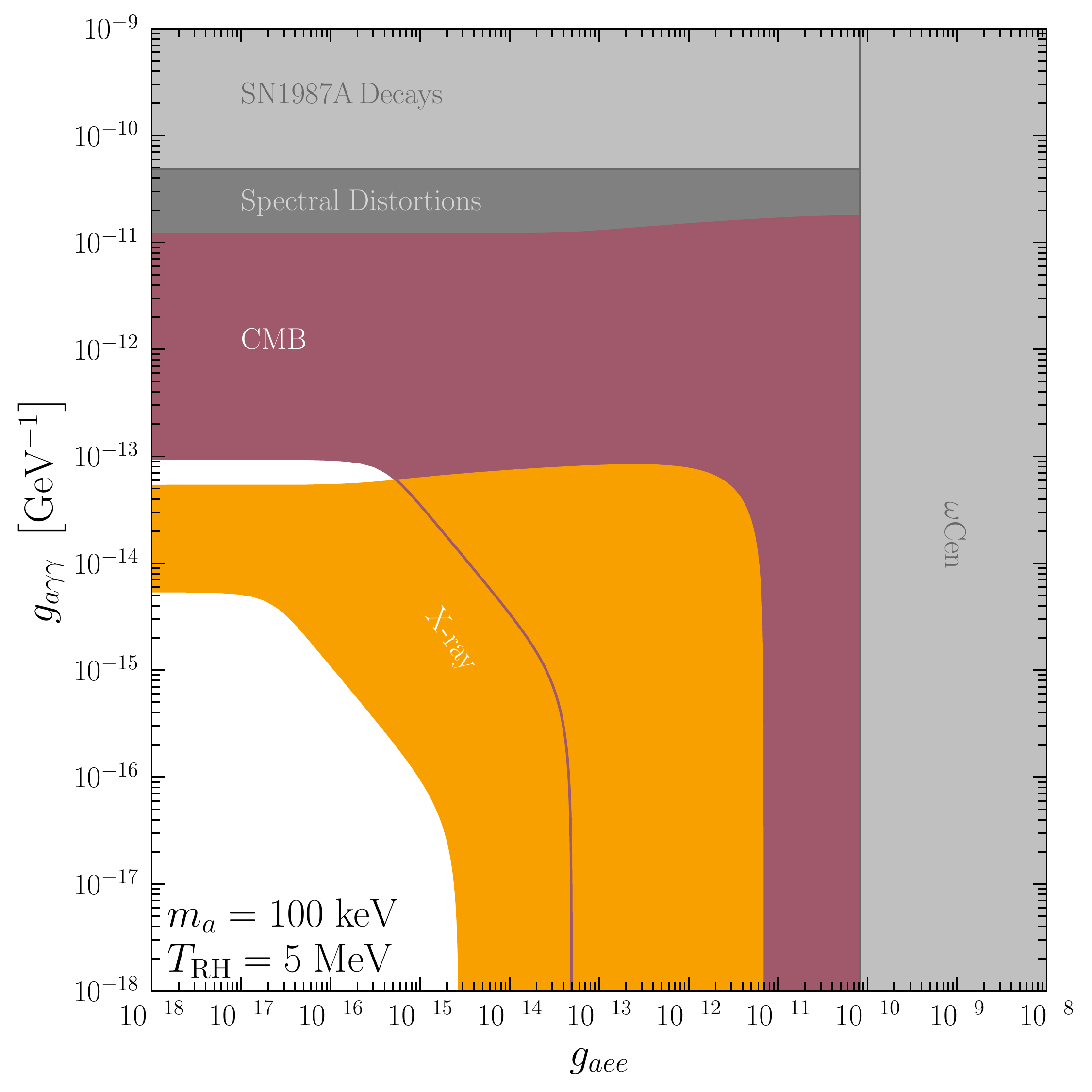}\\
\includegraphics[width=0.49\textwidth]{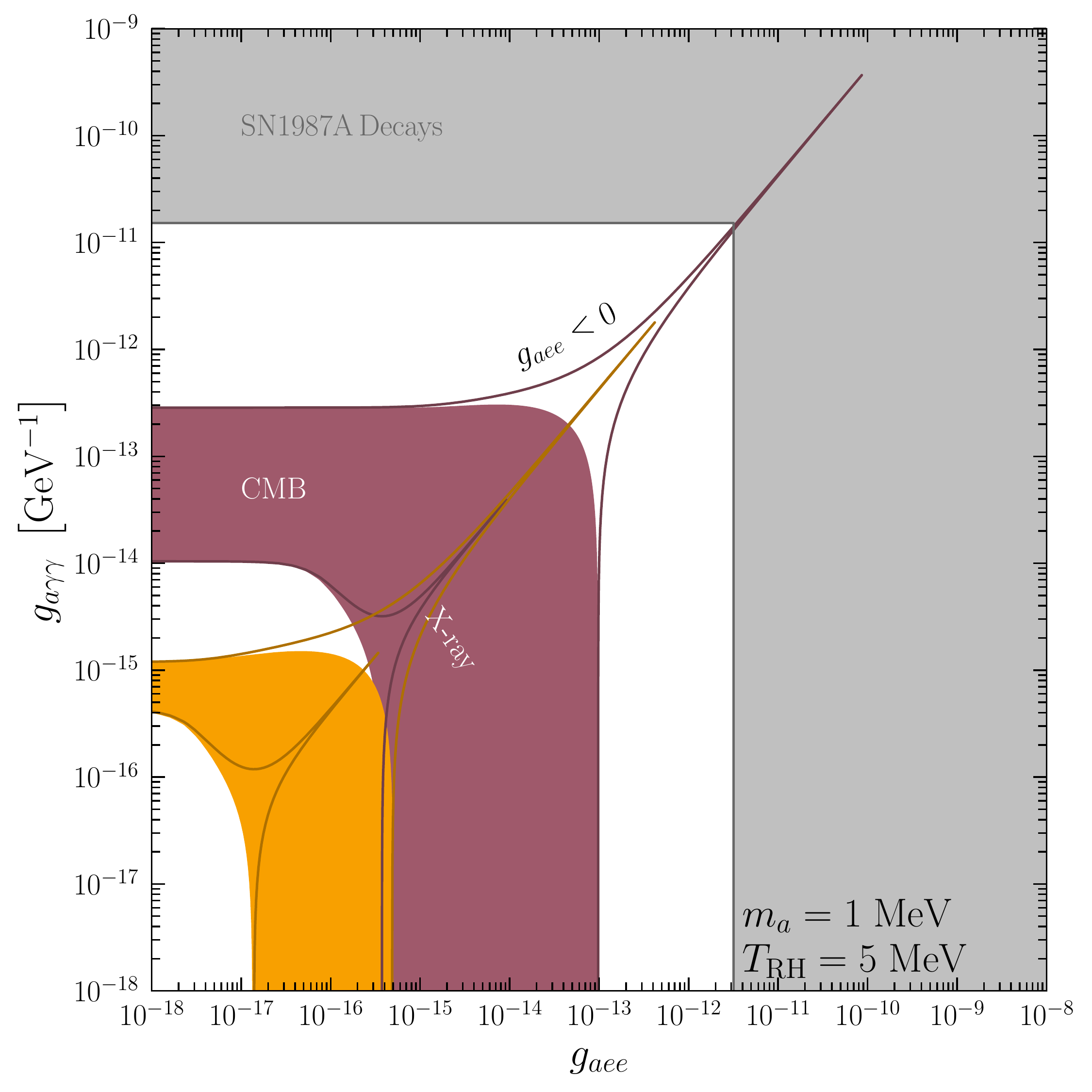}\;\;\includegraphics[width=0.49\textwidth]{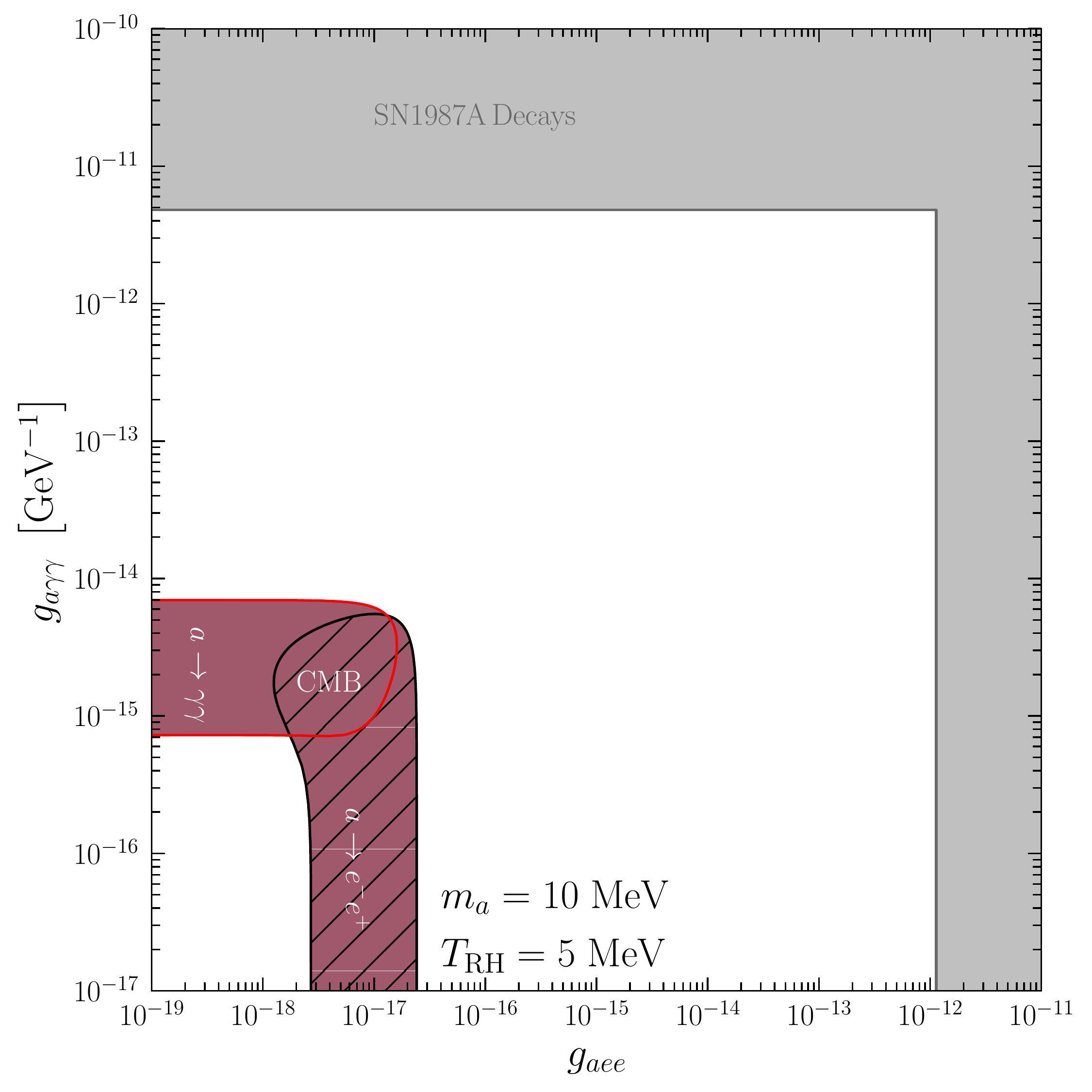}
\vspace{-0.4cm}
\caption{Constraints on axions with $\ga$ and $\gae$ floated independently, for four different fixed masses that scan either side of our fixed reheat temperature, $\TRH=5~{\rm MeV}$.
For $m_a = 1~{\rm MeV}$ we show both $\gae > 0$ and $\gae < 0$ (as indicated by the darker CMB and X-ray contours), for all other masses we only show the former.
In contrast to other figures, here we only show the constraints arising from DM X-ray searches, CMB anisotropies, in addition to the existing astrophysical bounds.
We note that for astrophysical bounds other than the basin, the sharp boundaries as one moves between $\ga$ and $\gae$ dominance are unphysical.
In reality there will be a transitional region where both contribute, which we have not attempted to include.}
\label{fig:Constraints_gae_gag}
\end{figure}

In general, the tree level couplings of an axion to a photon or electron can be independently determined in the UV.
In the main body of the paper we focused on the photophilic and photophobic axion, where one specific coupling dominates.
However, this is not the most general case where both couplings can be active.
For this reason we consider the impact of having both couplings contribute in this section.
The analysis proceeds almost identically.
As discussed in Sec.~\ref{sec:abundance}, the abundance is generated by the sum of all processes, and so we can simply include all possibilities now.
Again we include the one-loop correction to $\ga$ induced by $\gae$, the correction is unimportant for the production, but can play an important result on the late time decays and resulting constraints.

Firstly, we can imagine both couplings with ${\cal O}(1)$ coefficients in the UV, in which we expect $\ga$ and $\gae$ to both be simply determined by the axion decay constant, $f_a$, with characteristic values
\be
  \ga  = \frac{\alpha}{2\pi f_a},\hspace{0.5cm}  \gae = \frac{m_e}{f_a}.
\ee
This scenario is sometimes referred to as the set of \textit{universal couplings}.
Constraints in this case are shown \Fig{fig:Constraints_universal}, and as described there largely resemble the photophobic results of \Fig{fig:Constraints_gae}.

We can also consider scenario where we vary $\ga$ and $\gae$ independently at fixed mass, and these are shown in \Fig{fig:Constraints_gae_gag} for four different masses.
Here we only show the impact of our dominant X-ray and CMB anisotropy constraints.
As discussed below \eqref{eq:fullL}, there can be a relative sign between $\ga$ and $\gae$, which for example could be chosen to arrange for the decay rate in \Eq{eq:Gam_a2gamgam} to vanish.
the bounds in such a case are qualitatively very similar to those shown, except along lines in parameter space where the two photon decay rate is made to approximately vanish, and we show a single example in the bottom left panel of~\Fig{fig:Constraints_gae_gag} for $m_a = 1~{\rm MeV}$.
There are several features in these results worth commenting on.
Firstly, there are two different physical effects which determine the boundary of the constraints to the bottom left: production and decay.
For axions lighter than the reheat temperature, production proceeds dominantly through photon conversion (see \Fig{fig:FDM}).
In this case, it can be seen from \Eq{eq:F_a_approx} that when $\ga \gtrsim 3 \times 10^2\, \gae~{\rm GeV}^{-1}$, production is dominated by $\ga$, whereas in the opposite regime $\gae$ dominates production.
By comparison, for $m_a < 2 m_e$ (which is comparable to the reheat temperature), from \Eq{eq:Gam_a2gamgam} decays to photons are controlled by $\ga$ whenever $\ga \gtrsim \gae (m_a/\rm MeV)^2\, {\rm GeV}^{-1}$.
Given this, we see that for the results we show with $m_a = 10~\rm keV$ and $m_a = 100~\rm keV$, at low values of $\gae$ the production and decay is dominated by $\ga$.
But at larger $\gae$ values, the change in behavior is initiated by $\gae$ taking control of the production, while the decay remains dictated by $\ga$.
For heavier axions, the production is dominated by inverse decays, and the decay will be controlled by the direct analogues of these processes.
As usual, the cutoff on constraints at large couplings occurs whenever $\tau_a$ becomes shorter than the relevant process, be it $\tU$ or $t_\CMB$.
At the top-right and bottom-left panel of~\Fig{fig:Constraints_gae_gag}, at small $\ga$, the constraints is independent of $\ga$, although, the constraints shown are due to decays into photons. This is a result of the loop induced decay ($\propto \gae^2$) becoming larger than the tree level decay ($\propto \ga^2$) (cf. Eq.~\eqref{eq:Fulla2gg}).

%%%%%%%%%
\section{Accounting for Axion Production from Misalignment }
\label{sec:extension-Mis}
%%%%%%%%%

Throughout this work we have focused on axion production through freeze in.
That focus puts aside the canonical axion production mechanism: misalignment.
While the generic expectation is that misalignment will contribute -- in all likelihood dominantly -- it is not irreducible, and therefore we neglected it in our primary analysis.
In this section we discuss how to include misalignment in a relatively robust manner, and how this additional contribution to the axion density modifies our constraints.

At the outset, we note that misalignment production is expected to be highly complementary to freeze in.
As exemplified in \Eq{eq:F_a_approx}, on general grounds we expect the freeze-in contribution to be proportional to the axion-SM coupling squared, and therefore to scale proportionally to $f_a^{-2}$, peaking at smaller $f_a$ values.
In contrast, the axion energy density generated by misalignment will be proportional to $f_a^2$, the exact inverse of freeze in.
Despite this complementarity, the relic abundance derived from misalignment depends critically on the early Universe cosmology in two fundamental ways: on when the initial misalignment angle is set relative to inflation and on exactly how the Universe expanded after the axion field began to oscillate.
We will expand on both points below.

Consider first the effect of the time when the initial misalignment angle, $\theta_0$, was set relative to inflation.
If set before or during inflation, then $\theta_0$ will take on a single value throughout the observable Universe, and in principle could have been arbitrarily small, rendering the resulting contribution reducible.
If on the other hand the angle is set after inflation, then different spatial regions of the Universe will have obtained unique initial values; the total misalignment abundance can then be obtained by averaging over these regions, in detail $\langle \theta_0 \rangle \sim \pi/\sqrt{3} \sim 2$~\cite{Turner:1986tb}.
In this latter case there can also be significant axion production from cosmic string emission, where the strings emerge from the same U(1) breaking as the axion may have.
For details of axion generation in this case, we refer to, for example, \Refs{Gorghetto:2018myk,Gorghetto:2020qws,Buschmann:2019icd,Dine:2020pds,Buschmann:2021sdq}.
To avoid such details, we will focus on the first scenario, so that going forward $\theta_0$ is treated as a free parameter.

The second determinative aspect of the cosmology is the amount of expansion the Universe has undergone after the axion field began to oscillate, which occurs once $m_a \simeq H$.
While we know the Universe must have been radiation dominated during BBN, before this the expansion could have been considerably different.
Three potential pre-BBN cosmologies and their impact on axion misalignment production were studied in \Refc{Blinov:2019rhb}; in detail, they considered
\begin{enumerate}
  \item Standard cosmology where the Universe is radiation dominated;
  \item A period of early matter domination; and
  \item A period of early kination domination.
\end{enumerate}
As an example of the impact a modified early cosmology can have, misalignment combined with a period of early matter domination can naturally generate the correct relic DM abundance for $m_a \gtrsim 1~{\rm keV}$~\cite{Foster:2022ajl} (although there are other mechanisms to generate axion DM at such masses, see e.g. \Refc{Panci:2022wlc}).

\begin{figure}[!t]
\centering
\includegraphics[width=0.5\textwidth]{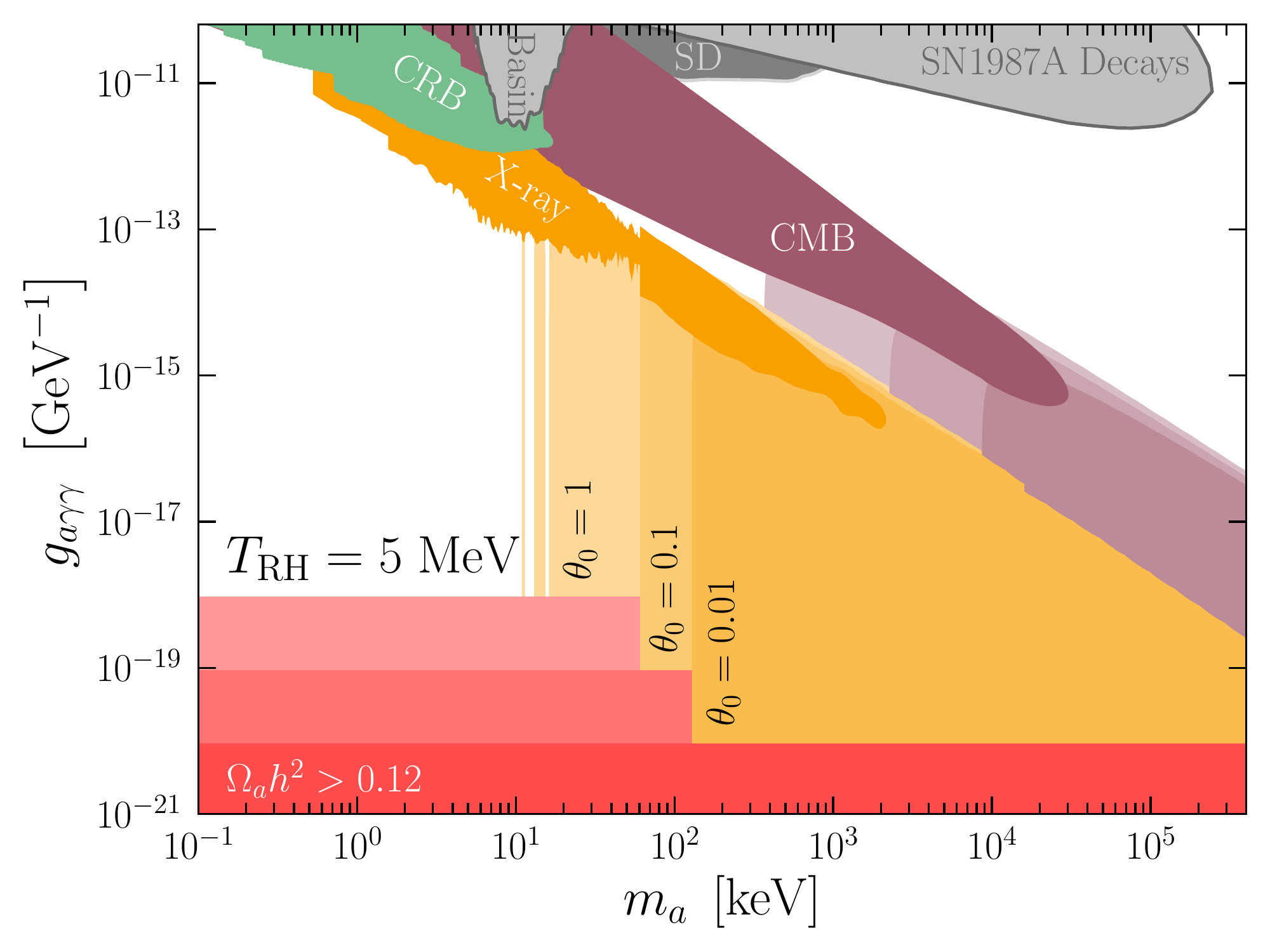}
\vspace{-0.4cm}
\caption{The analogue of \Fig{fig:Constraints} when the minimal misalignment contribution specified in \Eq{eq:Fa_mis} is combined with the irreducible freeze-in abundance.
Constraints are shown for $\theta_0 = 0$, $10^{-2}$, $10^{-1}$ and $1$ in decreasing levels of darkness.
The constraint from overclosing the Universe for these same three nonzero values of $\theta_0$ are shown in red.
}
\label{fig:Constraints_miss}
\vspace{-0.4cm}
\end{figure}

Once the axion field starts to oscillate, the axion energy density from misalignment will subsequently evolve as CDM.
To be explicit, $\rho_a^{\rm mis.} \propto R^{-3}$, where $R$ is the scale factor of the Universe.
The energy density in axions at the time of reheating will therefore be given by
\be
  \rho_{a,\RH}^{\rm mis.} = \frac{1}{2} m_a^2 f_a^2 \theta_0^2 \left(\frac{R_{\rm osc}}{R_\RH}\right)^3\!.
\ee
The dilution arising from the ratio of scale factors is highly dependent on the cosmological history between the time of initial oscillation and reheating.
Specifically, if we assume only a single form of energy dominates that era, and that its energy density scales proportional to $R^{-\beta}$, such that $R\propto H^{2/\beta}$, then the dilution factor is fixed by
\be
  \left( \frac{ R_{\rm osc} }{R_{\RH}} \right) 
  \simeq \left(\frac{m_a}{H(\TRH)}\right)^{-2/\beta}\!,
\ee
where we know $H(\TRH)$ since we know the energy density in radiation at this time.
From this analysis, we conclude that the minimal misalignment abundance arises for the smallest $\beta$.
For the three possible cosmologies enumerated above, the scale factor evolves according to $R^{-4}$, $R^{-3}$, and $R^{-6}$, respectively.
Therefore, the minimal abundance corresponds to a period of early matter domination before reheating.

A more detailed calculation in \Refc{Blinov:2019rhb}, found that for a period of early matter domination, the misalignment generated axion abundance is given by
\be
  {\cal F}^{\rm mis.}_a \simeq  \left(\frac{f_a \theta_0}{1.3\times 10^{15}~{\rm GeV}}\right)^2  \left(\frac{\TRH}{5~{\rm MeV}}\right)\!.
  \label{eq:Fa_mis}
\ee
If we assume the photon coupling has a characteristic value $\ga = \alpha/ 2\pi f_a$, then as discussed at the outset, we find that a larger misalignment contribution is generated for smaller couplings.
Misalignment would therefore allow us to probe smaller couplings, as shown in \Fig{fig:Constraints_miss}.
In particular, as ${\cal F}^{\rm mis.}_a \propto f_a^2$, whereas the decay rate is $\propto \ga^2 \propto f_a^{-2}$, the predicted flux only depends on $\ga$ through the exponential depletion factor $e^{-t/\tau_a}$, where $t$ corresponds to the specific probe, with $t=\tU$ or $t_\CMB$, for instance.
Once $\tau_a \gg t$, the dependence on $\ga$ vanishes entirely, other than the fact that eventually the axion energy density from \Eq{eq:Fa_mis} would overclose the Universe.

%%%%%%%%%
\section{Extension to Sterile Neutrinos and Other Light Dark Particles}
\label{sec:extension-Nu}
%%%%%%%%%

The argument outlined in this work is extremely general and can be readily extended to other states that could exist in the spectrum of our Universe.
In this section we will expand on this point, and consider the additional specific example of sterile neutrinos.

First, let us restate the central idea of our work.
Broadly, DM searches constrain a three dimensional parameter space spanned by mass, coupling to the SM, and DM fraction, $\{m,\,g,\,{\cal F}\}$.
Conventional approaches consider either a two-dimensional subset defined by the dominant DM of our Universe, ${\cal F} \sim 1$, or else consider smaller fractions.
Our argument is that ${\cal F}$ cannot be made arbitrarily small, and that there is an irreducible abundance (again assuming, for instance, that the state does not decay to invisible particles); there is a minimal abundance set by
\be
  {\cal F}_{\rm min} = \min_{\cal C} {\cal F}^{({\cal C})}(m,\,g),
\ee
where minimum is taken over the set of all cosmologies, ${\cal C}$, which match onto a conventional cosmology at $T = 5~{\rm MeV}$.
Any constraints obtained using the minimal fractional abundance are robust in the sense that for any other ${\cal C}$, the constraint will only ever strengthen.

We can extend this idea to constrain any other putative particle satisfying the following three conditions:
\begin{enumerate}
  \item The particle must be light ($m \lesssim 1~{\rm GeV}$);
  \item The particle must decay predominately to an observable SM final state; and
  \item The lifetime of the particle must be longer than the timescale associated with some observational probe.
\end{enumerate}
To expand on these, the first requirement arises as if the state is too heavy, its production will become heavily Boltzmann suppressed when the minimum $\TRH$ is adopted, whereas the second and third criteria are required in order to have an observable signature associated with the relic.
We note that in principle one can repurpose direct detection searches, in which case the particle need not decay (or if it does, must do so with $\tau_a \gtrsim \tU$), although then there must be a non-negligible scattering cross-section with the SM.

We now discuss how this method can be generalized to sterile neutrinos. 
Conventionally sterile neutrinos are generated by thermal freeze-out, through the decay of heavier particles, or via freeze-in through oscillations.
The irreducible abundance originates from freeze-in through a mass dependent combination of oscillations, inverse decay, and two-to-two production assuming the lowest possible reheating temperature. 
The explicit calculation of the relic density for this scenario was in fact already considered in \Refs{Gelmini:2008fq,Gelmini:2019wfp}, in the context of avoiding the strong cosmological and astrophysical constraints that exist on sterile neutrinos, and more generally studying the possible pre-BBN cosmology.
We will review the calculation of the abundance in those works, and constrain this irreducible abundance exactly as we did the axion in the main text, although we will not attempt an exhaustive analysis, discarding the possible constraints from BBN, $\Delta N_{\rm eff}$, and the CRB.

For simplicity, we focus on a sterile neutrino that mixes solely with $\nu_e$ (and we note that results can vary considerably when mixing with additional flavors is considered).
Accordingly, we consider only
\be
  \begin{pmatrix} \nu_e\\ \nu_{s} \end{pmatrix}
  = \begin{pmatrix} \cos \theta & \sin \theta \\ -\sin \theta & \cos \theta \end{pmatrix}
  \begin{pmatrix} \nu_1\\ \nu_2 \end{pmatrix}\!,
\ee
where $\nu_1$ and $\nu_2$ are the light and heavy mass eigenstates, respectively.
We will denote the mass of the mostly sterile mass eigenstate as $m_s$.
The dominant decay rates are the following~ \cite{Pal:1981rm,Essig:2013goa}:
\bea
  \Gamma(\nu_s \to 3\nu_a) =&~ 7.0\times 10^{-5}~{\rm s}^{-1}\,\sin^2(2\theta)\left(\frac{m_s}{\rm MeV}\right)^5\!,\\
  \Gamma(\nu_s \to \nu_e \gamma) =&~ 1.36\times 10^{-7}~{\rm s}^{-1}\sin^2(2\theta)\left(\frac{m_s}{\rm MeV}\right)^5\!, \\
  \Gamma(\nu_s \to \nu_e e^-e^+) =&~  4.2\times 10^{-5}~{\rm s}^{-1}\,\sin^2(2\theta)\left(\frac{m_s}{\rm MeV}\right)^5\!,
\eea
where we have ignored minor threshold corrections from the electron mass in the second decay channel and we have assumed a Majorana sterile neutrino.
The first two of the above decays are active throughout the entire mass range we consider, however the three body decay involving an electron-positron pair only opens for $m_s > 2m_e$.

\begin{figure}[!t]
    \centering
    \includegraphics[width=\textwidth]{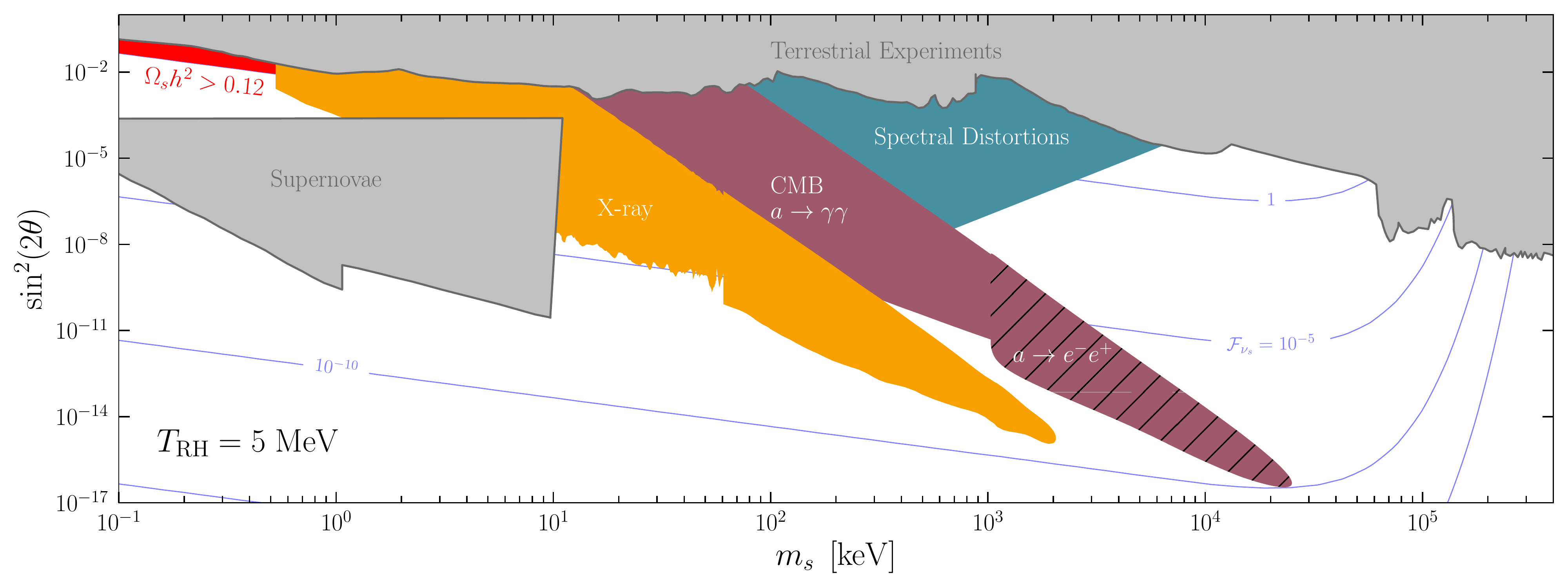}
    \vspace{-0.7cm}
    \caption{Irreducible constraints for sterile neutrinos.
    The {\color[rgb]{0.972549,0.627451,0.}\textbf{mustard}}, {\color[rgb]{0.623529, 0.34902, 0.419608}\textbf{velvet}}, and {\color[rgb]{0.275966, 0.563548, 0.632442}\textbf{turquoise}}  regions represent the same X-ray, CMB anisotropy, and CMB spectral distortions constraints.
    The first two of these are derived as for the axion, whereas for the spectral distortions we simply show the constraint from $\mu$ and $y$ distortions.
    We have also included constraints coming from over closure in {\textcolor{red}{\textbf{red}}}.
    The {\color[rgb]{0.552, 0.552, 0.552}\textbf{gray}} regions represents constraints obtained from terrestrial experiments~\cite{Abdurashitov:2017kka,Holzschuh:1999vy,Holzschuh:2000nj,Friedrich:2020nze,Borexino:2013bot,PIENU:2017wbj} and supernovae~\cite{Shi:1993ee}, both of which were taken from \Refc{Bolton:2019pcu}.
    Again, we have neglected bounds that would follow from BBN, $\Delta N_{\rm eff}$, and the CRB.
    }
    \label{fig:Sterile_Neutrinos_Constraints}
\end{figure}

An approximate analytical solution for the sterile neutrino abundance is provided in \Refc{Gelmini:2008fq}.
From that result, we can immediately compute ${\cal F}_{\nu_s} \equiv \max_T[\rho_{\nu_s}(T)/\rho_\DM(T)]$; contours of the result is shown in blue in \Fig{fig:Sterile_Neutrinos_Constraints}.
Before turning to the constraints, we emphasize two general points related to freeze-in via mixing.
\begin{enumerate}
  \item Generically, the production rate of sterile neutrinos through mixing peaks roughly at temperatures between 100 MeV and 1 GeV for sterile neutrino masses between 1 keV and 1 MeV~\cite{Gelmini:2019wfp}.
  In this sense, freeze-in through oscillations is already reasonably robust to modifications of standard cosmology for temperature above $\sim 1~{\rm GeV}$ since increasing the reheating temperature beyond this will not appreciably modify the abundance. 
  However, since the production rate after the peak scales as $T^3$, reheating at lower temperatures can reduce the abundance by several orders of magnitude.
  \item The production rate of sterile neutrinos through mixing strongly depends on any lepton asymmetries, as these allow Mikheyev–Smirnov–Wolfenstein (MSW) resonances to occur, thereby enhancing the production.
  Production enhanced by the MSW effect is referred to as the Shi-Fuller production mechanism~\cite{Shi:1998km}.
  We note that it is also possible for lepton asymmetries to lower the production rate; however, this will be irrelevant for the masses and temperatures we are considering, so we will focus on the most conservative scenario where there is no lepton asymmetry or resonant enhanced production.
  This is called the Dodelson-Widrow production mechanism~\cite{Dodelson:1993je}.
\end{enumerate}

The various constraints we considered for axions, outlined in Sec.~\ref{sec:constraints}, can be applied to sterile neutrinos up to two minor modifications.
First, the decay channels which lead to a detectable signal are the radiative decay $\nu_s \to \nu_e \gamma$, and, if $m_s > 2m_e$, the three-body decay, $\nu_s \to \nu_e e^+ e^-$.
However, the decay channel which determines the total lifetime of the sterile neutrino will be dominantly the unobserved decay to three active neutrinos.
The resulting irreducible constraints that arise from CMB anisotropies, X-ray measurements, and spectral distortions are shown in \Fig{fig:Sterile_Neutrinos_Constraints}.
The spectral distortion constraints are obtained only from $\mu$ and $y$ distortions, and we further have only considered the bound due to a radiative photon decay.
We note that constraints coming from decays to positron-electron pairs when $m_s> 2m_e$ will likely exclude additional parameter space from spectral distortions, although we have not included these here.
For CMB constraints, however, we have considered both observable decay channels; for this case, we have assumed only half of the sterile neutrinos energy goes into the electron positron pair as a conservative approximation.
Finally, we have included the constraints that arise from terrestrial experiments and supernovae which are independent of cosmology, both taken from \Refc{Bolton:2019pcu}.

\end{document}